\documentclass[9pt,conference]{IEEEtran}

\pdfoutput=1

\usepackage[hidelinks]{hyperref}
\usepackage[utf8]{inputenc}
\usepackage[T1]{fontenc}
\usepackage[english]{babel}
\usepackage[ruled, linesnumbered]{algorithm2e}
\usepackage{cite}
\usepackage{pgfplots}
\usepackage{xcolor}
\usepackage{algpseudocode}
\usepackage{array}
\usepackage{url}
\usepackage{colortbl}
\usepackage{tabularx}
\usepackage{multirow}
\usepackage{dcolumn}
\usepackage{calc}
\usepackage{pgf}
\usepackage{graphicx}
\usepackage{color}
\usepackage{tabto}
\usepackage{multicol}
\usepackage{ragged2e}
\usepackage{enumerate}
\usepackage{mathtools}
\usepackage{cuted}
\usepackage{bm}
\usepackage{lipsum}
\usepackage{siunitx}
\usepackage{placeins}
\usepackage{booktabs}
\usepackage{wrapfig}
\usepackage{empheq}
\usepackage[pdf]{graphviz}
\usepackage{hhline}
\usepackage{xargs}
\usepackage{todonotes}
\usepackage{boldline}
\usepackage{diagbox}
\usepackage[font=small]{subcaption}
\usepackage{pgfplotstable}
\usepackage{dblfloatfix}
\usepackage{pdfcomment}
\usepackage{capt-of}
\usepackage{censor}
\usepackage{cleveref,nameref}
\usepackage{amsmath}
\usepackage{amssymb}
\usepackage{amsthm}
\usepackage{csquotes}
\usepackage{pifont}
\usepackage{caption}
\usepackage[acronym,nohypertypes={acronym}]{glossaries}
\usepackage{pdfcomment}
\usepackage{setspace}
\let\labelindent\relax
\usepackage{enumitem}
\usepackage{titlecaps}
\usepackage{cancel}
\usepackage{arydshln}
\usepackage[multiple]{footmisc}
\usepackage{moreverb}
\usepackage{scalerel}
\usepackage{tikz}

\usetikzlibrary{svg.path}

\definecolor{orcidlogocol}{HTML}{A6CE39}
\tikzset{
  orcidlogo/.pic={
    \fill[orcidlogocol] svg{M256,128c0,70.7-57.3,128-128,128C57.3,256,0,198.7,0,128C0,57.3,57.3,0,128,0C198.7,0,256,57.3,256,128z};
    \fill[white] svg{M86.3,186.2H70.9V79.1h15.4v48.4V186.2z}
                 svg{M108.9,79.1h41.6c39.6,0,57,28.3,57,53.6c0,27.5-21.5,53.6-56.8,53.6h-41.8V79.1z M124.3,172.4h24.5c34.9,0,42.9-26.5,42.9-39.7c0-21.5-13.7-39.7-43.7-39.7h-23.7V172.4z}
                 svg{M88.7,56.8c0,5.5-4.5,10.1-10.1,10.1c-5.6,0-10.1-4.6-10.1-10.1c0-5.6,4.5-10.1,10.1-10.1C84.2,46.7,88.7,51.3,88.7,56.8z};
  }
}

\newcommand\orcid[1]{\href{https://orcid.org/#1}{\mbox{\scalerel*{
\begin{tikzpicture}[yscale=-1,transform shape]
	\scalebox{0.7}{\pic{orcidlogo};}
\end{tikzpicture}
}{|}}}\hspace{-1mm}}

\newtheorem{thm}{Theorem}

\crefname{lemma}{lemma}{lemmas}
\Crefname{lemma}{Lemma}{Lemmas}
\crefname{thm}{theorem}{theorems}
\Crefname{thm}{Theorem}{Theorems}

\makeatletter
\newcommand{\myref}[1]{\cref{#1}\mynameref{#1}{\csname r@#1\endcsname}}
\newcommand{\Myref}[1]{\Cref{#1}\mynameref{#1}{\csname r@#1\endcsname}}

\def\mynameref#1#2{%
  \begingroup
    \edef\@mytxt{#2}%
    \edef\@mytst{\expandafter\@thirdoffive\@mytxt}%
    \ifx\@mytst\empty\else
    \space(\nameref{#1})\fi
  \endgroup
}
\makeatother

\SetAlCapNameFnt{\footnotesize}
\SetAlCapFnt{\footnotesize}

\definecolor{heatmap1}{RGB}{67,149,187}
\definecolor{heatmap2}{RGB}{156,201,220}
\definecolor{heatmap3}{RGB}{255,255,255}
\definecolor{heatmap4}{RGB}{237,192,161}
\definecolor{heatmap5}{RGB}{218,132,75}

\definecolor{heatmap_imp_1}{RGB}{53,226,99}
\definecolor{heatmap_imp_2}{RGB}{215,246,15}
\definecolor{heatmap_imp_3}{RGB}{255,255,0}
\definecolor{heatmap_imp_4}{RGB}{255,179,0}
\definecolor{heatmap_imp_5}{RGB}{218,132,75}

\definecolor{mycherry}{RGB}{174,049,073}
\definecolor{mycallow}{RGB}{119,172,048}
\definecolor{myindigo}{RGB}{077,190,238}
\definecolor{myyellow}{RGB}{237,177,032}
\definecolor{mypurple}{RGB}{126,047,142}
\definecolor{myorange}{RGB}{217,083,025}

\definecolor{blue_fau}{rgb}{0.00,0.22,0.40}
\definecolor{gray_fau}{rgb}{0.59,0.64,0.68}
\definecolor{plot_fau}{rgb}{0.24,0.42,0.87}

\definecolor{airforceblue}{rgb}{0.36, 0.54, 0.66}		
\definecolor{aliceblue}{rgb}{0.94, 0.97, 1.0}			
\definecolor{alizarin}{rgb}{0.82, 0.1, 0.26}			
\definecolor{almond}{rgb}{0.94, 0.87, 0.8}			
\definecolor{amaranth}{rgb}{0.9, 0.17, 0.31}			
\definecolor{amber}{rgb}{1.0, 0.75, 0.0}			
\definecolor{amber(sae/ece)}{rgb}{1.0, 0.49, 0.0}		
\definecolor{americanrose}{rgb}{1.0, 0.01, 0.24}		
\definecolor{amethyst}{rgb}{0.6, 0.4, 0.8}			
\definecolor{anti-flashwhite}{rgb}{0.95, 0.95, 0.96}		
\definecolor{antiquebrass}{rgb}{0.8, 0.58, 0.46}		
\definecolor{antiquefuchsia}{rgb}{0.57, 0.36, 0.51}		
\definecolor{antiquewhite}{rgb}{0.98, 0.92, 0.84}		
\definecolor{ao}{rgb}{0.0, 0.0, 1.0}				
\definecolor{ao(english)}{rgb}{0.0, 0.5, 0.0}			
\definecolor{applegreen}{rgb}{0.55, 0.71, 0.0}			
\definecolor{apricot}{rgb}{0.98, 0.81, 0.69}			
\definecolor{aqua}{rgb}{0.0, 1.0, 1.0}				
\definecolor{aquamarine}{rgb}{0.5, 1.0, 0.83}			
\definecolor{armygreen}{rgb}{0.29, 0.33, 0.13}			
\definecolor{arsenic}{rgb}{0.23, 0.27, 0.29}			
\definecolor{arylideyellow}{rgb}{0.91, 0.84, 0.42}		
\definecolor{ashgrey}{rgb}{0.7, 0.75, 0.71}			
\definecolor{asparagus}{rgb}{0.53, 0.66, 0.42}			
\definecolor{atomictangerine}{rgb}{1.0, 0.6, 0.4}		
\definecolor{auburn}{rgb}{0.43, 0.21, 0.1}			
\definecolor{aureolin}{rgb}{0.99, 0.93, 0.0}			
\definecolor{aurometalsaurus}{rgb}{0.43, 0.5, 0.5}		
\definecolor{awesome}{rgb}{1.0, 0.13, 0.32}			
\definecolor{azure(colorwheel)}{rgb}{0.0, 0.5, 1.0}		
\definecolor{azure(web)(azuremist)}{rgb}{0.94, 1.0, 1.0}	
\definecolor{babyblue}{rgb}{0.54, 0.81, 0.94}			
\definecolor{babyblueeyes}{rgb}{0.63, 0.79, 0.95}		
\definecolor{babypink}{rgb}{0.96, 0.76, 0.76}			
\definecolor{ballblue}{rgb}{0.13, 0.67, 0.8}			
\definecolor{bananamania}{rgb}{0.98, 0.91, 0.71}		
\definecolor{bananayellow}{rgb}{1.0, 0.88, 0.21}		
\definecolor{battleshipgrey}{rgb}{0.52, 0.52, 0.51}		
\definecolor{bazaar}{rgb}{0.6, 0.47, 0.48}			
\definecolor{beaublue}{rgb}{0.74, 0.83, 0.9}			
\definecolor{beaver}{rgb}{0.62, 0.51, 0.44}			
\definecolor{beige}{rgb}{0.96, 0.96, 0.86}			
\definecolor{bisque}{rgb}{1.0, 0.89, 0.77}			
\definecolor{bistre}{rgb}{0.24, 0.17, 0.12}			
\definecolor{bittersweet}{rgb}{1.0, 0.44, 0.37}			
\definecolor{black}{rgb}{0.0, 0.0, 0.0}				
\definecolor{blanchedalmond}{rgb}{1.0, 0.92, 0.8}		
\definecolor{bleudefrance}{rgb}{0.19, 0.55, 0.91}		
\definecolor{blizzardblue}{rgb}{0.67, 0.9, 0.93}		
\definecolor{blond}{rgb}{0.98, 0.94, 0.75}			
\definecolor{blue}{rgb}{0.0, 0.0, 1.0}				
\definecolor{blue(munsell)}{rgb}{0.0, 0.5, 0.69}		
\definecolor{blue(ncs)}{rgb}{0.0, 0.53, 0.74}			
\definecolor{blue(pigment)}{rgb}{0.2, 0.2, 0.6}			
\definecolor{blue(ryb)}{rgb}{0.01, 0.28, 1.0}			
\definecolor{bluebell}{rgb}{0.64, 0.64, 0.82}			
\definecolor{bluegray}{rgb}{0.4, 0.6, 0.8}			
\definecolor{blue-green}{rgb}{0.0, 0.87, 0.87}			
\definecolor{blue-violet}{rgb}{0.54, 0.17, 0.89}		
\definecolor{blush}{rgb}{0.87, 0.36, 0.51}			
\definecolor{bole}{rgb}{0.47, 0.27, 0.23}			
\definecolor{bondiblue}{rgb}{0.0, 0.58, 0.71}			
\definecolor{bostonuniversityred}{rgb}{0.8, 0.0, 0.0}		
\definecolor{brandeisblue}{rgb}{0.0, 0.44, 1.0}			
\definecolor{brass}{rgb}{0.71, 0.65, 0.26}			
\definecolor{brickred}{rgb}{0.8, 0.25, 0.33}			
\definecolor{brightcerulean}{rgb}{0.11, 0.67, 0.84}		
\definecolor{brightgreen}{rgb}{0.4, 1.0, 0.0}			
\definecolor{brightlavender}{rgb}{0.75, 0.58, 0.89}		
\definecolor{brightmaroon}{rgb}{0.76, 0.13, 0.28}		
\definecolor{brightpink}{rgb}{1.0, 0.0, 0.5}			
\definecolor{brightturquoise}{rgb}{0.03, 0.91, 0.87}		
\definecolor{brightube}{rgb}{0.82, 0.62, 0.91}			
\definecolor{brilliantlavender}{rgb}{0.96, 0.73, 1.0}		
\definecolor{brilliantrose}{rgb}{1.0, 0.33, 0.64}		
\definecolor{brinkpink}{rgb}{0.98, 0.38, 0.5}			
\definecolor{britishracinggreen}{rgb}{0.0, 0.26, 0.15}		
\definecolor{bronze}{rgb}{0.8, 0.5, 0.2}			
\definecolor{brown(traditional)}{rgb}{0.59, 0.29, 0.0}		
\definecolor{brown(web)}{rgb}{0.65, 0.16, 0.16}			
\definecolor{bubblegum}{rgb}{0.99, 0.76, 0.8}			
\definecolor{bubbles}{rgb}{0.91, 1.0, 1.0}			
\definecolor{buff}{rgb}{0.94, 0.86, 0.51}			
\definecolor{bulgarianrose}{rgb}{0.28, 0.02, 0.03}		
\definecolor{burgundy}{rgb}{0.5, 0.0, 0.13}			
\definecolor{burlywood}{rgb}{0.87, 0.72, 0.53}			
\definecolor{burntorange}{rgb}{0.8, 0.33, 0.0}			
\definecolor{burntsienna}{rgb}{0.91, 0.45, 0.32}		
\definecolor{burntumber}{rgb}{0.54, 0.2, 0.14}			
\definecolor{byzantine}{rgb}{0.74, 0.2, 0.64}			
\definecolor{byzantium}{rgb}{0.44, 0.16, 0.39}			
\definecolor{cadet}{rgb}{0.33, 0.41, 0.47}			
\definecolor{cadetblue}{rgb}{0.37, 0.62, 0.63}			
\definecolor{cadetgrey}{rgb}{0.57, 0.64, 0.69}			
\definecolor{cadmiumgreen}{rgb}{0.0, 0.42, 0.24}		
\definecolor{cadmiumorange}{rgb}{0.93, 0.53, 0.18}		
\definecolor{cadmiumred}{rgb}{0.89, 0.0, 0.13}			
\definecolor{cadmiumyellow}{rgb}{1.0, 0.96, 0.0}		
\definecolor{calpolypomonagreen}{rgb}{0.12, 0.3, 0.17}		
\definecolor{cambridgeblue}{rgb}{0.64, 0.76, 0.68}		
\definecolor{camel}{rgb}{0.76, 0.6, 0.42}			
\definecolor{camouflagegreen}{rgb}{0.47, 0.53, 0.42}		
\definecolor{canaryyellow}{rgb}{1.0, 0.94, 0.0}			
\definecolor{candyapplered}{rgb}{1.0, 0.03, 0.0}		
\definecolor{candypink}{rgb}{0.89, 0.44, 0.48}			
\definecolor{capri}{rgb}{0.0, 0.75, 1.0}			
\definecolor{caputmortuum}{rgb}{0.35, 0.15, 0.13}		
\definecolor{cardinal}{rgb}{0.77, 0.12, 0.23}			
\definecolor{caribbeangreen}{rgb}{0.0, 0.8, 0.6}		
\definecolor{carmine}{rgb}{0.59, 0.0, 0.09}			
\definecolor{carminepink}{rgb}{0.92, 0.3, 0.26}			
\definecolor{carminered}{rgb}{1.0, 0.0, 0.22}			
\definecolor{carnationpink}{rgb}{1.0, 0.65, 0.79}		
\definecolor{carnelian}{rgb}{0.7, 0.11, 0.11}			
\definecolor{carolinablue}{rgb}{0.6, 0.73, 0.89}		
\definecolor{carrotorange}{rgb}{0.93, 0.57, 0.13}		
\definecolor{ceil}{rgb}{0.57, 0.63, 0.81}			
\definecolor{celadon}{rgb}{0.67, 0.88, 0.69}			
\definecolor{celestialblue}{rgb}{0.29, 0.59, 0.82}		
\definecolor{cerise}{rgb}{0.87, 0.19, 0.39}			
\definecolor{cerisepink}{rgb}{0.93, 0.23, 0.51}			
\definecolor{cerulean}{rgb}{0.0, 0.48, 0.65}			
\definecolor{ceruleanblue}{rgb}{0.16, 0.32, 0.75}		
\definecolor{chamoisee}{rgb}{0.63, 0.47, 0.35}			
\definecolor{champagne}{rgb}{0.97, 0.91, 0.81}			
\definecolor{charcoal}{rgb}{0.21, 0.27, 0.31}			
\definecolor{chartreuse(traditional)}{rgb}{0.87, 1.0, 0.0}	
\definecolor{chartreuse(web)}{rgb}{0.5, 1.0, 0.0}		
\definecolor{cherryblossompink}{rgb}{1.0, 0.72, 0.77}		
\definecolor{chestnut}{rgb}{0.8, 0.36, 0.36}			
\definecolor{chocolate(traditional)}{rgb}{0.48, 0.25, 0.0}	
\definecolor{chocolate(web)}{rgb}{0.82, 0.41, 0.12}		
\definecolor{chromeyellow}{rgb}{1.0, 0.65, 0.0}			
\definecolor{cinereous}{rgb}{0.6, 0.51, 0.48}			
\definecolor{cinnabar}{rgb}{0.89, 0.26, 0.2}			
\definecolor{cinnamon}{rgb}{0.82, 0.41, 0.12}			
\definecolor{citrine}{rgb}{0.89, 0.82, 0.04}			
\definecolor{classicrose}{rgb}{0.98, 0.8, 0.91}			
\definecolor{cobalt}{rgb}{0.0, 0.28, 0.67}			
\definecolor{cocoabrown}{rgb}{0.82, 0.41, 0.12}			
\definecolor{columbiablue}{rgb}{0.61, 0.87, 1.0}		
\definecolor{coolblack}{rgb}{0.0, 0.18, 0.39}			
\definecolor{coolgrey}{rgb}{0.55, 0.57, 0.67}			
\definecolor{copper}{rgb}{0.72, 0.45, 0.2}			
\definecolor{copperrose}{rgb}{0.6, 0.4, 0.4}			
\definecolor{coquelicot}{rgb}{1.0, 0.22, 0.0}			
\definecolor{coral}{rgb}{1.0, 0.5, 0.31}			
\definecolor{coralpink}{rgb}{0.97, 0.51, 0.47}			
\definecolor{coralred}{rgb}{1.0, 0.25, 0.25}			
\definecolor{cordovan}{rgb}{0.54, 0.25, 0.27}			
\definecolor{corn}{rgb}{0.98, 0.93, 0.36}			
\definecolor{cornellred}{rgb}{0.7, 0.11, 0.11}			
\definecolor{cornflowerblue}{rgb}{0.39, 0.58, 0.93}		
\definecolor{cornsilk}{rgb}{1.0, 0.97, 0.86}			
\definecolor{cosmiclatte}{rgb}{1.0, 0.97, 0.91}			
\definecolor{cottoncandy}{rgb}{1.0, 0.74, 0.85}			
\definecolor{cream}{rgb}{1.0, 0.99, 0.82}			
\definecolor{crimson}{rgb}{0.86, 0.08, 0.24}			
\definecolor{crimsonglory}{rgb}{0.75, 0.0, 0.2}			
\definecolor{cyan}{rgb}{0.0, 1.0, 1.0}				
\definecolor{cyan(process)}{rgb}{0.0, 0.72, 0.92}		
\definecolor{daffodil}{rgb}{1.0, 1.0, 0.19}			
\definecolor{dandelion}{rgb}{0.94, 0.88, 0.19}			
\definecolor{darkblue}{rgb}{0.0, 0.0, 0.55}			
\definecolor{darkbrown}{rgb}{0.4, 0.26, 0.13}			
\definecolor{darkbyzantium}{rgb}{0.36, 0.22, 0.33}		
\definecolor{darkcandyapplered}{rgb}{0.64, 0.0, 0.0}		
\definecolor{darkcerulean}{rgb}{0.03, 0.27, 0.49}		
\definecolor{darkchampagne}{rgb}{0.76, 0.7, 0.5}		
\definecolor{darkchestnut}{rgb}{0.6, 0.41, 0.38}		
\definecolor{darkcoral}{rgb}{0.8, 0.36, 0.27}			
\definecolor{darkcyan}{rgb}{0.0, 0.55, 0.55}			
\definecolor{darkelectricblue}{rgb}{0.33, 0.41, 0.47}		
\definecolor{darkgoldenrod}{rgb}{0.72, 0.53, 0.04}		
\definecolor{darkgray}{rgb}{0.66, 0.66, 0.66}			
\definecolor{darkgreen}{rgb}{0.0, 0.2, 0.13}			
\definecolor{darkjunglegreen}{rgb}{0.1, 0.14, 0.13}		
\definecolor{darkkhaki}{rgb}{0.74, 0.72, 0.42}			
\definecolor{darklava}{rgb}{0.28, 0.24, 0.2}			
\definecolor{darklavender}{rgb}{0.45, 0.31, 0.59}		
\definecolor{darkmagenta}{rgb}{0.55, 0.0, 0.55}			
\definecolor{darkmidnightblue}{rgb}{0.0, 0.2, 0.4}		
\definecolor{darkolivegreen}{rgb}{0.33, 0.42, 0.18}		
\definecolor{darkorange}{rgb}{1.0, 0.55, 0.0}			
\definecolor{darkorchid}{rgb}{0.6, 0.2, 0.8}			
\definecolor{darkpastelblue}{rgb}{0.47, 0.62, 0.8}		
\definecolor{darkpastelgreen}{rgb}{0.01, 0.75, 0.24}		
\definecolor{darkpastelpurple}{rgb}{0.59, 0.44, 0.84}		
\definecolor{darkpastelred}{rgb}{0.76, 0.23, 0.13}		
\definecolor{darkpink}{rgb}{0.91, 0.33, 0.5}			
\definecolor{darkpowderblue}{rgb}{0.0, 0.2, 0.6}		
\definecolor{darkraspberry}{rgb}{0.53, 0.15, 0.34}		
\definecolor{darkred}{rgb}{0.55, 0.0, 0.0}			
\definecolor{darksalmon}{rgb}{0.91, 0.59, 0.48}			
\definecolor{darkscarlet}{rgb}{0.34, 0.01, 0.1}			
\definecolor{darkseagreen}{rgb}{0.56, 0.74, 0.56}		
\definecolor{darksienna}{rgb}{0.24, 0.08, 0.08}			
\definecolor{darkslateblue}{rgb}{0.28, 0.24, 0.55}		
\definecolor{darkslategray}{rgb}{0.18, 0.31, 0.31}		
\definecolor{darkspringgreen}{rgb}{0.09, 0.45, 0.27}		
\definecolor{darktan}{rgb}{0.57, 0.51, 0.32}			
\definecolor{darktangerine}{rgb}{1.0, 0.66, 0.07}		
\definecolor{darktaupe}{rgb}{0.28, 0.24, 0.2}			
\definecolor{darkterracotta}{rgb}{0.8, 0.31, 0.36}		
\definecolor{darkturquoise}{rgb}{0.0, 0.81, 0.82}		
\definecolor{darkviolet}{rgb}{0.58, 0.0, 0.83}			
\definecolor{dartmouthgreen}{rgb}{0.05, 0.5, 0.06}		
\definecolor{davy\'sgrey}{rgb}{0.33, 0.33, 0.33}		
\definecolor{debianred}{rgb}{0.84, 0.04, 0.33}			
\definecolor{deepcarmine}{rgb}{0.66, 0.13, 0.24}		
\definecolor{deepcarminepink}{rgb}{0.94, 0.19, 0.22}		
\definecolor{deepcarrotorange}{rgb}{0.91, 0.41, 0.17}		
\definecolor{deepcerise}{rgb}{0.85, 0.2, 0.53}			
\definecolor{deepchampagne}{rgb}{0.98, 0.84, 0.65}		
\definecolor{deepchestnut}{rgb}{0.73, 0.31, 0.28}		
\definecolor{deepfuchsia}{rgb}{0.76, 0.33, 0.76}		
\definecolor{deepjunglegreen}{rgb}{0.0, 0.29, 0.29}		
\definecolor{deeplilac}{rgb}{0.6, 0.33, 0.73}			
\definecolor{deepmagenta}{rgb}{0.8, 0.0, 0.8}			
\definecolor{deeppeach}{rgb}{1.0, 0.8, 0.64}			
\definecolor{deeppink}{rgb}{1.0, 0.08, 0.58}			
\definecolor{deepsaffron}{rgb}{1.0, 0.6, 0.2}			
\definecolor{deepskyblue}{rgb}{0.0, 0.75, 1.0}			
\definecolor{denim}{rgb}{0.08, 0.38, 0.74}			
\definecolor{desert}{rgb}{0.76, 0.6, 0.42}			
\definecolor{desertsand}{rgb}{0.93, 0.79, 0.69}			
\definecolor{dimgray}{rgb}{0.41, 0.41, 0.41}			
\definecolor{dodgerblue}{rgb}{0.12, 0.56, 1.0}			
\definecolor{dogwoodrose}{rgb}{0.84, 0.09, 0.41}		
\definecolor{dollarbill}{rgb}{0.52, 0.73, 0.4}			
\definecolor{drab}{rgb}{0.59, 0.44, 0.09}			
\definecolor{dukeblue}{rgb}{0.0, 0.0, 0.61}			
\definecolor{earthyellow}{rgb}{0.88, 0.66, 0.37}		
\definecolor{ecru}{rgb}{0.76, 0.7, 0.5}				
\definecolor{eggplant}{rgb}{0.38, 0.25, 0.32}			
\definecolor{eggshell}{rgb}{0.94, 0.92, 0.84}			
\definecolor{egyptianblue}{rgb}{0.06, 0.2, 0.65}		
\definecolor{electricblue}{rgb}{0.49, 0.98, 1.0}		
\definecolor{electriccrimson}{rgb}{1.0, 0.0, 0.25}		
\definecolor{electriccyan}{rgb}{0.0, 1.0, 1.0}			
\definecolor{electricgreen}{rgb}{0.0, 1.0, 0.0}			
\definecolor{electricindigo}{rgb}{0.44, 0.0, 1.0}		
\definecolor{electriclavender}{rgb}{0.96, 0.73, 1.0}		
\definecolor{electriclime}{rgb}{0.8, 1.0, 0.0}			
\definecolor{electricpurple}{rgb}{0.75, 0.0, 1.0}		
\definecolor{electricultramarine}{rgb}{0.25, 0.0, 1.0}		
\definecolor{electricviolet}{rgb}{0.56, 0.0, 1.0}		
\definecolor{electricyellow}{rgb}{1.0, 1.0, 0.0}		
\definecolor{emerald}{rgb}{0.31, 0.78, 0.47}			
\definecolor{etonblue}{rgb}{0.59, 0.78, 0.64}			
\definecolor{fallow}{rgb}{0.76, 0.6, 0.42}			
\definecolor{falured}{rgb}{0.5, 0.09, 0.09}			
\definecolor{fandango}{rgb}{0.71, 0.2, 0.54}			
\definecolor{fashionfuchsia}{rgb}{0.96, 0.0, 0.63}		
\definecolor{fawn}{rgb}{0.9, 0.67, 0.44}			
\definecolor{feldgrau}{rgb}{0.3, 0.36, 0.33}			
\definecolor{ferngreen}{rgb}{0.31, 0.47, 0.26}			
\definecolor{ferrarired}{rgb}{1.0, 0.11, 0.0}			
\definecolor{fielddrab}{rgb}{0.42, 0.33, 0.12}			
\definecolor{firebrick}{rgb}{0.7, 0.13, 0.13}			
\definecolor{fireenginered}{rgb}{0.81, 0.09, 0.13}		
\definecolor{flame}{rgb}{0.89, 0.35, 0.13}			
\definecolor{flamingopink}{rgb}{0.99, 0.56, 0.67}		
\definecolor{flavescent}{rgb}{0.97, 0.91, 0.56}			
\definecolor{flax}{rgb}{0.93, 0.86, 0.51}			
\definecolor{floralwhite}{rgb}{1.0, 0.98, 0.94}			
\definecolor{fluorescentorange}{rgb}{1.0, 0.75, 0.0}		
\definecolor{fluorescentpink}{rgb}{1.0, 0.08, 0.58}		
\definecolor{fluorescentyellow}{rgb}{0.8, 1.0, 0.0}		
\definecolor{folly}{rgb}{1.0, 0.0, 0.31}			
\definecolor{forestgreen(traditional)}{rgb}{0.0, 0.27, 0.13}	
\definecolor{forestgreen(web)}{rgb}{0.13, 0.55, 0.13}		
\definecolor{frenchbeige}{rgb}{0.65, 0.48, 0.36}		
\definecolor{frenchblue}{rgb}{0.0, 0.45, 0.73}			
\definecolor{frenchlilac}{rgb}{0.53, 0.38, 0.56}		
\definecolor{frenchrose}{rgb}{0.96, 0.29, 0.54}			
\definecolor{fuchsia}{rgb}{1.0, 0.0, 1.0}			
\definecolor{fuchsiapink}{rgb}{1.0, 0.47, 1.0}			
\definecolor{fulvous}{rgb}{0.86, 0.52, 0.0}			
\definecolor{fuzzywuzzy}{rgb}{0.8, 0.4, 0.4}			
\definecolor{gainsboro}{rgb}{0.86, 0.86, 0.86}			
\definecolor{gamboge}{rgb}{0.89, 0.61, 0.06}			
\definecolor{ghostwhite}{rgb}{0.97, 0.97, 1.0}			
\definecolor{ginger}{rgb}{0.69, 0.4, 0.0}			
\definecolor{glaucous}{rgb}{0.38, 0.51, 0.71}			
\definecolor{gold(metallic)}{rgb}{0.83, 0.69, 0.22}		
\definecolor{gold(web)(golden)}{rgb}{1.0, 0.84, 0.0}		
\definecolor{goldenbrown}{rgb}{0.6, 0.4, 0.08}			
\definecolor{goldenpoppy}{rgb}{0.99, 0.76, 0.0}			
\definecolor{goldenyellow}{rgb}{1.0, 0.87, 0.0}			
\definecolor{goldenrod}{rgb}{0.85, 0.65, 0.13}			
\definecolor{grannysmithapple}{rgb}{0.66, 0.89, 0.63}		
\definecolor{gray}{rgb}{0.5, 0.5, 0.5}				
\definecolor{gray(html/cssgray)}{rgb}{0.5, 0.5, 0.5}		
\definecolor{gray(x11gray)}{rgb}{0.75, 0.75, 0.75}		
\definecolor{gray-asparagus}{rgb}{0.27, 0.35, 0.27}		
\definecolor{green(colorwheel)(x11green)}{rgb}{0.0, 1.0, 0.0}	
\definecolor{green(html/cssgreen)}{rgb}{0.0, 0.5, 0.0}		
\definecolor{green(munsell)}{rgb}{0.0, 0.66, 0.47}		
\definecolor{green(ncs)}{rgb}{0.0, 0.62, 0.42}			
\definecolor{green(pigment)}{rgb}{0.0, 0.65, 0.31}		
\definecolor{green(ryb)}{rgb}{0.4, 0.69, 0.2}			
\definecolor{green-yellow}{rgb}{0.68, 1.0, 0.18}		
\definecolor{grullo}{rgb}{0.66, 0.6, 0.53}			
\definecolor{guppiegreen}{rgb}{0.0, 1.0, 0.5}			
\definecolor{halayaube}{rgb}{0.4, 0.22, 0.33}			
\definecolor{hanblue}{rgb}{0.27, 0.42, 0.81}			
\definecolor{hanpurple}{rgb}{0.32, 0.09, 0.98}			
\definecolor{hansayellow}{rgb}{0.91, 0.84, 0.42}		
\definecolor{harlequin}{rgb}{0.25, 1.0, 0.0}			
\definecolor{harvardcrimson}{rgb}{0.79, 0.0, 0.09}		
\definecolor{harvestgold}{rgb}{0.85, 0.57, 0.0}			
\definecolor{heartgold}{rgb}{0.5, 0.5, 0.0}			
\definecolor{heliotrope}{rgb}{0.87, 0.45, 1.0}			
\definecolor{hollywoodcerise}{rgb}{0.96, 0.0, 0.63}		
\definecolor{honeydew}{rgb}{0.94, 1.0, 0.94}			
\definecolor{hooker\'sgreen}{rgb}{0.0, 0.44, 0.0}		
\definecolor{hotmagenta}{rgb}{1.0, 0.11, 0.81}			
\definecolor{hotpink}{rgb}{1.0, 0.41, 0.71}			
\definecolor{huntergreen}{rgb}{0.21, 0.37, 0.23}		
\definecolor{iceberg}{rgb}{0.44, 0.65, 0.82}			
\definecolor{icterine}{rgb}{0.99, 0.97, 0.37}			
\definecolor{inchworm}{rgb}{0.7, 0.93, 0.36}			
\definecolor{indiagreen}{rgb}{0.07, 0.53, 0.03}			
\definecolor{indianred}{rgb}{0.8, 0.36, 0.36}			
\definecolor{indianyellow}{rgb}{0.89, 0.66, 0.34}		
\definecolor{indigo(dye)}{rgb}{0.0, 0.25, 0.42}			
\definecolor{indigo(web)}{rgb}{0.29, 0.0, 0.51}			
\definecolor{internationalkleinblue}{rgb}{0.0, 0.18, 0.65}	
\definecolor{internationalorange}{rgb}{1.0, 0.31, 0.0}		
\definecolor{iris}{rgb}{0.35, 0.31, 0.81}			
\definecolor{isabelline}{rgb}{0.96, 0.94, 0.93}			
\definecolor{islamicgreen}{rgb}{0.0, 0.56, 0.0}			
\definecolor{ivory}{rgb}{1.0, 1.0, 0.94}			
\definecolor{jade}{rgb}{0.0, 0.66, 0.42}			
\definecolor{jasper}{rgb}{0.84, 0.23, 0.24}			
\definecolor{jazzberryjam}{rgb}{0.65, 0.04, 0.37}		
\definecolor{jonquil}{rgb}{0.98, 0.85, 0.37}			
\definecolor{junebud}{rgb}{0.74, 0.85, 0.34}			
\definecolor{junglegreen}{rgb}{0.16, 0.67, 0.53}		
\definecolor{kellygreen}{rgb}{0.3, 0.73, 0.09}			
\definecolor{khaki(html/css)(khaki)}{rgb}{0.76, 0.69, 0.57}	
\definecolor{khaki(x11)(lightkhaki)}{rgb}{0.94, 0.9, 0.55}	
\definecolor{lasallegreen}{rgb}{0.03, 0.47, 0.19}		
\definecolor{languidlavender}{rgb}{0.84, 0.79, 0.87}		
\definecolor{lapislazuli}{rgb}{0.15, 0.38, 0.61}		
\definecolor{laserlemon}{rgb}{1.0, 1.0, 0.13}			
\definecolor{lava}{rgb}{0.81, 0.06, 0.13}			
\definecolor{lavender(floral)}{rgb}{0.71, 0.49, 0.86}		
\definecolor{lavender(web)}{rgb}{0.9, 0.9, 0.98}		
\definecolor{lavenderblue}{rgb}{0.8, 0.8, 1.0}			
\definecolor{lavenderblush}{rgb}{1.0, 0.94, 0.96}		
\definecolor{lavendergray}{rgb}{0.77, 0.76, 0.82}		
\definecolor{lavenderindigo}{rgb}{0.58, 0.34, 0.92}		
\definecolor{lavendermagenta}{rgb}{0.93, 0.51, 0.93}		
\definecolor{lavendermist}{rgb}{0.9, 0.9, 0.98}			
\definecolor{lavenderpink}{rgb}{0.98, 0.68, 0.82}		
\definecolor{lavenderpurple}{rgb}{0.59, 0.48, 0.71}		
\definecolor{lavenderrose}{rgb}{0.98, 0.63, 0.89}		
\definecolor{lawngreen}{rgb}{0.49, 0.99, 0.0}			
\definecolor{lemon}{rgb}{1.0, 0.97, 0.0}			
\definecolor{lemonchiffon}{rgb}{1.0, 0.98, 0.8}			
\definecolor{lightapricot}{rgb}{0.99, 0.84, 0.69}		
\definecolor{lightblue}{rgb}{0.68, 0.85, 0.9}			
\definecolor{lightbrown}{rgb}{0.71, 0.4, 0.11}			
\definecolor{lightcarminepink}{rgb}{0.9, 0.4, 0.38}		
\definecolor{lightcoral}{rgb}{0.94, 0.5, 0.5}			
\definecolor{lightcornflowerblue}{rgb}{0.6, 0.81, 0.93}		
\definecolor{lightcyan}{rgb}{0.88, 1.0, 1.0}			
\definecolor{lightfuchsiapink}{rgb}{0.98, 0.52, 0.9}		
\definecolor{lightgoldenrodyellow}{rgb}{0.98, 0.98, 0.82}	
\definecolor{lightgray}{rgb}{0.83, 0.83, 0.83}			
\definecolor{lightgreen}{rgb}{0.56, 0.93, 0.56}			
\definecolor{lightkhaki}{rgb}{0.94, 0.9, 0.55}			
\definecolor{lightmauve}{rgb}{0.86, 0.82, 1.0}			
\definecolor{lightpastelpurple}{rgb}{0.69, 0.61, 0.85}		
\definecolor{lightpink}{rgb}{1.0, 0.71, 0.76}			
\definecolor{lightsalmon}{rgb}{1.0, 0.63, 0.48}			
\definecolor{lightsalmonpink}{rgb}{1.0, 0.6, 0.6}		
\definecolor{lightseagreen}{rgb}{0.13, 0.7, 0.67}		
\definecolor{lightskyblue}{rgb}{0.53, 0.81, 0.98}		
\definecolor{lightslategray}{rgb}{0.47, 0.53, 0.6}		
\definecolor{lighttaupe}{rgb}{0.7, 0.55, 0.43}			
\definecolor{lightthulianpink}{rgb}{0.9, 0.56, 0.67}		
\definecolor{lightyellow}{rgb}{1.0, 1.0, 0.88}			
\definecolor{lilac}{rgb}{0.78, 0.64, 0.78}			
\definecolor{lime(colorwheel)}{rgb}{0.75, 1.0, 0.0}		
\definecolor{lime(web)(x11green)}{rgb}{0.0, 1.0, 0.0}		
\definecolor{limegreen}{rgb}{0.2, 0.8, 0.2}			
\definecolor{lincolngreen}{rgb}{0.11, 0.35, 0.02}		
\definecolor{linen}{rgb}{0.98, 0.94, 0.9}			
\definecolor{liver}{rgb}{0.33, 0.29, 0.31}			
\definecolor{lust}{rgb}{0.9, 0.13, 0.13}			
\definecolor{macaroniandcheese}{rgb}{1.0, 0.74, 0.53}		
\definecolor{magenta}{rgb}{1.0, 0.0, 1.0}			
\definecolor{magenta(dye)}{rgb}{0.79, 0.08, 0.48}		
\definecolor{magenta(process)}{rgb}{1.0, 0.0, 0.56}		
\definecolor{magicmint}{rgb}{0.67, 0.94, 0.82}			
\definecolor{magnolia}{rgb}{0.97, 0.96, 1.0}			
\definecolor{mahogany}{rgb}{0.75, 0.25, 0.0}			
\definecolor{maize}{rgb}{0.98, 0.93, 0.37}			
\definecolor{majorelleblue}{rgb}{0.38, 0.31, 0.86}		
\definecolor{malachite}{rgb}{0.04, 0.85, 0.32}			
\definecolor{manatee}{rgb}{0.59, 0.6, 0.67}			
\definecolor{mangotango}{rgb}{1.0, 0.51, 0.26}			
\definecolor{maroon(html/css)}{rgb}{0.5, 0.0, 0.0}		
\definecolor{maroon(x11)}{rgb}{0.69, 0.19, 0.38}		
\definecolor{mauve}{rgb}{0.88, 0.69, 1.0}			
\definecolor{mauvetaupe}{rgb}{0.57, 0.37, 0.43}			
\definecolor{mauvelous}{rgb}{0.94, 0.6, 0.67}			
\definecolor{mayablue}{rgb}{0.45, 0.76, 0.98}			
\definecolor{meatbrown}{rgb}{0.9, 0.72, 0.23}			
\definecolor{mediumaquamarine}{rgb}{0.4, 0.8, 0.67}		
\definecolor{mediumblue}{rgb}{0.0, 0.0, 0.8}			
\definecolor{mediumcandyapplered}{rgb}{0.89, 0.02, 0.17}	
\definecolor{mediumcarmine}{rgb}{0.69, 0.25, 0.21}		
\definecolor{mediumchampagne}{rgb}{0.95, 0.9, 0.67}		
\definecolor{mediumelectricblue}{rgb}{0.01, 0.31, 0.59}		
\definecolor{mediumjunglegreen}{rgb}{0.11, 0.21, 0.18}		
\definecolor{mediumlavendermagenta}{rgb}{0.8, 0.6, 0.8}		
\definecolor{mediumorchid}{rgb}{0.73, 0.33, 0.83}		
\definecolor{mediumpersianblue}{rgb}{0.0, 0.4, 0.65}		
\definecolor{mediumpurple}{rgb}{0.58, 0.44, 0.86}		
\definecolor{mediumred-violet}{rgb}{0.73, 0.2, 0.52}		
\definecolor{mediumseagreen}{rgb}{0.24, 0.7, 0.44}		
\definecolor{mediumslateblue}{rgb}{0.48, 0.41, 0.93}		
\definecolor{mediumspringbud}{rgb}{0.79, 0.86, 0.54}		
\definecolor{mediumspringgreen}{rgb}{0.0, 0.98, 0.6}		
\definecolor{mediumtaupe}{rgb}{0.4, 0.3, 0.28}			
\definecolor{mediumtealblue}{rgb}{0.0, 0.33, 0.71}		
\definecolor{mediumturquoise}{rgb}{0.28, 0.82, 0.8}		
\definecolor{mediumviolet-red}{rgb}{0.78, 0.08, 0.52}		
\definecolor{melon}{rgb}{0.99, 0.74, 0.71}			
\definecolor{midnightblue}{rgb}{0.1, 0.1, 0.44}			
\definecolor{midnightgreen(eaglegreen)}{rgb}{0.0, 0.29, 0.33}	
\definecolor{mikadoyellow}{rgb}{1.0, 0.77, 0.05}		
\definecolor{mint}{rgb}{0.24, 0.71, 0.54}			
\definecolor{mintcream}{rgb}{0.96, 1.0, 0.98}			
\definecolor{mintgreen}{rgb}{0.6, 1.0, 0.6}			
\definecolor{mistyrose}{rgb}{1.0, 0.89, 0.88}			
\definecolor{moccasin}{rgb}{0.98, 0.92, 0.84}			
\definecolor{modebeige}{rgb}{0.59, 0.44, 0.09}			
\definecolor{moonstoneblue}{rgb}{0.45, 0.66, 0.76}		
\definecolor{mordantred19}{rgb}{0.68, 0.05, 0.0}		
\definecolor{mossgreen}{rgb}{0.68, 0.87, 0.68}			
\definecolor{mountainmeadow}{rgb}{0.19, 0.73, 0.56}		
\definecolor{mountbattenpink}{rgb}{0.6, 0.48, 0.55}		
\definecolor{mulberry}{rgb}{0.77, 0.29, 0.55}			
\definecolor{mustard}{rgb}{1.0, 0.86, 0.35}			
\definecolor{myrtle}{rgb}{0.13, 0.26, 0.12}			
\definecolor{msugreen}{rgb}{0.09, 0.27, 0.23}			
\definecolor{nadeshikopink}{rgb}{0.96, 0.68, 0.78}		
\definecolor{napiergreen}{rgb}{0.16, 0.5, 0.0}			
\definecolor{naplesyellow}{rgb}{0.98, 0.85, 0.37}		
\definecolor{navajowhite}{rgb}{1.0, 0.87, 0.68}			
\definecolor{navyblue}{rgb}{0.0, 0.0, 0.5}			
\definecolor{neoncarrot}{rgb}{1.0, 0.64, 0.26}			
\definecolor{neonfuchsia}{rgb}{1.0, 0.25, 0.39}			
\definecolor{neongreen}{rgb}{0.22, 0.88, 0.08}			
\definecolor{non-photoblue}{rgb}{0.64, 0.87, 0.93}		
\definecolor{oceanboatblue}{rgb}{0.0, 0.47, 0.75}		
\definecolor{ochre}{rgb}{0.8, 0.47, 0.13}			
\definecolor{officegreen}{rgb}{0.0, 0.5, 0.0}			
\definecolor{oldgold}{rgb}{0.81, 0.71, 0.23}			
\definecolor{oldlace}{rgb}{0.99, 0.96, 0.9}			
\definecolor{oldlavender}{rgb}{0.47, 0.41, 0.47}		
\definecolor{oldmauve}{rgb}{0.4, 0.19, 0.28}			
\definecolor{oldrose}{rgb}{0.75, 0.5, 0.51}			
\definecolor{olive}{rgb}{0.5, 0.5, 0.0}				
\definecolor{olivedrab(web)(olivedrab3)}{rgb}{0.42, 0.56, 0.14}	
\definecolor{olivedrab7}{rgb}{0.24, 0.2, 0.12}			
\definecolor{olivine}{rgb}{0.6, 0.73, 0.45}			
\definecolor{onyx}{rgb}{0.06, 0.06, 0.06}			
\definecolor{operamauve}{rgb}{0.72, 0.52, 0.65}			
\definecolor{orange(colorwheel)}{rgb}{1.0, 0.5, 0.0}		
\definecolor{orange(ryb)}{rgb}{0.98, 0.6, 0.01}			
\definecolor{orange(webcolor)}{rgb}{1.0, 0.65, 0.0}		
\definecolor{orangepeel}{rgb}{1.0, 0.62, 0.0}			
\definecolor{orange-red}{rgb}{1.0, 0.27, 0.0}			
\definecolor{orchid}{rgb}{0.85, 0.44, 0.84}			
\definecolor{otterbrown}{rgb}{0.4, 0.26, 0.13}			
\definecolor{outerspace}{rgb}{0.25, 0.29, 0.3}			
\definecolor{outrageousorange}{rgb}{1.0, 0.43, 0.29}		
\definecolor{oxfordblue}{rgb}{0.0, 0.13, 0.28}			
\definecolor{oucrimsonred}{rgb}{0.6, 0.0, 0.0}			
\definecolor{pakistangreen}{rgb}{0.0, 0.4, 0.0}			
\definecolor{palatinateblue}{rgb}{0.15, 0.23, 0.89}		
\definecolor{palatinatepurple}{rgb}{0.41, 0.16, 0.38}		
\definecolor{paleaqua}{rgb}{0.74, 0.83, 0.9}			
\definecolor{paleblue}{rgb}{0.69, 0.93, 0.93}			
\definecolor{palebrown}{rgb}{0.6, 0.46, 0.33}			
\definecolor{palecarmine}{rgb}{0.69, 0.25, 0.21}		
\definecolor{palecerulean}{rgb}{0.61, 0.77, 0.89}		
\definecolor{palechestnut}{rgb}{0.87, 0.68, 0.69}		
\definecolor{palecopper}{rgb}{0.85, 0.54, 0.4}			
\definecolor{palecornflowerblue}{rgb}{0.67, 0.8, 0.94}		
\definecolor{palegold}{rgb}{0.9, 0.75, 0.54}			
\definecolor{palegoldenrod}{rgb}{0.93, 0.91, 0.67}		
\definecolor{palegreen}{rgb}{0.6, 0.98, 0.6}			
\definecolor{palemagenta}{rgb}{0.98, 0.52, 0.9}			
\definecolor{palepink}{rgb}{0.98, 0.85, 0.87}			
\definecolor{paleplum}{rgb}{0.8, 0.6, 0.8}			
\definecolor{palered-violet}{rgb}{0.86, 0.44, 0.58}		
\definecolor{palerobineggblue}{rgb}{0.59, 0.87, 0.82}		
\definecolor{palesilver}{rgb}{0.79, 0.75, 0.73}			
\definecolor{palespringbud}{rgb}{0.93, 0.92, 0.74}		
\definecolor{paletaupe}{rgb}{0.74, 0.6, 0.49}			
\definecolor{paleviolet-red}{rgb}{0.86, 0.44, 0.58}		
\definecolor{pansypurple}{rgb}{0.47, 0.09, 0.29}		
\definecolor{papayawhip}{rgb}{1.0, 0.94, 0.84}			
\definecolor{parisgreen}{rgb}{0.31, 0.78, 0.47}			
\definecolor{pastelblue}{rgb}{0.68, 0.78, 0.81}			
\definecolor{pastelbrown}{rgb}{0.51, 0.41, 0.33}		
\definecolor{pastelgray}{rgb}{0.81, 0.81, 0.77}			
\definecolor{pastelgreen}{rgb}{0.47, 0.87, 0.47}		
\definecolor{pastelmagenta}{rgb}{0.96, 0.6, 0.76}		
\definecolor{pastelorange}{rgb}{1.0, 0.7, 0.28}			
\definecolor{pastelpink}{rgb}{1.0, 0.82, 0.86}			
\definecolor{pastelpurple}{rgb}{0.7, 0.62, 0.71}		
\definecolor{pastelred}{rgb}{1.0, 0.41, 0.38}			
\definecolor{pastelviolet}{rgb}{0.8, 0.6, 0.79}			
\definecolor{pastelyellow}{rgb}{0.99, 0.99, 0.59}		
\definecolor{patriarch}{rgb}{0.5, 0.0, 0.5}			
\definecolor{payne\'sgrey}{rgb}{0.25, 0.25, 0.28}		
\definecolor{peach}{rgb}{1.0, 0.9, 0.71}			
\definecolor{peach-orange}{rgb}{1.0, 0.8, 0.6}			
\definecolor{peachpuff}{rgb}{1.0, 0.85, 0.73}			
\definecolor{peach-yellow}{rgb}{0.98, 0.87, 0.68}		
\definecolor{pear}{rgb}{0.82, 0.89, 0.19}			
\definecolor{pearl}{rgb}{0.94, 0.92, 0.84}			
\definecolor{peridot}{rgb}{0.9, 0.89, 0.0}			
\definecolor{periwinkle}{rgb}{0.8, 0.8, 1.0}			
\definecolor{persianblue}{rgb}{0.11, 0.22, 0.73}		
\definecolor{persiangreen}{rgb}{0.0, 0.65, 0.58}		
\definecolor{persianindigo}{rgb}{0.2, 0.07, 0.48}		
\definecolor{persianorange}{rgb}{0.85, 0.56, 0.35}		
\definecolor{peru}{rgb}{0.8, 0.52, 0.25}			
\definecolor{persianpink}{rgb}{0.97, 0.5, 0.75}			
\definecolor{persianplum}{rgb}{0.44, 0.11, 0.11}		
\definecolor{persianred}{rgb}{0.8, 0.2, 0.2}			
\definecolor{persianrose}{rgb}{1.0, 0.16, 0.64}			
\definecolor{persimmon}{rgb}{0.93, 0.35, 0.0}			
\definecolor{phlox}{rgb}{0.87, 0.0, 1.0}			
\definecolor{phthaloblue}{rgb}{0.0, 0.06, 0.54}			
\definecolor{phthalogreen}{rgb}{0.07, 0.21, 0.14}		
\definecolor{piggypink}{rgb}{0.99, 0.87, 0.9}			
\definecolor{pinegreen}{rgb}{0.0, 0.47, 0.44}			
\definecolor{pink}{rgb}{1.0, 0.75, 0.8}				
\definecolor{pink-orange}{rgb}{1.0, 0.6, 0.4}			
\definecolor{pinkpearl}{rgb}{0.91, 0.67, 0.81}			
\definecolor{pinksherbet}{rgb}{0.97, 0.56, 0.65}		
\definecolor{pistachio}{rgb}{0.58, 0.77, 0.45}			
\definecolor{platinum}{rgb}{0.9, 0.89, 0.89}			
\definecolor{plum(traditional)}{rgb}{0.56, 0.27, 0.52}		
\definecolor{plum(web)}{rgb}{0.8, 0.6, 0.8}			
\definecolor{portlandorange}{rgb}{1.0, 0.35, 0.21}		
\definecolor{powderblue(web)}{rgb}{0.69, 0.88, 0.9}		
\definecolor{princetonorange}{rgb}{1.0, 0.56, 0.0}		
\definecolor{prune}{rgb}{0.44, 0.11, 0.11}			
\definecolor{prussianblue}{rgb}{0.0, 0.19, 0.33}		
\definecolor{psychedelicpurple}{rgb}{0.87, 0.0, 1.0}		
\definecolor{puce}{rgb}{0.8, 0.53, 0.6}				
\definecolor{pumpkin}{rgb}{1.0, 0.46, 0.09}			
\definecolor{purple(html/css)}{rgb}{0.5, 0.0, 0.5}		
\definecolor{purple(munsell)}{rgb}{0.62, 0.0, 0.77}		
\definecolor{purple(x11)}{rgb}{0.63, 0.36, 0.94}		
\definecolor{purpleheart}{rgb}{0.41, 0.21, 0.61}		
\definecolor{purplemountainmajesty}{rgb}{0.59, 0.47, 0.71}	
\definecolor{purplepizzazz}{rgb}{1.0, 0.31, 0.85}		
\definecolor{purpletaupe}{rgb}{0.31, 0.25, 0.3}			
\definecolor{radicalred}{rgb}{1.0, 0.21, 0.37}			
\definecolor{raspberry}{rgb}{0.89, 0.04, 0.36}			
\definecolor{raspberryglace}{rgb}{0.57, 0.37, 0.43}		
\definecolor{raspberrypink}{rgb}{0.89, 0.31, 0.61}		
\definecolor{raspberryrose}{rgb}{0.7, 0.27, 0.42}		
\definecolor{rawumber}{rgb}{0.51, 0.4, 0.27}			
\definecolor{razzledazzlerose}{rgb}{1.0, 0.2, 0.8}		
\definecolor{razzmatazz}{rgb}{0.89, 0.15, 0.42}			
\definecolor{red}{rgb}{1.0, 0.0, 0.0}				
\definecolor{red(munsell)}{rgb}{0.95, 0.0, 0.24}		
\definecolor{red(ncs)}{rgb}{0.77, 0.01, 0.2}			
\definecolor{red(pigment)}{rgb}{0.93, 0.11, 0.14}		
\definecolor{red(ryb)}{rgb}{1.0, 0.15, 0.07}			
\definecolor{red-brown}{rgb}{0.65, 0.16, 0.16}			
\definecolor{red-violet}{rgb}{0.78, 0.08, 0.52}			
\definecolor{redwood}{rgb}{0.67, 0.31, 0.32}			
\definecolor{regalia}{rgb}{0.32, 0.18, 0.5}			
\definecolor{richblack}{rgb}{0.0, 0.25, 0.25}			
\definecolor{richbrilliantlavender}{rgb}{0.95, 0.65, 1.0}	
\definecolor{richcarmine}{rgb}{0.84, 0.0, 0.25}			
\definecolor{richelectricblue}{rgb}{0.03, 0.57, 0.82}		
\definecolor{richlavender}{rgb}{0.67, 0.38, 0.8}		
\definecolor{richlilac}{rgb}{0.71, 0.4, 0.82}			
\definecolor{richmaroon}{rgb}{0.69, 0.19, 0.38}			
\definecolor{riflegreen}{rgb}{0.25, 0.28, 0.2}			
\definecolor{robineggblue}{rgb}{0.0, 0.8, 0.8}			
\definecolor{rose}{rgb}{1.0, 0.0, 0.5}				
\definecolor{rosebonbon}{rgb}{0.98, 0.26, 0.62}			
\definecolor{roseebony}{rgb}{0.4, 0.3, 0.28}			
\definecolor{rosegold}{rgb}{0.72, 0.43, 0.47}			
\definecolor{rosemadder}{rgb}{0.89, 0.15, 0.21}			
\definecolor{rosepink}{rgb}{1.0, 0.4, 0.8}			
\definecolor{rosequartz}{rgb}{0.67, 0.6, 0.66}			
\definecolor{rosetaupe}{rgb}{0.56, 0.36, 0.36}			
\definecolor{rosevale}{rgb}{0.67, 0.31, 0.32}			
\definecolor{rosewood}{rgb}{0.4, 0.0, 0.04}			
\definecolor{rossocorsa}{rgb}{0.83, 0.0, 0.0}			
\definecolor{rosybrown}{rgb}{0.74, 0.56, 0.56}			
\definecolor{royalazure}{rgb}{0.0, 0.22, 0.66}			
\definecolor{royalblue(traditional)}{rgb}{0.0, 0.14, 0.4}	
\definecolor{royalblue(web)}{rgb}{0.25, 0.41, 0.88}		
\definecolor{royalfuchsia}{rgb}{0.79, 0.17, 0.57}		
\definecolor{royalpurple}{rgb}{0.47, 0.32, 0.66}		
\definecolor{ruby}{rgb}{0.88, 0.07, 0.37}			
\definecolor{ruddy}{rgb}{1.0, 0.0, 0.16}			
\definecolor{ruddybrown}{rgb}{0.73, 0.4, 0.16}			
\definecolor{ruddypink}{rgb}{0.88, 0.56, 0.59}			
\definecolor{rufous}{rgb}{0.66, 0.11, 0.03}			
\definecolor{russet}{rgb}{0.5, 0.27, 0.11}			
\definecolor{rust}{rgb}{0.72, 0.25, 0.05}			
\definecolor{sacramentostategreen}{rgb}{0.0, 0.34, 0.25}	
\definecolor{saddlebrown}{rgb}{0.55, 0.27, 0.07}		
\definecolor{safetyorange(blazeorange)}{rgb}{1.0, 0.4, 0.0}	
\definecolor{saffron}{rgb}{0.96, 0.77, 0.19}			
\definecolor{st.patrick\'sblue}{rgb}{0.14, 0.16, 0.48}		
\definecolor{salmon}{rgb}{1.0, 0.55, 0.41}			
\definecolor{salmonpink}{rgb}{1.0, 0.57, 0.64}			
\definecolor{sand}{rgb}{0.76, 0.7, 0.5}				
\definecolor{sanddune}{rgb}{0.59, 0.44, 0.09}			
\definecolor{sandstorm}{rgb}{0.93, 0.84, 0.25}			
\definecolor{sandybrown}{rgb}{0.96, 0.64, 0.38}			
\definecolor{sandytaupe}{rgb}{0.59, 0.44, 0.09}			
\definecolor{sangria}{rgb}{0.57, 0.0, 0.04}			
\definecolor{sapgreen}{rgb}{0.31, 0.49, 0.16}			
\definecolor{sapphire}{rgb}{0.03, 0.15, 0.4}			
\definecolor{satinsheengold}{rgb}{0.8, 0.63, 0.21}		
\definecolor{scarlet}{rgb}{1.0, 0.13, 0.0}			
\definecolor{schoolbusyellow}{rgb}{1.0, 0.85, 0.0}		
\definecolor{screamin\'green}{rgb}{0.46, 1.0, 0.44}		
\definecolor{seagreen}{rgb}{0.18, 0.55, 0.34}			
\definecolor{sealbrown}{rgb}{0.2, 0.08, 0.08}			
\definecolor{seashell}{rgb}{1.0, 0.96, 0.93}			
\definecolor{selectiveyellow}{rgb}{1.0, 0.73, 0.0}		
\definecolor{sepia}{rgb}{0.44, 0.26, 0.08}			
\definecolor{shadow}{rgb}{0.54, 0.47, 0.36}			
\definecolor{shamrockgreen}{rgb}{0.0, 0.62, 0.38}		
\definecolor{shockingpink}{rgb}{0.99, 0.06, 0.75}		
\definecolor{sienna}{rgb}{0.53, 0.18, 0.09}			
\definecolor{silver}{rgb}{0.75, 0.75, 0.75}			
\definecolor{sinopia}{rgb}{0.8, 0.25, 0.04}			
\definecolor{skobeloff}{rgb}{0.0, 0.48, 0.45}			
\definecolor{skyblue}{rgb}{0.53, 0.81, 0.92}			
\definecolor{skymagenta}{rgb}{0.81, 0.44, 0.69}			
\definecolor{slateblue}{rgb}{0.42, 0.35, 0.8}			
\definecolor{slategray}{rgb}{0.44, 0.5, 0.56}			
\definecolor{smalt(darkpowderblue)}{rgb}{0.0, 0.2, 0.6}		
\definecolor{smokeytopaz}{rgb}{0.58, 0.25, 0.03}		
\definecolor{smokyblack}{rgb}{0.06, 0.05, 0.03}			
\definecolor{snow}{rgb}{1.0, 0.98, 0.98}			
\definecolor{spirodiscoball}{rgb}{0.06, 0.75, 0.99}		
\definecolor{splashedwhite}{rgb}{1.0, 0.99, 1.0}		
\definecolor{springbud}{rgb}{0.65, 0.99, 0.0}			
\definecolor{springgreen}{rgb}{0.0, 1.0, 0.5}			
\definecolor{steelblue}{rgb}{0.27, 0.51, 0.71}			
\definecolor{stildegrainyellow}{rgb}{0.98, 0.85, 0.37}		
\definecolor{straw}{rgb}{0.89, 0.85, 0.44}			
\definecolor{sunglow}{rgb}{1.0, 0.8, 0.2}			
\definecolor{sunset}{rgb}{0.98, 0.84, 0.65}			
\definecolor{tan}{rgb}{0.82, 0.71, 0.55}			
\definecolor{tangelo}{rgb}{0.98, 0.3, 0.0}			
\definecolor{tangerine}{rgb}{0.95, 0.52, 0.0}			
\definecolor{tangerineyellow}{rgb}{1.0, 0.8, 0.0}		
\definecolor{taupe}{rgb}{0.28, 0.24, 0.2}			
\definecolor{taupegray}{rgb}{0.55, 0.52, 0.54}			
\definecolor{teagreen}{rgb}{0.82, 0.94, 0.75}			
\definecolor{tearose(orange)}{rgb}{0.97, 0.51, 0.47}		
\definecolor{tearose(rose)}{rgb}{0.96, 0.76, 0.76}		
\definecolor{teal}{rgb}{0.0, 0.5, 0.5}				
\definecolor{tealblue}{rgb}{0.21, 0.46, 0.53}			
\definecolor{tealgreen}{rgb}{0.0, 0.51, 0.5}			
\definecolor{tenné(tawny)}{rgb}{0.8, 0.34, 0.0}			
\definecolor{terracotta}{rgb}{0.89, 0.45, 0.36}			
\definecolor{thistle}{rgb}{0.85, 0.75, 0.85}			
\definecolor{thulianpink}{rgb}{0.87, 0.44, 0.63}		
\definecolor{ticklemepink}{rgb}{0.99, 0.54, 0.67}		
\definecolor{tiffanyblue}{rgb}{0.04, 0.73, 0.71}		
\definecolor{tiger\'seye}{rgb}{0.88, 0.55, 0.24}		
\definecolor{timberwolf}{rgb}{0.86, 0.84, 0.82}			
\definecolor{titaniumyellow}{rgb}{0.93, 0.9, 0.0}		
\definecolor{tomato}{rgb}{1.0, 0.39, 0.28}			
\definecolor{toolbox}{rgb}{0.45, 0.42, 0.75}			
\definecolor{tractorred}{rgb}{0.99, 0.05, 0.21}			
\definecolor{trolleygrey}{rgb}{0.5, 0.5, 0.5}			
\definecolor{tropicalrainforest}{rgb}{0.0, 0.46, 0.37}		
\definecolor{trueblue}{rgb}{0.0, 0.45, 0.81}			
\definecolor{tuftsblue}{rgb}{0.28, 0.57, 0.81}			
\definecolor{tumbleweed}{rgb}{0.87, 0.67, 0.53}			
\definecolor{turkishrose}{rgb}{0.71, 0.45, 0.51}		
\definecolor{turquoise}{rgb}{0.19, 0.84, 0.78}			
\definecolor{turquoiseblue}{rgb}{0.0, 1.0, 0.94}		
\definecolor{turquoisegreen}{rgb}{0.63, 0.84, 0.71}		
\definecolor{tuscanred}{rgb}{0.51, 0.21, 0.21}			
\definecolor{twilightlavender}{rgb}{0.54, 0.29, 0.42}		
\definecolor{tyrianpurple}{rgb}{0.4, 0.01, 0.24}		
\definecolor{uablue}{rgb}{0.0, 0.2, 0.67}			
\definecolor{uared}{rgb}{0.85, 0.0, 0.3}			
\definecolor{ube}{rgb}{0.53, 0.47, 0.76}			
\definecolor{uclablue}{rgb}{0.33, 0.41, 0.58}			
\definecolor{uclagold}{rgb}{1.0, 0.7, 0.0}			
\definecolor{ufogreen}{rgb}{0.24, 0.82, 0.44}			
\definecolor{ultramarine}{rgb}{0.07, 0.04, 0.56}		
\definecolor{ultramarineblue}{rgb}{0.25, 0.4, 0.96}		
\definecolor{ultrapink}{rgb}{1.0, 0.44, 1.0}			
\definecolor{umber}{rgb}{0.39, 0.32, 0.28}			
\definecolor{unitednationsblue}{rgb}{0.36, 0.57, 0.9}		
\definecolor{unmellowyellow}{rgb}{1.0, 1.0, 0.4}		
\definecolor{upforestgreen}{rgb}{0.0, 0.27, 0.13}		
\definecolor{upmaroon}{rgb}{0.48, 0.07, 0.07}			
\definecolor{upsdellred}{rgb}{0.68, 0.09, 0.13}			
\definecolor{urobilin}{rgb}{0.88, 0.68, 0.13}			
\definecolor{usccardinal}{rgb}{0.6, 0.0, 0.0}			
\definecolor{uscgold}{rgb}{1.0, 0.8, 0.0}			
\definecolor{utahcrimson}{rgb}{0.83, 0.0, 0.25}			
\definecolor{vanilla}{rgb}{0.95, 0.9, 0.67}			
\definecolor{vegasgold}{rgb}{0.77, 0.7, 0.35}			
\definecolor{venetianred}{rgb}{0.78, 0.03, 0.08}		
\definecolor{verdigris}{rgb}{0.26, 0.7, 0.68}			
\definecolor{vermilion}{rgb}{0.89, 0.26, 0.2}			
\definecolor{veronica}{rgb}{0.63, 0.36, 0.94}			
\definecolor{violet}{rgb}{0.56, 0.0, 1.0}			
\definecolor{violet(colorwheel)}{rgb}{0.5, 0.0, 1.0}		
\definecolor{violet(ryb)}{rgb}{0.53, 0.0, 0.69}			
\definecolor{violet(web)}{rgb}{0.93, 0.51, 0.93}		
\definecolor{viridian}{rgb}{0.25, 0.51, 0.43}			
\definecolor{vividauburn}{rgb}{0.58, 0.15, 0.14}		
\definecolor{vividburgundy}{rgb}{0.62, 0.11, 0.21}		
\definecolor{vividcerise}{rgb}{0.85, 0.11, 0.51}		
\definecolor{vividtangerine}{rgb}{1.0, 0.63, 0.54}		
\definecolor{vividviolet}{rgb}{0.62, 0.0, 1.0}			
\definecolor{warmblack}{rgb}{0.0, 0.26, 0.26}			
\definecolor{wenge}{rgb}{0.39, 0.33, 0.32}			
\definecolor{wheat}{rgb}{0.96, 0.87, 0.7}			
\definecolor{white}{rgb}{1.0, 1.0, 1.0}				
\definecolor{whitesmoke}{rgb}{0.96, 0.96, 0.96}			
\definecolor{wildblueyonder}{rgb}{0.64, 0.68, 0.82}		
\definecolor{wildstrawberry}{rgb}{1.0, 0.26, 0.64}		
\definecolor{wildwatermelon}{rgb}{0.99, 0.42, 0.52}		
\definecolor{wisteria}{rgb}{0.79, 0.63, 0.86}			
\definecolor{xanadu}{rgb}{0.45, 0.53, 0.47}			
\definecolor{yaleblue}{rgb}{0.06, 0.3, 0.57}			
\definecolor{yellow}{rgb}{1.0, 1.0, 0.0}			
\definecolor{yellow(munsell)}{rgb}{0.94, 0.8, 0.0}		
\definecolor{yellow(ncs)}{rgb}{1.0, 0.83, 0.0}			
\definecolor{yellow(process)}{rgb}{1.0, 0.94, 0.0}		
\definecolor{yellow(ryb)}{rgb}{1.0, 1.0, 0.2}			
\definecolor{yellow-green}{rgb}{0.6, 0.8, 0.2}			
\definecolor{zaffre}{rgb}{0.0, 0.08, 0.66}			
\definecolor{zinnwalditebrown}{rgb}{0.17, 0.09, 0.03}		

\definecolor{my_orange}{RGB}{255,151,46}
\definecolor{my_blue}{RGB}{38,125,184}
\definecolor{my_green}{RGB}{38,184,125}

\pgfplotsset{compat=1.9}
\usepgfplotslibrary{groupplots}
\usetikzlibrary{%
	shapes.geometric, 
	positioning, 
	calc, 
	arrows, 
	tikzmark, 
	shapes.gates.logic.US, 
	shapes.gates.logic.IEC,
	decorations.pathreplacing,
	decorations.markings,
	shadows,
	patterns,
	arrows.meta,
	plotmarks,
}

\newlength\algowd

\newcommand{\rc}[1]{{\color{blue_fau!75}#1}}
\newcommand{\gc}[1]{{\color{blue_fau!75}#1}}

\newcommand{\nosemic}{\renewcommand{\@endalgocfline}{\relax}}
\newcommand{\dosemic}{\renewcommand{\@endalgocfline}{\algocf@endline}}
\let\oldnl\nl
\newcommand{\nonl}{\renewcommand{\nl}{\let\nl\oldnl}}
\makeatother

\newcommand{\eqmark}[1]{\tikz[baseline,remember picture] \coordinate (#1) {};}

\newcommand{\fixlacc}[1]{\mathrlap{#1}\hphantom{\left(\left(S_{i+1}^{j-1} \oplus (a_i \mint b_j) \right) \mint C_{i-1}^{j}\right)}}
\newcommand{\fixlapp}[1]{\mathrlap{#1}\hphantom{\left(\left(\hat{S}_{i+1}^{j-1} \oplus (a_i \mint b_j) \right) \mint \hat{C}_{i-1}^{j}\right)}}
\newcommand{\fixmacc}[1]{\mathrlap{#1}\hphantom{S_{i-n+1}^{n-1}}}
\newcommand{\fixmapp}[1]{\mathrlap{#1}\hphantom{\hat{S}_{i-n+1}^{n-1}}}
\newcommand{\fixtapp}[1]{\mathrlap{#1}\hphantom{[t{+}n,2n{-}1]}}

\newcommand{\mint}{{\scriptstyle\!\!~\wedge~\!\!}}
\newcommand{\maxt}{{\scriptstyle\!\!~\vee~\!\!}}

\makeatletter
\newcount\my@repeat@count
\newcommand{\bs}[1]{%
  \begingroup
  \my@repeat@count=\z@
  \@whilenum\my@repeat@count<#1\do{\!\advance\my@repeat@count\@ne}%
  \endgroup
}
\makeatother

\makeatletter
\newcount\my@repeat@count
\newcommand{\ns}[1]{%
  \begingroup
  \my@repeat@count=\z@
  \@whilenum\my@repeat@count<#1\do{~\advance\my@repeat@count\@ne}%
  \endgroup
}
\makeatother

\newcommand{\es}[1]{\hspace{#1mm}}%

\newcommand{\rhu}[1]{\lceil\hspace{-2.5pt}\lvert#1\lvert\hspace{-2.5pt}\rceil}

\SetNlSkip{.2em}
\SetAlCapSkip{1em}
\SetAlCapSty{}

\makeatletter
\def\footnoterule{\kern-3\p@
  \hrule \@width 2in \kern 2.6\p@}
\makeatother

\newacronym{ac}{AC}{approximate computing}
\newacronym{als}{ALS}{approximate logic synthesis}
\newacronym{lat}{LAT}{local approximate transformation}
\newacronym{ber}{BER}{bit error rate}
\newacronym{po}{PO}{primary output}
\newacronym{pi}{PI}{primary input}
\newacronym{dse}{DSE}{design space exploration}
\newacronym{mc}{MC}{Monte-Carlo}
\newacronym{bd}{BD}{Boolean difference}
\newacronym[firstplural=change propagation matrices (CPMs)]{cpm}{CPM}{change propagation matrix}
\newacronym{op}{OP}{operating system}
\newacronym{ea}{EA}{evolutionary algorithm}
\newacronym{dfs}{DFS}{depth-first search}
\newacronym{bdd}{BDD}{binary decision diagram}
\newacronym{aig}{AIG}{AND-inverter graph}
\newacronym{dnf}{DNF}{disjunctive normal form}
\newacronym{sop}{SoP}{sum-of-products}
\newacronym{sat}{SAT}{boolean satisfiability problem}
\newacronym{me}{ME}{maximum error}
\newacronym{nme}{NME}{normalized maximum error}
\newacronym{med}{MED}{mean error distance}
\newacronym{mred}{MRED}{mean relative error distance}
\newacronym{red}{RED}{relative error distance}
\newacronym{nmed}{NMED}{normalized mean error distance}
\newacronym{er}{ER}{arithmetic error rate}
\newacronym{aer}{AER}{arithmetic error rate}
\newacronym{ed}{ED}{error distance}
\newacronym{mao}{MAO}{maximum accurate output}
\newacronym{mae}{MAE}{maximum absolute error}
\newacronym{psnr}{PSNR}{peak signal to noise radio}
\newacronym[firstplural=structural similarity indices (SSIMs)]{ssim}{SSIM}{structural similarity index}
\newacronym{cp}{CP}{critical path}
\newacronym{fpga}{FPGA}{field-programmable gate array}
\newacronym{dsp}{DSP}{digital signal processing}
\newacronym{lut}{LUT}{look-up table}
\newacronym{lsb}{LSB}{least significant bit}
\newacronym{msb}{MSB}{most significant bit}
\newacronym{lsp}{LSP}{least significant part}
\newacronym{msp}{MSP}{most significant part}
\newacronym{qor}{QoR}{quality of results}
\newacronym{cdf}{CDF}{cumulative distribution function}
\newacronym{pdp}{PDP}{power-delay product}
\newacronym{pdf}{PDF}{probability density function}
\newacronym{aca}{ACA}{Almost Correct Adder}
\newacronym{gda}{GDA}{Accuracy Gracefully-Degrading Adder}
\newacronym{gear}{GeAr}{Generic Accuracy Configurable Adder}
\newacronym{mac}{MAC}{multiply and accumulate}
\newacronym{gmed}{GMED}{generalized mean error distance}

\pdfpageattr {/Group << /S /Transparency /I true /CS /DeviceRGB>>}

\newcommand{\op}[1]{{~#1~}}

\newcommand{\rb}[1]{\color{my_orange}#1}

\newcommand{\capmark}[1]{\protect\tikz{\protect\filldraw[white] (0,0) -- (0,0);\protect\filldraw[#1,fill opacity = 0.15] (0,0.075) circle (0.75mm);}}
\newcommand{\trimark}[1]{\protect\tikz{\protect\filldraw[white] (0,0) -- (0,0);\protect\filldraw[#1,fill opacity = 0.15] (-0.09,0) -- (0,0.17) -- (0.09,0) -- cycle;}}
\newcommand{\diamark}[1]{\protect\tikz{\protect\filldraw[white] (0,0) -- (0,0);\protect\filldraw[#1,fill opacity = 0.15] (-0.06,0.07) -- (0,0.14) -- (0.06,0.07) -- (0,0) -- cycle;}}
\newcommand{\trimarkinv}[1]{\protect\tikz{\protect\filldraw[white] (0,0) -- (0,0);\protect\filldraw[#1,fill opacity = 0.15,rotate=180] (-0.06,0) -- (0,0.12) -- (0.06,0) -- cycle;}}

\newcolumntype{P}[1]{>{\centering\arraybackslash}p{#1}}
\newcolumntype{R}[1]{>{\raggedleft\arraybackslash}m{#1}}

\newcommand{\Bs}{\mathbb{B}}
\newcommand{\absv}[1]{\left\lvert #1 \right\rvert}

\DeclareMathOperator*{\er}{ER}
\DeclareMathOperator*{\ber}{BER}
\DeclareMathOperator*{\ed}{ED}
\DeclareMathOperator*{\med}{MED}
\DeclareMathOperator*{\mae}{MAE}

\DeclareMathOperator*{\mred}{MRED}
\DeclareMathOperator*{\nmed}{NMED}
\DeclareMathOperator*{\gmed}{GMED}

\title{On the Approximation of Accuracy-configurable Sequential Multipliers via Segmented Carry Chains}

\author
{
    \IEEEauthorblockN{Jorge Echavarria\orcid{0000-0002-3751-5273}, Stefan Wildermann, Oliver Keszocze, Faramarz Khosravi, Andreas Becher, J\"urgen Teich}
    \IEEEauthorblockA{Department of Computer Science, Friedrich-Alexander-Universit\"at Erlangen-N\"urnberg (FAU), 91058 Erlangen, Germany\\
    Email:\{jorge.echavarria, stefan.wildermann, oliver.keszoecze, faramarz.khosravi, andreas.becher, juergen.teich\}@fau.de}
}

\begin{document}
	
	\maketitle
	
	\begin{abstract}
		In this paper, we present a multiplier based on a sequence of approximated accumulations.
		According to a given splitting point of the carry chains, the technique herein introduced allows varying the quality of the accumulations and, consequently, the overall product.
		Our approximate multiplier trades-off accuracy for a reduced latency---with respect to an accurate sequential multiplier---and exploits the inherent area savings of sequential over combinatorial approaches.
		We implemented multiple versions with different bit-width and accuracy configurations, targeting an FPGA and a 45nm ASIC to estimate resources, power consumption, and latency.
		We also present two error analyses of the proposed design based on closed-form analysis and simulations.
	\end{abstract}
	
	\newcommand{\FIGexacccomseq}[1]{
    \begin{table}[#1]
        \vspace{-5pt}
		\begin{minipage}[t]{\columnwidth}
			\begin{subtable}[t]{0.5\textwidth}
                \begin{minipage}[t][5.25cm][t]{0.5\columnwidth}
    {\small
    \begin{equation*}
        \begin{aligned}
            &\text{\scriptsize Multiplier}   && \es{4}\:1\:0\:1\:1                 && \es{-18}                                          \text{\color{gray}\tiny 1}  \\[-2pt]
            &\text{\scriptsize Multiplicand} && \es{4}\:1\:1\:0\:1 \eqmark{m1}     && \es{-18}                                          \text{\color{gray}\tiny 2}  \\[-2pt]
            &                                && \es{-5}\left.\begin{aligned}
                                                                                                     & \gc{1\:0\:1\:}1           & \es{-2}\text{\color{gray}\tiny 3}  \\[-2pt]
                                                                                            \gc{0\:} & \gc{0\:0\:0}  \eqmark{m8} & \es{-2}\text{\color{gray}\tiny 4}  \\[-2pt]
                                                                       \tikzmark{P1}\rc{0}\:\gc{0\:} & \gc{1\:0\:1\:}1           & \es{-2}\text{\color{gray}\tiny 5}  \\
                                                                                         \gc{1\:0\:} & \gc{1\:}1                 & \es{-2}\text{\color{gray}\tiny 6}  \\[-2pt]
                                                                                      \gc{1\:0\:1\:} & \gc{1}    \eqmark{m9}     & \es{-2}\text{\color{gray}\tiny 7}  \\[-2pt]
                                                                 \tikzmark{P2}\rc{1}\:\gc{0\:0\:0\:} & \gc{0\:}1                 & \es{-2}\text{\color{gray}\tiny 8}  \\
                                                                            \tikzmark{P3}\gc{0\:0\:} & \gc{1\:0\:}1\:1           & \es{-2}\text{\color{gray}\tiny 9}  \\[-2pt]
                                                                      \tikzmark{P4}\gc{1\:0\:0\:0\:} & \gc{0\:1}   \eqmark{m10}  & \es{-2}\text{\color{gray}\tiny 10} \\[-2pt]
                                                                                   \gc{1\:0\:0\:0\:} & \gc{1\:1\:}1\:1           & \es{-2}\text{\color{gray}\tiny 11}
                                                             \end{aligned}\es{2}\right\rbrace\begin{aligned}
                                                                                                 &\text{\scriptsize Partial} \\[-2pt]
                                                                                                 &\text{\scriptsize products}
                                                                                             \end{aligned}&&\\[-2pt]
            &\text{\scriptsize Product}      && \es{-4}1\:0\:0\:0\:1\:1\:1\:1 \eqmark{m2} && \es{-19}                                        \text{\color{gray}\tiny 12}
        \end{aligned}
        \tikz[overlay,remember picture, dashed]{\draw([xshift=- 8mm, yshift=-0.75mm]m1.south)  -- ([xshift=0mm, yshift=-0.75mm]m1.south);}
        \tikz[overlay,remember picture        ]{\draw([xshift=- 8mm, yshift=-0.75mm]m8.south)  -- ([xshift=2mm, yshift=-0.75mm]m8.south);}
        \tikz[overlay,remember picture        ]{\draw([xshift=- 8mm, yshift=-0.75mm]m9.south)  -- ([xshift=2mm, yshift=-0.75mm]m9.south);}
        \tikz[overlay,remember picture        ]{\draw([xshift=-12mm, yshift=-0.75mm]m10.south) -- ([xshift=4mm, yshift=-0.75mm]m10.south);}
    \end{equation*}}
    \begin{tikzpicture}[overlay,remember picture]
        \scalebox{0.5}{\draw[gray,preaction={-triangle 90,draw,ultra thin,gray}] ([xshift=40pt, yshift=95pt]{pic cs:P1}) to[out=180,in=180] ([xshift=40pt, yshift=47pt]{pic cs:P3});}
        \scalebox{0.5}{\draw[gray,preaction={-triangle 90,draw,ultra thin,gray}] ([xshift=30pt, yshift=60pt]{pic cs:P2}) to[out=180,in=180] ([xshift=30pt, yshift=35pt]{pic cs:P4});}
    \end{tikzpicture}
\end{minipage}

                \parbox{\linewidth}{\vspace*{8mm}\caption{}\label{subtab:ex_acc_com}}
			\end{subtable}\hspace{20pt}
			\begin{subtable}[t]{0.5\textwidth}
                \begin{minipage}[t][5.25cm][t]{0.45\columnwidth}
    {\small
    \begin{equation*}
        \begin{aligned}
            &\text{\scriptsize Multiplier}         & \text{\color{gray}\tiny 1}  \es{-2} &&            &1\:0\:1\:1                  \\[-2pt]
            &\text{\scriptsize Multiplicand}\bs{5} & \text{\color{gray}\tiny 2}  \es{-2} &&            &1\:1\:0\:1\eqmark{m3}       \\[-2pt]
            &                                      & \text{\color{gray}\tiny 3}  \es{-2} &&            &\gc{0\:0\:0\:0\:}           \\[-2pt]
            &                                      & \text{\color{gray}\tiny 4}  \es{-2} &&            &\gc{1\:0\:1\:1\:}\eqmark{m4}\\[-2pt]
            &                                      & \text{\color{gray}\tiny 5}  \es{-2} &&    \rc{0\:}&\gc{1\:0\:1\:1\:}           \\[-2pt]
            &                                      & \text{\color{gray}\tiny 6}  \es{-2} && \eqmark{S1}&\gc{0\:1\:0\:1\:}1          \\[-2pt]
            &                                      & \text{\color{gray}\tiny 7}  \es{-2} &&            &\gc{0\:0\:0\:0\:}\eqmark{m5}\\[-2pt]
            &                                      & \text{\color{gray}\tiny 8}  \es{-2} &&    \rc{0\:}&\gc{0\:1\:0\:1\:}1          \\[-2pt]
            &                                      & \text{\color{gray}\tiny 9}  \es{-2} && \eqmark{S2}&\gc{0\:0\:1\:0\:}1\:1       \\[-2pt]
            &                                      & \text{\color{gray}\tiny 10} \es{-2} &&            &\gc{1\:0\:1\:1\:}\eqmark{m6}\\[-2pt]
            &                                      & \text{\color{gray}\tiny 11} \es{-2} &&    \rc{0\:}&\gc{1\:1\:0\:1\:}1\:1       \\[-2pt]
            &                                      & \text{\color{gray}\tiny 12} \es{-2} && \eqmark{S3}&\gc{0\:1\:1\:0\:}1\:1\:1    \\[-2pt]
            &                                      & \text{\color{gray}\tiny 13} \es{-2} &&            &\gc{1\:0\:1\:1\:}\eqmark{m7}\\[-2pt]
            &                                      & \text{\color{gray}\tiny 14} \es{-2} &&    \rc{1\:}&\gc{0\:0\:0\:1\:}1\:1\:1    \\[-2pt]
            &\text{\scriptsize Product}            & \text{\color{gray}\tiny 15} \es{-2} && \eqmark{S4}&1\:0\:0\:0\:1\:1\:1\:1
        \end{aligned}
        \tikz[overlay,remember picture, dashed]{\draw([xshift=-8 mm, yshift=-0.75mm]m3.south) -- ([xshift=-0.5mm, yshift=-0.75mm]m3.south);}
        \tikz[overlay,remember picture        ]{\draw([xshift=-11mm, yshift=-0.75mm]m4.south) -- ([xshift=-0.5mm, yshift=-0.75mm]m4.south);}
        \tikz[overlay,remember picture        ]{\draw([xshift=-11mm, yshift=-0.75mm]m5.south) -- ([xshift=-0.5mm, yshift=-0.75mm]m5.south);}
        \tikz[overlay,remember picture        ]{\draw([xshift=-11mm, yshift=-0.75mm]m6.south) -- ([xshift=-0.5mm, yshift=-0.75mm]m6.south);}
        \tikz[overlay,remember picture        ]{\draw([xshift=-11mm, yshift=-0.75mm]m7.south) -- ([xshift=-0.5mm, yshift=-0.75mm]m7.south);}
        \tikz[overlay,remember picture]{\scalebox{0.5}{\draw[gray,preaction={-triangle 90,draw,ultra thin,gray}] ([xshift=-57pt, yshift= 41pt]S1.south) |- ([xshift=-48pt, yshift= 26pt]S1.south);}}
        \tikz[overlay,remember picture]{\scalebox{0.5}{\draw[gray,preaction={-triangle 90,draw,ultra thin,gray}] ([xshift=-57pt, yshift=  9pt]S2.south) |- ([xshift=-48pt, yshift=- 6pt]S2.south);}}
        \tikz[overlay,remember picture]{\scalebox{0.5}{\draw[gray,preaction={-triangle 90,draw,ultra thin,gray}] ([xshift=-57pt, yshift=-22pt]S3.south) |- ([xshift=-48pt, yshift=-37pt]S3.south);}}
        \tikz[overlay,remember picture]{\scalebox{0.5}{\draw[gray,preaction={-triangle 90,draw,ultra thin,gray}] ([xshift=-57pt, yshift=-55pt]S4.south) |- ([xshift=-48pt, yshift=-70pt]S4.south);}}
    \end{equation*}}
\end{minipage}

                \hspace*{-5mm}
                \parbox{\linewidth}{\vspace*{8mm}\caption{}\label{subtab:ex_acc_seq}}
			\end{subtable}
        \end{minipage}
		\parbox{\linewidth}{\vspace*{-1mm}\caption{Accurate (\protect\subref{subtab:ex_acc_com}) combinatorial and (\protect\subref{subtab:ex_acc_seq}) sequential multiplication.}}
    \end{table}
}

\newcommand{\FIGschacccom}{
    \begin{figure*}
        \centering
		\resizebox{\linewidth}{!}{\input{figures/sch_acc_com_multiplier}}
		\caption{Schematic of a combinatorial circuit implementing the multiplication strategy shown in \Cref{subtab:ex_acc_com}. Note that this design may be pipelined with $\log_2n$ levels, each with $\frac{n}{2^k}$ binary $(n+2\rhu{2^{k-2}}){\cdot}n$-bit adders, this would require $n{-}1$ clock cycles. However, the theoretical \gls{cp} of the outermost level would be equivalent to $n{+}\frac{n}{2}$, with a register of a similar width.}
		\label{fig:com_schematic}
    \end{figure*}
}

\newcommand{\FIGschseq}{
    \begin{figure*}
        \centering
        \begin{subfigure}[t]{0.425\textwidth}
			\centering
			\resizebox{\linewidth}{!}{\begin{tikzpicture}
	[
        >          = stealth,
	    alu/.style = {
            trapezium,
            trapezium angle      = 65,
            shape border rotate  = 180,
            minimum width        = 4cm,
            minimum height       = 1.75cm,
            trapezium stretches  = true,
            append after command = {
                \pgfextra
                    \draw[fill=lightgray!25]    (\tikzlastnode.top    left  corner)
                                             -- (\tikzlastnode.top    right corner)
                                             -- (\tikzlastnode.bottom right corner)
                                             -- ($(\tikzlastnode.bottom right corner)!.800!(\tikzlastnode.bottom side)$)
                                             -- ([yshift=-5.5mm]\tikzlastnode.bottom side)
                                             -- ($(\tikzlastnode.bottom side)!.200!(\tikzlastnode.bottom left corner)$)
                                             -- (\tikzlastnode.bottom left  corner)
                                             -- (\tikzlastnode.top    left  corner);
                \endpgfextra
            },
	    },
	    empty_alu/.style = {
            trapezium,
            trapezium angle      = 65,
            shape border rotate  = 180,
            minimum width        = 4cm,
            minimum height       = 1.75cm,
            trapezium stretches  = true,
	    },
	]
	
	\node[alu] at (0,0) (acc_alu) {};
	\node[align=center] at ([yshift = -0.3cm]acc_alu) (label) {\Large Adder\\[1mm]\large +};
	\node[align=center] at ([xshift = -1.3cm, yshift =  0.3cm]acc_alu) (Cin_MMSP)  {\large $\text{C}_{\text{in}}$};
	\node[align=center] at ([xshift = -0.8cm, yshift = -0.6cm]acc_alu) (Cout_MMSP) {\large $\text{C}_{\text{out}}$};
	
	\node[empty_alu] at (4.75,0) (phantom_alu) {};
	\node at ([xshift = -1.25cm]acc_alu.bottom right corner) (cin_reg_TL) {};
	\node at ([xshift =  0.8cm ]cin_reg_TL) (cin_reg_BR) {};
    
	\node at ([yshift = -1cm   ]acc_alu.top right corner) (reg_MSP_LT) {};
	\node at ([yshift = -2.25cm]acc_alu.top left  corner) (reg_MSP_RB) {};
	\node at ([yshift =  2.5cm ]$(reg_MSP_LT) + (reg_MSP_RB)$) (reg_MSP_C) {};
	\draw[fill=lightgray!25] ([xshift = -0.8cm]reg_MSP_LT) rectangle ([xshift = 0.8cm]reg_MSP_RB);
	\node[align=center] at (reg_MSP_C) (shift_MSP) {\large Shift Reg A};
	
	\node at ([yshift = -1cm   ]phantom_alu.top right corner) (reg_LSP_LT) {};
	\node at ([yshift = -2.25cm]phantom_alu.top left  corner) (reg_LSP_RB) {};
	\node at ([yshift =  2.5cm ]$(reg_LSP_LT) + (reg_LSP_RB) - (5,0)$) (reg_LSP_C) {};
	\draw[fill=lightgray!25] ([xshift = -0.8cm]$(reg_LSP_LT) - (0.25,0)$) rectangle ([xshift = 0.8cm]$(reg_LSP_RB) - (0.25,0)$);
	\node[align=center] at (reg_LSP_C) (shift_LSP) {\large Shift Reg B};
	
	\node at ([xshift = -2.25cm]reg_MSP_LT) (cout_reg_TL) {};
	\node at ([xshift =  0.8 cm]cout_reg_TL) (cout_reg_BR) {};
	\draw[fill=lightgray!25] (cout_reg_TL) rectangle ([yshift = -1.25cm]cout_reg_BR);
	
	
	\draw[-triangle 90] ($(Cout_MMSP) - (0.4,0)$) -- ($(Cout_MMSP) - (3  ,0  )$) |- ($(cout_reg_TL) - (0,0.25)$);
	\draw[-triangle 90] ($(cout_reg_BR) + (0.0,-0.25)$) -- ($(cout_reg_BR) + (0.25,-0.25)$) |- ([xshift = 1.7cm]$(cout_reg_BR) - (1.05,0.6)$);

	\draw[-triangle 90] ([xshift = -6.75cm, yshift = -1.5cm]$(cin_reg_BR)  + (7,0.5)$) |- ([xshift = -6.3cm, yshift = -1.5cm]$(cin_reg_BR) + (7,1)$);
	
	\path[
        solid,
        draw       = black,
        line width = 1mm,
        preaction  = {
            -triangle 90,
            thin,
            draw,
            shorten > = -1mm
        },
    ] (acc_alu.south) -- ([yshift = -0.9cm]acc_alu.south);
    
	\path[
        solid,
        draw       = black,
        line width = 1mm,
        preaction  = {
            -triangle 90,
            thin,
            draw,
            shorten > = -1mm
        },
    ] ([xshift = -0.25cm, yshift = 2.75cm]phantom_alu.south) -- ([xshift = -0.25cm, yshift = -0.9cm]phantom_alu.south);
	
	\path[
        solid,
        draw       = black,
        line width = 1mm,
        preaction  = {
            -triangle 90,
            thin,
            draw,
            shorten > = -1mm
        },
    ] ([yshift = -2.25cm]acc_alu.south) -- ([yshift = -3.25cm]acc_alu.south);
    
	\path[
        solid,
        draw       = black,
        line width = 1mm,
        preaction  = {
            -triangle 90,
            thin,
            draw,
            shorten > = -1mm
        },
    ] ([xshift = -0.25cm, yshift = -2.25cm]phantom_alu.south) -- ([xshift = -0.25cm, yshift = -3.25cm]phantom_alu.south);
    
    \path[
        solid,
        draw       = black,
        line width = 1mm,
        preaction  = {
            -triangle 90,
            thin,
            draw,
            shorten > = -1mm
        },
    ]([yshift = -2.55cm]acc_alu.south) -| ($(cin_reg_TL) + (-1.1,0.5)$  ) -| ($(acc_alu.bottom right corner) + (0.85,0.1)$);
    
	\draw[-triangle 90] ([yshift =-2.55cm]acc_alu.south) -| ($(reg_LSP_LT) - (1.5,0.61)$) -- ($(reg_LSP_LT) - (1.05,0.61)$);
	\draw[-triangle 90] ([xshift = -0.25cm, yshift =-2.55cm]phantom_alu.south) -- ([xshift = 1.5cm, yshift =-2.55cm]phantom_alu.south);

	\path[
        solid,
        draw       = black,
        line width = 1mm,
        preaction  = {
            -triangle 90,
            thin,
            draw,
            shorten > = -1mm
        },
    ] ($(acc_alu.bottom left corner) + (-0.85,1)$) -- ($(acc_alu.bottom left corner) + (-0.85,0.1)$);
    
	
	\draw[black,fill=black] ([yshift = -2.55cm]acc_alu.south) circle (.75ex);
	\draw[black,fill=black] ([xshift = -0.25cm, yshift = -2.55cm]phantom_alu.south) circle (.75ex);
	
	
	\node[align=center] at ([xshift = -6.75cm, yshift = -1.5cm]$(cin_reg_BR) + (7,0.3)$) (foo) {\selectfont $0$};
	
	
	\draw ($(cout_reg_TL) + (0,-0.9)$) -- ($(cout_reg_TL) + (0.15,-1)$) -- ($(cout_reg_TL) + (0,-1.1)$);
	\draw ($(shift_MSP.south west) - (0.85,0.2)$) -- ($(shift_MSP.south west) - (0.7 ,0.1)$) -- ($(shift_MSP.south west) + (-0.85 ,0)$);
    \draw ($(shift_LSP.south west) - (0.86,0.2)$) -- ($(shift_LSP.south west) - (0.71,0.1)$) -- ($(shift_LSP.south west) + (-0.86 ,0)$);
	

    \draw ([xshift =-3.75cm, yshift = 2.2 cm]phantom_alu.south) -- ([xshift =-3.45cm, yshift = 2.5cm]phantom_alu.south);
    \draw ([xshift =-0.4 cm, yshift = 2.2 cm]phantom_alu.south) -- ([xshift =-0.1 cm, yshift = 2.5cm]phantom_alu.south);
    
	\draw ([xshift =-4.5 cm, yshift =-0.6 cm]acc_alu.south)     -- ([xshift =-4.2 cm, yshift =-0.3cm]acc_alu.south);
	\draw ([xshift =-0.15cm, yshift =-0.6 cm]acc_alu.south)     -- ([xshift = 0.15cm, yshift =-0.3cm]acc_alu.south);
	
	\draw ([xshift =-0.15cm, yshift =-3.05cm]acc_alu.south)     -- ([xshift = 0.15cm, yshift =-2.75cm]acc_alu.south);
	\draw ([xshift =-0.4 cm, yshift =-3.05cm]phantom_alu.south) -- ([xshift =-0.1 cm, yshift =-2.75cm]phantom_alu.south);
	
	
	\node[align=center] at ($(phantom_alu.bottom left corner) + (-5.9 , 0.5)$) (foo) {\selectfont $n$};
	\node[align=center] at ($(phantom_alu.bottom left corner) + (-2.55, 0.5)$) (foo) {\selectfont $n$};
	
	\node[align=center] at ($(acc_alu.bottom left corner)     - (6.7 , 2.2)$) (foo) {\selectfont $n$};
	\node[align=center] at ($(acc_alu.bottom left corner)     - (2.3 , 2.2)$) (foo) {\selectfont $n$};
	
	\node[align=center] at ([xshift =-0.325cm, yshift =-2.95cm]acc_alu.south) (foo) {\selectfont $n$};
	\node[align=center] at ([xshift = 4.2cm , yshift =-2.95cm]acc_alu.south) (foo) {\selectfont $n$};
	
	\node[align=center] at ($(acc_alu.bottom left corner) + (-0.95,1.25)$) (MSP_multiplicand) {$B_{lsb} \wedge (a_{n-1} \ldots a_0)$};
	\node[align=center] at ($(phantom_alu.south) + (-0.2,3.02)$) (multiplier) {$b_{n-1} \ldots b_0$};

	\node[align=center] at ($(cout_reg_TL.north west) + (0.3,-0.4)$) (ff_d_out) {$D$};
	\node[align=center] at ($(cout_reg_TL.north east) + (0.5,-0.4)$) (ff_q_out) {$Q$};
	
	\node[align=center] at ([yshift =-3.75cm]acc_alu.south) (MMSP_product) {$p_{2n-1} \ldots p_n$};
	\node[align=center] at ([yshift =-3.75cm]phantom_alu.south) (MLSP_product) {$p_{n-1} \ldots p_0$};
	
	\node[align=center] at ([xshift =-3.45cm, yshift = -2.78cm]phantom_alu.south) (MLSP_product) {$A_{lsb}$};
	\node[align=center] at ([xshift = 1.8 cm, yshift = -2.6 cm]phantom_alu.south) (MLSP_product) {$B_{lsb}$};
	
\end{tikzpicture}}
			\parbox{\linewidth}{\vspace*{-0mm}\caption{}\label{subfig:sch_acc_seq}}
		\end{subfigure}\hfill%
		\begin{subfigure}[t]{0.55\textwidth}
			\centering
			\resizebox{\linewidth}{!}{\input{figures/sch_app_seq_multiplier}}
			\parbox{\linewidth}{\vspace*{-0mm}\caption{}\label{subfig:sch_app_seq}}
		\end{subfigure}\\[3mm]
		\parbox{\linewidth}{\vspace*{-3mm}\caption{Schematics of sequential circuits implementing the (\protect\subref{subfig:sch_acc_seq}) accurate and the (\protect\subref{subfig:sch_app_seq}) approximate multiplication strategies shown in \Cref{subtab:ex_acc_seq,subfig:ex_app_seq}, respectively. The D flip-flops have asynchronous \textit{clear} inputs. The controllers and clock lines are not shown. The shift registers have synchronous inputs for parallel \textit{load}, \textit{shifting} to the right with left serial input, and \textit{clear} to set to 0---see \mbox{Line 3} of \Cref{subtab:ex_acc_seq,subfig:ex_app_seq}, respectively.  Whilst not shown in (\protect\subref{subfig:sch_acc_seq}), in (\protect\subref{subfig:sch_app_seq}) a decrement unit, which informs the controller about sequence completion, and enables the multiplexing of the least significant $n+t$ bits according to the carry-out of the last accumulation is shown.}
		\label{fig:seq_schematics}}
    \end{figure*}
}

\newcommand{\FIGindicesex}[1]{
	\begin{table*}
		\begin{subtable}[t]{0.5\textwidth}    
			\centering
			\small
\renewcommand{\arraystretch}{1.1}
\newcommand{\tgray}[1]{\leavevmode\color{gray}\tiny#1}
\begin{tabular}{l@{}P{15pt}@{}P{12pt}@{}P{25pt}@{}P{25pt}@{}P{25pt}@{}P{25pt}@{}P{20pt}@{}P{20pt}@{}P{20pt}@{}P{20pt}}
    \scriptsize Multiplier   & {\tgray{1}}  &                & $a_3$                & $a_2$                & $a_1$                & $a_0$                &            &            &            &   \\
    \scriptsize Multiplicand & {\tgray{2}}  &                & $b_3$                & $b_2$                & $b_1$                & $b_0$                &            &            &            &            \\
    \cdashline{4-7}\\[-8pt]
                             & {\tgray{3}}  &                & $\gc{0}$             & $\gc{0}$             & $\gc{0}$             & $\gc{0}$             &            &            &            &            \\[-2pt]
                             & {\tgray{4}}  &                & $\gc{a_3{\mint}b_0}$ & $\gc{a_2{\mint}b_0}$ & $\gc{a_1{\mint}b_0}$ & $\gc{a_0{\mint}b_0}$ &            &            &            &            \\
    \cline{3-7}\\[-8pt]
                             & {\tgray{5}}  & $\gc{S_4^0}$   & $\gc{S_3^0}$         & $\gc{S_2^0}$         & $\gc{S_1^0}$         & $\gc{S_0^0}$         & $\rb{b_3}$ & $\rb{b_2}$ & $\rb{b_1}$ & $\rb{b_0}$ \\
                             & {\tgray{6}}  & \tikzmark{S40} & $\gc{S_4^0}$         & $\gc{S_3^0}$         & $\gc{S_2^0}$         & $\gc{S_1^0}$         & $S_0^0$    & $\rb{b_3}$ & $\rb{b_2}$ & $\rb{b_1}$ \\
                             & {\tgray{7}}  &                & $\gc{a_3{\mint}b_1}$ & $\gc{a_2{\mint}b_1}$ & $\gc{a_1{\mint}b_1}$ & $\gc{a_0{\mint}b_1}$ &            &            &            &            \\
    \cline{3-7}\\[-8pt]
                             & {\tgray{8}}  & $\gc{S_4^1}$   & $\gc{S_3^1}$         & $\gc{S_2^1}$         & $\gc{S_1^1}$         & $\gc{S_0^1}$         & $S_0^0$    & $\rb{b_3}$ & $\rb{b_2}$ & $\rb{b_1}$ \\
                             & {\tgray{9}}  & \tikzmark{S41} & $\gc{S_4^1}$         & $\gc{S_3^1}$         & $\gc{S_2^1}$         & $\gc{S_1^1}$         & $S_0^1$    & $S_0^0$    & $\rb{b_3}$ & $\rb{b_2}$ \\
                             & {\tgray{10}} &                & $\gc{a_3{\mint}b_2}$ & $\gc{a_2{\mint}b_2}$ & $\gc{a_1{\mint}b_2}$ & $\gc{a_0{\mint}b_2}$ &            &            &            &            \\
    \cline{3-7}\\[-8pt]
                             & {\tgray{11}} & $\gc{S_4^2}$   & $\gc{S_3^2}$         & $\gc{S_2^2}$         & $\gc{S_1^2}$         & $\gc{S_0^2}$         & $S_0^1$    & $S_0^0$    & $\rb{b_3}$ & $\rb{b_2}$ \\
                             & {\tgray{12}} & \tikzmark{S42} & $\gc{S_4^2}$         & $\gc{S_3^2}$         & $\gc{S_2^2}$         & $\gc{S_1^2}$         & $S_0^2$    & $S_0^1$    & $S_0^0$    & $\rb{b_3}$ \\
                             & {\tgray{13}} &                & $\gc{a_3{\mint}b_3}$ & $\gc{a_2{\mint}b_3}$ & $\gc{a_1{\mint}b_3}$ & $\gc{a_0{\mint}b_3}$ &            &            &            &            \\
    \cline{3-7}\\[-8pt]
                             & {\tgray{14}} & $\gc{S_4^3}$   & $\gc{S_3^3}$         & $\gc{S_2^3}$         & $\gc{S_1^3}$         & $\gc{S_0^3}$         & $S_0^2$    & $S_0^1$    & $S_0^0$    & $\rb{b_3}$ \\
    \scriptsize Product      & {\tgray{15}} & \tikzmark{S43} & $S_4^3$              & $S_3^3$              & $S_2^3$              & $S_1^3$              & $S_0^3$    & $S_0^2$    & $S_0^1$    & $S_0^0$
\end{tabular}
\begin{tikzpicture}[overlay,remember picture]
    \scalebox{0.65}{\draw[gray,preaction={-triangle 90,draw,ultra thin,gray}] ([xshift=-105pt, yshift= 24pt]{pic cs:S40}) |- ([xshift=-84pt, yshift= 17pt]{pic cs:S40});}
    \scalebox{0.65}{\draw[gray,preaction={-triangle 90,draw,ultra thin,gray}] ([xshift=-105pt, yshift= 04pt]{pic cs:S41}) |- ([xshift=-84pt, yshift=-03pt]{pic cs:S41});}
    \scalebox{0.65}{\draw[gray,preaction={-triangle 90,draw,ultra thin,gray}] ([xshift=-105pt, yshift=-14pt]{pic cs:S42}) |- ([xshift=-84pt, yshift=-21pt]{pic cs:S42});}
    \scalebox{0.65}{\draw[gray,preaction={-triangle 90,draw,ultra thin,gray}] ([xshift=-105pt, yshift=-34pt]{pic cs:S43}) |- ([xshift=-84pt, yshift=-41pt]{pic cs:S43});}
\end{tikzpicture}

			\caption{}
			\label{subfig:indices}
		\end{subtable}%
		\begin{subtable}[t]{0.5\textwidth}
			\centering
			\vspace{-3.4cm}
			\begin{minipage}[t][0cm][t]{0.45\columnwidth}
    {\small
    \begin{equation*}
        \begin{aligned}
            &\text{\scriptsize Multiplier}         & \text{\color{gray}\tiny 1}  \es{-2} &&            &1\:0\:1\:1                  \\[-0.75pt]
            &\text{\scriptsize Multiplicand}\bs{5} & \text{\color{gray}\tiny 2}  \es{-2} &&            &1\:1\:0\:1\eqmark{q3}       \\[-0.75pt]
            &                                      & \text{\color{gray}\tiny 3}  \es{-2} &&            &\gc{0\:0\:0\:0\:}           \\[-0.75pt]
            &                                      & \text{\color{gray}\tiny 4}  \es{-2} &&            &\gc{1\:0\:1\:1\:}\eqmark{q4}\\[-0.75pt]
            &                                      & \text{\color{gray}\tiny 5}  \es{-2} &&    \rc{0\:}&\gc{1\:0\:1\:1\:}           \\[-0.75pt]
            &                                      & \text{\color{gray}\tiny 6}  \es{-2} && \eqmark{A1}&\gc{0\:1\:0\:1\:}1          \\[-0.75pt]
            &                                      & \text{\color{gray}\tiny 7}  \es{-2} &&            &\gc{0\:0\:0\:0\:}\eqmark{q5}\\[-0.75pt]
            &                                      & \text{\color{gray}\tiny 8}  \es{-2} &&    \rc{0\:}&\gc{0\:1\:0\:1\:}1          \\[-0.75pt]
            &                                      & \text{\color{gray}\tiny 9}  \es{-2} && \eqmark{A2}&\gc{0\:0\:1\:0\:}1\:1       \\[-0.75pt]
            &                                      & \text{\color{gray}\tiny 10} \es{-2} &&            &\gc{1\:0\:1\:1\:}\eqmark{q6}\\[-0.75pt]
            &                                      & \text{\color{gray}\tiny 11} \es{-2} &&    \rc{0\:}&\gc{1\:0\:0\:1\:}1\:1       \\[-0.75pt]
            &                                      & \text{\color{gray}\tiny 12} \es{-2} && \eqmark{A3}&\gc{0\:1\:0\:0\:}1\:1\:1    \\[-0.75pt]
            &                                      & \text{\color{gray}\tiny 13} \es{-2} &&            &\gc{1\:0\:1\:1\:}\eqmark{q7}\\[-0.75pt]
            &                                      & \text{\color{gray}\tiny 14} \es{-2} &&    \rc{1\:}&\gc{0\:0\:1\:1\:}1\:1\:1    \\[-0.75pt]
            &\text{\scriptsize Product}            & \text{\color{gray}\tiny 15} \es{-2} && \eqmark{A4}&1\:0\:0\:1\:1\:1\:1\:1
        \end{aligned}
        \tikz[overlay,remember picture, draw=my_orange, opacity=0.5]{\draw[fill=my_orange, opacity=0.5]([xshift=-4.7mm, yshift=6.5mm]q6.south) rectangle ([xshift=-2.5mm, yshift=-0.3mm]q6.south);}
        \tikz[overlay,remember picture, draw=my_blue  , opacity=0.5]{\draw[fill=my_blue  , opacity=0.5]([xshift=-6.9mm, yshift=6.5mm]q7.south) rectangle ([xshift=-4.7mm, yshift=-0.3mm]q7.south);}
        %
        %
        \tikz[overlay,remember picture, dashed]{\draw([xshift=- 8  mm, yshift=-0.75mm]q3.south) -- ([xshift=-0.55mm, yshift=-0.75mm]q3.south);}
        \tikz[overlay,remember picture        ]{\draw([xshift=-10.5mm, yshift=-0.75mm]q4.south) -- ([xshift=-0.55mm, yshift=-0.75mm]q4.south);}
        \tikz[overlay,remember picture        ]{\draw([xshift=-10.5mm, yshift=-0.75mm]q5.south) -- ([xshift=-0.55mm, yshift=-0.75mm]q5.south);}
        \tikz[overlay,remember picture        ]{\draw([xshift=-10.5mm, yshift=-0.75mm]q6.south) -- ([xshift=-0.55mm, yshift=-0.75mm]q6.south);}
        \tikz[overlay,remember picture        ]{\draw([xshift=-10.5mm, yshift=-0.75mm]q7.south) -- ([xshift=-0.55mm, yshift=-0.75mm]q7.south);}
        \tikz[overlay,remember picture]{\scalebox{0.5}{\draw[gray,preaction={-triangle 90,draw,ultra thin,gray}] ([xshift=-56pt, yshift= 45pt]A1.south) |- ([xshift=-47pt, yshift= 27pt]A1.south);}}
        \tikz[overlay,remember picture]{\scalebox{0.5}{\draw[gray,preaction={-triangle 90,draw,ultra thin,gray}] ([xshift=-56pt, yshift= 10pt]A2.south) |- ([xshift=-47pt, yshift=- 5pt]A2.south);}}
        \tikz[overlay,remember picture]{\scalebox{0.5}{\draw[gray,preaction={-triangle 90,draw,ultra thin,gray}] ([xshift=-56pt, yshift=-26pt]A3.south) |- ([xshift=-47pt, yshift=-41pt]A3.south);}}
        \tikz[overlay,remember picture]{\scalebox{0.5}{\draw[gray,preaction={-triangle 90,draw,ultra thin,gray}] ([xshift=-56pt, yshift=-62pt]A4.south) |- ([xshift=-47pt, yshift=-77pt]A4.south);}}
    \end{equation*}}
\end{minipage}

			\vspace{6.47cm} 
			\caption{}
			\label{subfig:ex_app_seq}
		\end{subtable}
		\caption{(\protect\subref{subfig:indices}) Indexing of a sequential multiplication algorithm. (\protect\subref{subfig:ex_app_seq}) Approximate sequential multiplier with two 2-bit adders. That is, $t=2$.}
		\label{fig:indices_ex}
	\end{table*}
}

\newcommand{\FIGexappseq}[1]{
    \begin{table}[#1]
        \centering
        
        \vspace{10mm} 
        \caption{Approximate sequential multiplier with two 2-bit adders. That is, $t=2$.}
        \label{fig:ex_app_seq}
    \end{table}
}

\newcommand{\FIGplotsems}[1]{
    \begin{figure}[#1]
        \addtocounter{figure}{1}
        \pgfplotsset{yticklabel style = {text width = 3em,align = right}}
        \addtocounter{subfigure}{-2}
        \captionsetup[subfigure]{labelformat=parens}
        \centering
        \begin{tikzpicture}
    \begin{semilogxaxis}[
        width            = 0.75\columnwidth,
        height           = 4cm,
        xlabel           = {},
        ylabel           = {$\er$},
        ylabel style     = {yshift = -2mm},
        grid             = both,
        grid style       = dotted,
        ytick            = {0.0, 0.2, 0.4, 0.6, 0.8, 1},
        xtick            = {4, 8, 16, 32},
        xticklabels      = {4, 8, 16, 32},
        enlarge x limits = {abs = 0.2cm},
        enlarge y limits = {abs = 0.2cm},
        xmin             = 4,
        xmax             = 32,
        ymin             = 0.0,
        ymax             = 1,
        name             = ER_plot,
        set layers,
    ]
    
        \addplot[
            only marks,
            solid,
            mark         = triangle*,
            draw         = my_blue,
            fill         = my_blue,
            fill opacity = 0.15,
        ] table [
            x = n,
            y = ER,
            col sep = tab,
        ] {results/results_metrics_4_16.dat};
        
        \addplot[
            only marks,
            every mark/.append style = {
                solid,
                draw         = my_blue,
                fill         = my_blue,
                fill opacity = 0.15,
                rotate       = 180,
            },
            mark = triangle*,
        ] table [
            x = n,
            y = ER,
            col sep = tab,
        ] {results/results_sim_metrics_18_32.dat};
        
        \foreach \i in {0.05,0.2,0.6,0.65,0.82,0.85,0.88,0.89}
        {
            \edef\temp{\noexpand\filldraw[amethyst,fill opacity = 0.15] (axis cs:8,\i) circle (0.5mm);}\temp
        }
        
        \foreach \i in {0.9309,0.9465,0.9563,0.9602,0.9743,0.9806,0.9901}
        {
            \edef\temp{\noexpand\filldraw[red!75,fill opacity = 0.15] (axis cs:8,\i) circle (0.5mm);}\temp
        }
        
        \foreach \i in {0.9977,0.998,0.9995,0.9997}
        {
            \edef\temp{\noexpand\filldraw[red!75,fill opacity = 0.15] (axis cs:16,\i) circle (0.5mm);}\temp
        }
        
    \end{semilogxaxis}
    
    \node[align=center] at ([xshift=-0.51\columnwidth,yshift=-2.5mm]ER_plot.center) (foo) {\parbox{0\linewidth}{(a)}};
    
\end{tikzpicture}   \\
        \begin{tikzpicture}
    \begin{loglogaxis}[
        width            = 0.75\columnwidth,
        height           = 4cm,
        xlabel           = {},
        ylabel           = {$\mae$},
        ylabel style     = {yshift = -0.2cm},
        grid             = both,
        grid style       = dotted,
        ytick            = {1e+1, 1e+5, 1e+9, 1e+13, 1e+18},
        xtick            = {4, 8, 16, 32},
        xticklabels      = {4, 8, 16, 32},
        enlarge x limits = {abs = 0.2cm},
        enlarge y limits = {abs = 0.2cm},
        xmin             = 4,
        xmax             = 32,
        ymin             = 1e+1,
        ymax             = 1e+18,
        name             = MAE_plot,
        set layers,
    ]
    
        \addplot[
            only marks,
            solid,
            mark         = triangle*,
            draw         = my_blue,
            fill         = my_blue,
            fill opacity = 0.15,
        ] table [
            x = n,
            y = MAE,
            col sep = tab,
        ] {results/results_metrics_4_16.dat};
        
        \addplot[
            only marks,
            solid,
            mark         = triangle*,
            draw         = my_blue,
            fill         = my_blue,
            fill opacity = 0.15,
        ] table [
            x = n,
            y = MAE,
            col sep = tab,
        ] {results/results_sim_metrics_18_32.dat};

        \foreach \i in {2, 10, 18, 50, 82, 114, 242, 370, 498, 626, 1138, 1650, 2162, 4210, 6258, 14450}
        {
            \edef\temp{\noexpand\filldraw[my_green,fill opacity = 0.15] (axis cs:8,\i) circle (0.5mm);}\temp
        }
        
        \foreach \i in {4096,4225,5369,9953,16320,4288,5440,10048,16320,4223,4605,5369,15104,28416}
        {
            \edef\temp{\noexpand\filldraw[red!75,fill opacity = 0.15] (axis cs:8,\i) circle (0.5mm);}\temp
        }
        
        \foreach \i in {2.8e+1,1.02e+3,2.50e+4,1.08e+5,4.56e+5,14.89e+5,53.17e+5,594.85e+5,5949.93e+5}
        {
            \edef\temp{\noexpand\filldraw[my_orange,fill opacity = 0.15] (axis cs:16,\i) circle (0.5mm);}\temp
        }
        
        \foreach \i in {268.44e+6,274.5e+6,280.89e+6,293.47e+6,318.63e+6,469.45e+6,1073.73e+6,520.03e+6,1878.89e+6}
        {
            \edef\temp{\noexpand\filldraw[red!75,fill opacity = 0.15] (axis cs:16,\i) circle (0.5mm);}\temp
        }
        
        \foreach \i in {1.02e+3,4.82e+5,1.43e+8,2.53e+9,3.99e+10,6.88e+11,1.54e+14,2.13e+16,1.62e+18}
        {
            \edef\temp{\noexpand\filldraw[my_orange,fill opacity = 0.15] (axis cs:32,\i) circle (0.5mm);}\temp
        }
        
    \end{loglogaxis}
    
    \node[align=center] at ([xshift=-0.51\columnwidth,yshift=-2.5mm]ER_plot.center) (foo) {\parbox{0\linewidth}{(b)}}; 
    
\end{tikzpicture}  \\
        \begin{tikzpicture}
    \begin{loglogaxis}[
        width            = 0.75\columnwidth,
        height           = 4cm,
        xlabel           = {},
        ylabel           = {$\med$},
        ylabel style     = {yshift = -0.2cm},
        grid             = both,
        grid style       = dotted,
        ytick            = {1e+1,1e+4,1e+7,1e+10,1e+13},
        xtick            = {4, 8, 16, 32},
        xticklabels      = {4, 8, 16, 32},
        enlarge x limits = {abs = 0.2cm},
        enlarge y limits = {abs = 0.2cm},
        xmin             = 4,
        xmax             = 32,
        ymin             = 1e+1,
        ymax             = 1e+13,
        name             = MED_plot,
        set layers,
    ]
    
        \addplot[
            only marks,
            solid,
            mark         = triangle*,
            draw         = my_blue,
            fill         = my_blue,
            fill opacity = 0.15,
        ] table [
            x = n,
            y = MED,
            col sep = tab,
        ] {results/results_metrics_4_16.dat};
        
        \addplot[
           only marks,
            every mark/.append style = {
                solid,
                draw         = my_blue,
                fill         = my_blue,
                fill opacity = 0.15,
                rotate       = 180,
            },
            mark = triangle*,
        ] table [
            x = n,
            y = MED,
            col sep = tab,
        ] {results/results_sim_metrics_18_32.dat};
        
        \foreach \i in {0.3e+2,0.31e+2,6e+2,7.5e+2,12.5e+2,14e+2,37.5e+2,38.5e+2}
        {
            \edef\temp{\noexpand\filldraw[amethyst,fill opacity = 0.15] (axis cs:8,\i) circle (0.5mm);}\temp
        }
        
        \foreach \i in {4.60e+2,2.70e+2,6.00e+1,6.00e+2,7.00e+1}
        {
            \edef\temp{\noexpand\filldraw[pink,fill opacity = 0.15] (axis cs:8,\i) circle (0.5mm);}\temp
        }
        
        \foreach \i in {1.4e+8,4.38e+6,6.68e+6,5.87e+5,1.54e+7,2.44e+7}
        {
            \edef\temp{\noexpand\filldraw[pink,fill opacity = 0.15] (axis cs:16,\i) circle (0.5mm);}\temp
        }
        
        \foreach \i in {12884508.675,639071630.28,8589672.45,165351194.6625,112954192.7175,66999445.11,52826485.5675,42089395.005}
        {
            \edef\temp{\noexpand\filldraw[brown,fill opacity = 0.15] (axis cs:16,\i) circle (0.5mm);}\temp
        }
        
        
    \end{loglogaxis}
    
    \node[align=center] at ([xshift=-0.51\columnwidth,yshift=-2.5mm]ER_plot.center) (foo) {\parbox{0\linewidth}{(c)}}; 
    
\end{tikzpicture}  \\
        \begin{tikzpicture}
    \begin{loglogaxis}[
        width            = 0.75\columnwidth,
        height           = 4cm,
        xlabel           = {},
        ylabel           = {$\nmed$},
        ylabel style     = {yshift = -0.2cm},
        grid             = both,
        grid style       = dotted,
        ytick            = {1e-17,1e-13,1e-9,1e-5,1e-1},
        xtick            = {4, 8, 16, 32},
        xticklabels      = {4, 8, 16, 32},
        enlarge x limits = {abs = 0.2cm},
        enlarge y limits = {abs = 0.2cm},
        xmin             = 4,
        xmax             = 32,
        ymin             = 1e-17,
        ymax             = 1e-1,
        name             = NMED_plot,
        set layers,
    ]
    
        \addplot[
            only marks,
            solid,
            mark         = triangle*,
            draw         = my_blue,
            fill         = my_blue,
            fill opacity = 0.15,
        ] table [
            x = n,
            y = NMED,
            col sep = tab,
        ] {results/results_metrics_4_16.dat};
        
        \addplot[
            only marks,
            every mark/.append style = {
                solid,
                draw         = my_blue,
                fill         = my_blue,
                fill opacity = 0.15,
                rotate       = 180,
            },
            mark = triangle*,
        ] table [
            x = n,
            y = NMED,
            col sep = tab,
        ] {results/results_sim_metrics_18_32.dat};
        
        \foreach \i in {7.69e-6,3.84e-5,1.92e-4,4.38e-4,9.30e-4,2.41e-3,6.34e-3,8.31e-3,5.54e-2}
        {
            \edef\temp{\noexpand\filldraw[my_green,fill opacity = 0.15] (axis cs:8,\i) circle (0.5mm);}\temp
        }
        
        \foreach \i in {0.93e-2,0.9e-2,1.01e-2,2.16e-2,4.11e-2,0.87e-2,1.01e-2,2.22e-2,4.15e-2,0.81e-2,0.78e-2,0.85e-2,2.61e-2,5.46e-2}
        {
            \edef\temp{\noexpand\filldraw[red!75,fill opacity = 0.15] (axis cs:8,\i) circle (0.5mm);}\temp
        }
        
        \foreach \i in {0.007,0.0042,0.0028,0.0016,0.0009,0.0092,0.0053,0.0033,0.0019,0.001}
        {
            \edef\temp{\noexpand\filldraw[pink,fill opacity = 0.15] (axis cs:8,\i) circle (0.5mm);}\temp
        }
        
        \foreach \i in {2.79e-9,8.43e-8,1.68e-6,1.16e-5,3.62e-5,13.3e-5,50.8e-5,703e-5,4420e-5}
        {
            \edef\temp{\noexpand\filldraw[my_orange,fill opacity = 0.15] (axis cs:16,\i) circle (0.5mm);}\temp
        }
        
        \foreach \i in {9.256e-3,13.634e-3,41.636e-3,54.596e-3}
        {
            \edef\temp{\noexpand\filldraw[red!75,fill opacity = 0.15] (axis cs:16,\i) circle (0.5mm);}\temp
        }
        
        \foreach \i in {3.59e-3,1.02e-3,3.43e-4,3.02e-5,5.1e-4,1.24e-3}
        {
            \edef\temp{\noexpand\filldraw[pink,fill opacity = 0.15] (axis cs:16,\i) circle (0.5mm);}\temp
        }
        
        \foreach \i in {1.93e-17,7.92e-15,2.54e-12,4.62e-11,6.78e-10,103e-10,21400e-10,4.16e-4,1.05e-1}
        {
            \edef\temp{\noexpand\filldraw[my_orange,fill opacity = 0.15] (axis cs:32,\i) circle (0.5mm);}\temp
        }
        
        \foreach \i in {3.59e-3,3.42e-4,3.03e-5,5.1e-4,1e-3,4e-4,3.87e-3}
        {
            \edef\temp{\noexpand\filldraw[pink,fill opacity = 0.15] (axis cs:32,\i) circle (0.5mm);}\temp
        }
        
    \end{loglogaxis}
    
    \node[align=center] at ([xshift=-0.51\columnwidth,yshift=-2.5mm]ER_plot.center) (foo) {\parbox{0\linewidth}{(d)}}; 
    
\end{tikzpicture} \\
        \begin{tikzpicture}
    \begin{loglogaxis}[
        width            = 0.75\columnwidth,
        height           = 4cm,
        xlabel           = {$\text{Bitwidth }n$},
        ylabel           = {$\mred$},
        ylabel style     = {yshift = -0.2cm},
        grid             = both,
        grid style       = dotted,
        ytick            = {1e-11, 1e-7, 1e-3, 1e+1, 1e+5},
        xtick            = {4, 8, 16, 32},
        xticklabels      = {4, 8, 16, 32},
        enlarge x limits = {abs = 0.2cm},
        enlarge y limits = {abs = 0.2cm},
        xmin             = 4,
        xmax             = 32,
        ymin             = 1e-11,
        ymax             = 1e+5,
        name             = MRED_plot,
        set layers,
    ]
    
        \addplot[
            only marks,
            solid,
            mark         = triangle*,
            draw         = my_blue,
            fill         = my_blue,
            fill opacity = 0.15,
        ] table [
            x = n,
            y = MRED,
            col sep = tab,
        ] {results/results_metrics_4_16.dat};
        
        \addplot[
            only marks,
            every mark/.append style = {
                solid,
                draw         = my_blue,
                fill         = my_blue,
                fill opacity = 0.15,
                rotate       = 180,
            },
            mark = triangle*,
        ] table [
            x = n,
            y = MRED,
            col sep = tab,
        ] {results/results_sim_metrics_18_32.dat};
        
        \foreach \i in {0.15,0.2,5,6.5,11,12,20,22}
        {
            \edef\temp{\noexpand\filldraw[amethyst,fill opacity = 0.15] (axis cs:8,\i) circle (0.5mm);}\temp
        }
        
        \foreach \i in {3.76e-2,7.88e-2,14.42e-2,16.79e-2,23.28e-2}
        {
            \edef\temp{\noexpand\filldraw[red!75,fill opacity = 0.15] (axis cs:8,\i) circle (0.5mm);}\temp
        }
        
        \foreach \i in {0.0283,0.0218,0.017,0.0117,0.0079,0.0384,0.028,0.0209,0.0141,0.0096}
        {
            \edef\temp{\noexpand\filldraw[pink,fill opacity = 0.15] (axis cs:8,\i) circle (0.5mm);}\temp
        }
        
        \foreach \i in {7.15e-7,2.49e-4,3.37e-3,1.39e-2,9.94e-2,12.4e-2,52e-2,1021e-2,1133e-2}
        {
            \edef\temp{\noexpand\filldraw[my_orange,fill opacity = 0.15] (axis cs:16,\i) circle (0.5mm);}\temp
        }
        
        \foreach \i in {3.85e-2,5.43e-2,15.1e-2,16.98e-2,22.43e-2}
        {
            \edef\temp{\noexpand\filldraw[red!75,fill opacity = 0.15] (axis cs:16,\i) circle (0.5mm);}\temp
        }
        
        \foreach \i in {1.80e-2,7.55e-3,5.06e-4,3.45e-3}
        {
            \edef\temp{\noexpand\filldraw[pink,fill opacity = 0.15] (axis cs:16,\i) circle (0.5mm);}\temp
        }
        
        \foreach \i in {7.01e-12,2.52e-9,1.16e-6,1.42e-5,1.83e-4,23.2e-4,5350e-4,5.79e+1,6.92e+4}
        {
            \edef\temp{\noexpand\filldraw[my_orange,fill opacity = 0.15] (axis cs:32,\i) circle (0.5mm);}\temp
        }
        
    \end{loglogaxis}
    
    \node[align=center] at ([xshift=-0.51\columnwidth,yshift=-2.5mm]ER_plot.center) (foo) {\parbox{0\linewidth}{(e)}}; 
    
\end{tikzpicture} \\
        \addtocounter{figure}{-1}
        \caption{Relations: \capmark{my_green} \cite{Liu:19.2}; \capmark{pink} \cite{Toan:20}; \capmark{my_orange} \cite{Liu:19.1}; \capmark{red!75} \cite{Liu:18}; \capmark{amethyst} \cite{Guo:20}; \capmark{brown} \cite{Ebrahimi:20}. The symbol \trimark{my_blue} marks our findings determined using exhaustive experimentation. The symbol \trimarkinv{my_blue} marks our findings evaluated using \gls{mc} simulations. Note that multiple points taken from the same source are shown as the referenced works report multiple variations of their architectures for the same bitwidths. Similarly, the multiple markings showcasing our design correspond to designs with different chosen splitting points $t$, $t\op{\in} \{2,\ldots,n/2\}$ in the carry chain.}
        \label{fig:plots_ems}
    \end{figure}
}

\newcommand{\FIGcdfs}[1]{
    \begin{figure}[#1]
        \centering
        \begin{tikzpicture}
    \begin{axis}[
        width             = 0.75\columnwidth,
        height            = 4cm,
        xlabel            = {$\widehat{ED}(p,\hat{p})$},
        ylabel            = {Percent},
        ylabel style      = {yshift = -0.2cm},
        grid              = both,
        grid style        = dotted,
        xtick             = {0, 0.2, 0.4, 0.6, 0.8, 1},
        extra x ticks     = {0.15},
        extra x tick style = {
            grid              = none,
            font              = \tiny,
            tick align        = outside,
            tick pos          = left,
            major tick length = 0.75\baselineskip,
            tick style        = {
                black!25,
                line width = 0.15mm,
            },
        },
        ytick             = {0, 0.25, 0.5, 0.75, 1},
        yticklabels       = {0, 25, 50, 75, 100},
        enlarge x limits  = {abs = 0.2cm},
        enlarge y limits  = {abs = 0.2cm},
        xmin              = 0,
        xmax              = 1,
        ymin              = 0,
        ymax              = 1,
        legend style      = {
			at     = {(0.78,0.05)},
			anchor = south,
			draw   = none,
			nodes  = {
                scale  = 0.65, 
                anchor = west,
                transform shape,
            },
            legend image post style = {scale = 0.3},
		},
		legend columns = 1,
        set layers,
    ]

        \addplot[black!25,forget plot] coordinates {( 0.15, 1.1) (0.15,-0.1)};
        
        \addplot[
            my_blue,
        ] table [
            x = x,
            y = y,
            col sep = tab,
        ] {results/CDF_6_3.dat};
        \addlegendentry{$n=\phantom{1}6,~~t=3$};
        
        \addplot[
            red,
        ] table [
            x = x,
            y = y,
            col sep = tab,
        ] {results/CDF_8_4.dat};
        \addlegendentry{$n=\phantom{1}8,~~t=4$};
        
        \addplot[
            pink,
        ] table [
            x = x,
            y = y,
            col sep = tab,
        ] {results/CDF_10_5.dat};
        \addlegendentry{$n=10,~~t=5$};
        
        \addplot[
            purple,
        ] table [
            x = x,
            y = y,
            col sep = tab,
        ] {results/CDF_12_6.dat};
        \addlegendentry{$n=12,~~t=6$};
        
        \addplot[
            brown,
        ] table [
            x = x,
            y = y,
            col sep = tab,
        ] {results/CDF_14_7.dat};
        \addlegendentry{$n=14,~~t=7$};
        
        \addplot[
            green,
        ] table [
            x = x,
            y = y,
            col sep = tab,
        ] {results/CDF_16_8.dat};
        \addlegendentry{$n=16,~~t=8$};

    \end{axis}
\end{tikzpicture}
        \caption{Plots of the normalized empirical \glspl{cdf} of $n$-bit approximate multipliers, with $n\op{=}6,8,{\ldots},16$ and $t\op{=}\frac{n}{2}$.}
        \label{fig:cdfs}
    \end{figure}
}

\newcommand{\FIGplotpfront}[1]{
    \begin{figure}[#1]
        \centering
        \begin{tikzpicture}
    \begin{semilogyaxis}[
        view               = {0}{90},
        width              = 0.7\columnwidth,
        height             = 4.5cm,
        colorbar           = true,
        colorbar style     = {
            ylabel style   = {rotate = 180},
            ylabel         = NMED,
            ytick          = {1e-6,3e-6,5e-6,7e-6},
            scaled y ticks = {base 10:6},
            every y tick scale label/.style = {
                at     = {(yticklabel* cs:1.01,0cm)},
                anchor = near yticklabel
            },
        },
        ytick              = {10e+8,10e+9,10e+10,10e+11,10e+12,10e+13},
        xmin               = 0.69,
        xmax               = 1.01,
        ymin               = 10e+8,
        ymax               = 10e+13,
        xlabel             = ER,
        ylabel             = $\log(\text{MAE})$,
        grid               = both,
        grid style         = dotted,
        colormap           = {newmap}{
            color(0cm)     = (heatmap1);
            color(1cm)     = (heatmap2);
            color(2cm)     = (heatmap3);
            color(3cm)     = (heatmap4);
            color(4cm)     = (heatmap5);
        },
        set layers,
    ]

        \addplot3[
            scatter,
            mark = diamond*,
            fill opacity = 0.25,
            only marks,
        ] table [
            x = ER,
            y = MAE,
            z = NMED,
            col sep = tab,
        ] {results/results_sim_metrics_28.dat};
        
        \addplot3[
            scatter,
            mark = *,
            fill opacity = 0.25,
            only marks,
        ] table [
            x = ER,
            y = MAE,
            z = NMED,
            col sep = tab,
        ] {results/results_sim_metrics_30.dat};
    
        \addplot3[
            scatter,
            mark = triangle*,
            fill opacity = 0.25,
            only marks,
        ] table [
            x = ER,
            y = MAE,
            z = NMED,
            col sep = tab,
        ] {results/results_sim_metrics_32.dat};
    
    \end{semilogyaxis}
\end{tikzpicture}
        \caption{Pareto fronts comparing the \gls{mae}, \gls{er}, and \gls{nmed} of 28-, 30- and 32-bit approximate multipliers with halved-carry-chain adders represented by marks \diamark{my_blue}, \capmark{my_blue}, and \trimark{my_blue}, respectively. The ordinate is shown in logarithmic scale.}
        \label{fig:plot_pfront}
    \end{figure}
}

\newcommand{\FIGbaboon}[1]{
    \begin{figure}[#1]
        \centering
        \begin{subfigure}[t]{0.425\columnwidth}
            \centering 
            \includegraphics[width=0.7\columnwidth]{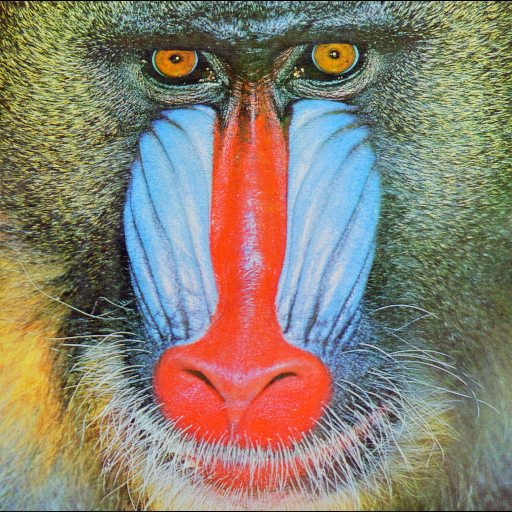}
            \caption{Accurate result of multiplying an image with itself.}
            \label{subfig:baboon_acc}
        \end{subfigure}\qquad
        \begin{subfigure}[t]{0.425\columnwidth}
            \centering
            \includegraphics[width=0.7\columnwidth]{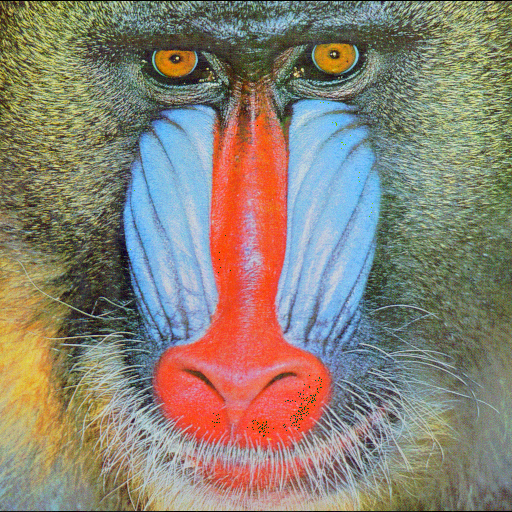}
            \caption{Approximate result using 8-bit approximate sequential multipliers with halved-carry-chain accumulations.}\vspace{3mm}
            \label{subfig:baboon_app}
        \end{subfigure}\\
        \caption{Liu et al. report 5 different tests in \cite{Liu:19.2} obtaining \glspl{ssim} and \glspl{psnr} within the ranges of [0.8472,0.9753] and [35.44,47.52] dB, respectively. Our approach achieves an \gls{ssim} of 0.9627 and a \gls{psnr} of 38.6456 dB.}\vspace{2mm}
        \label{fig:babbon_comparison}
    \end{figure}
}

\newcommand{\FIGresourcescomparison}[1]{
    \begin{figure}[#1]
        \centering
        \begin{minipage}[b]{0.6\columnwidth}
            \begin{subfigure}[t]{\columnwidth}
                \centering
                \includegraphics[width=0.9\columnwidth]{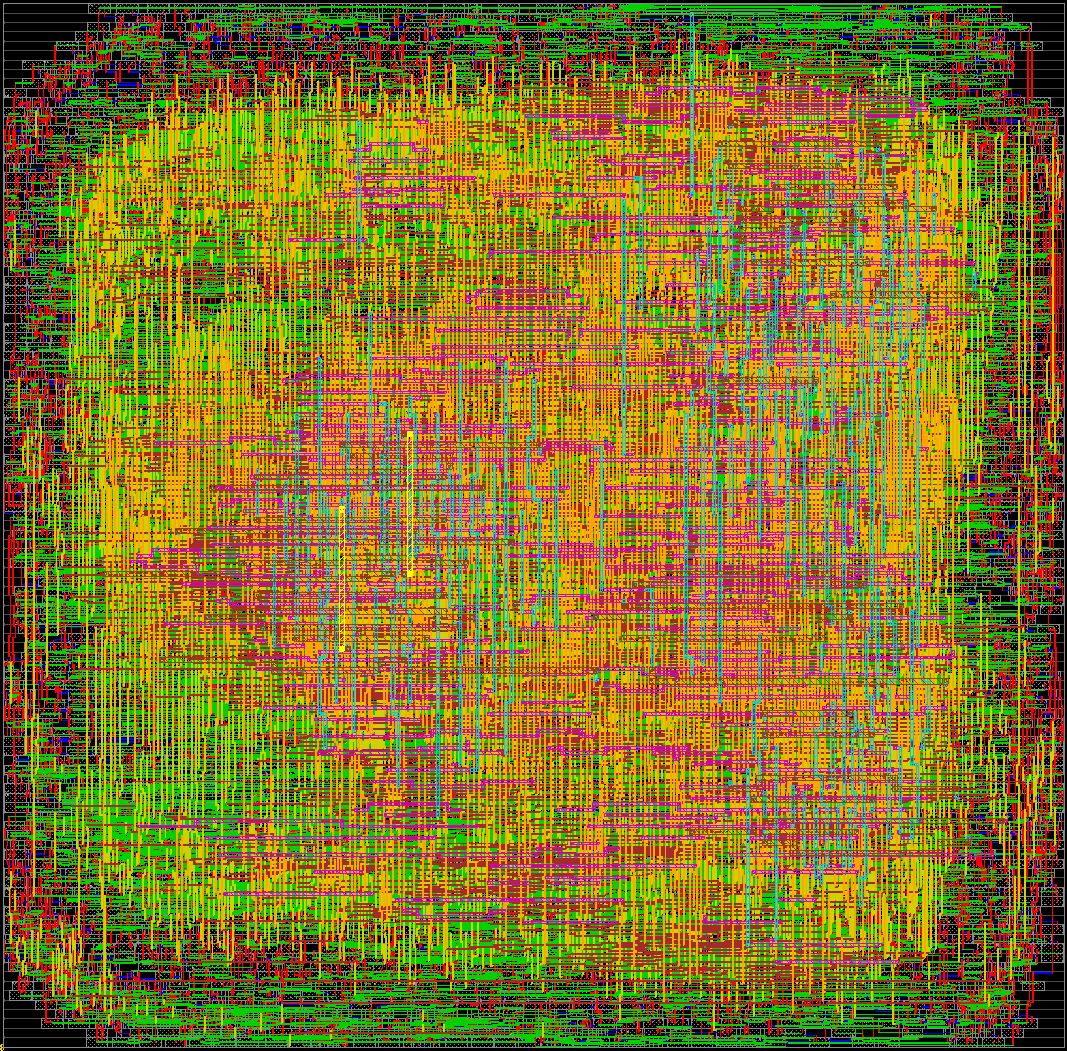}
                \caption{Accurate combinatorial.}
                \label{subfig:Com_acc_64}
            \end{subfigure}
        \end{minipage}\hfill
        \begin{minipage}[b]{0.4\columnwidth}
            \begin{subfigure}[t]{\columnwidth}
                \centering
                \includegraphics[width=0.55\columnwidth]{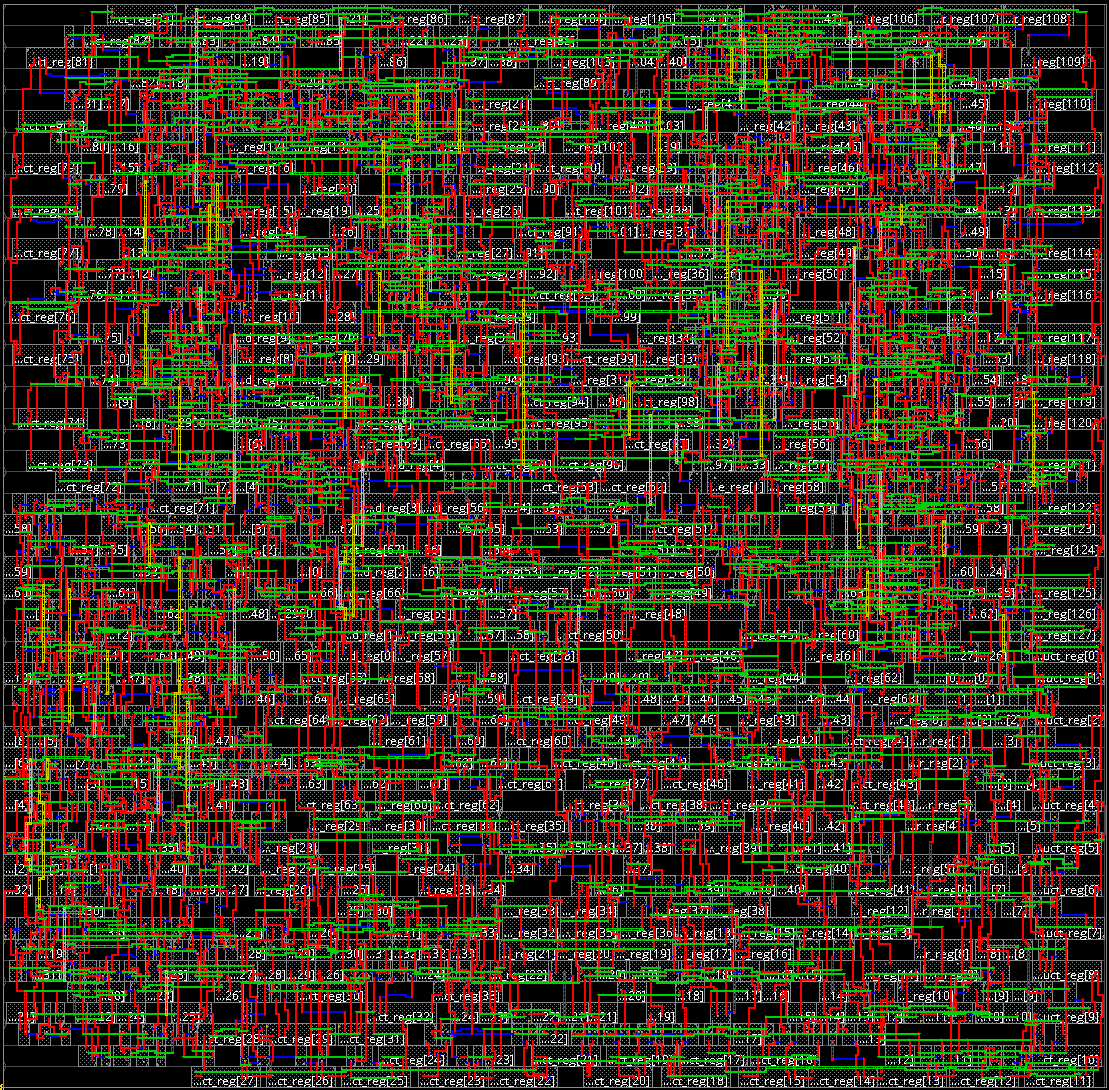}
                \caption{Accurate sequential.}\vspace{3.5mm}
                \label{subfig:Seq_acc_64}
            \end{subfigure}\\[5pt]
            \begin{subfigure}[t]{1\columnwidth}
                \centering
                \includegraphics[width=0.6\columnwidth]{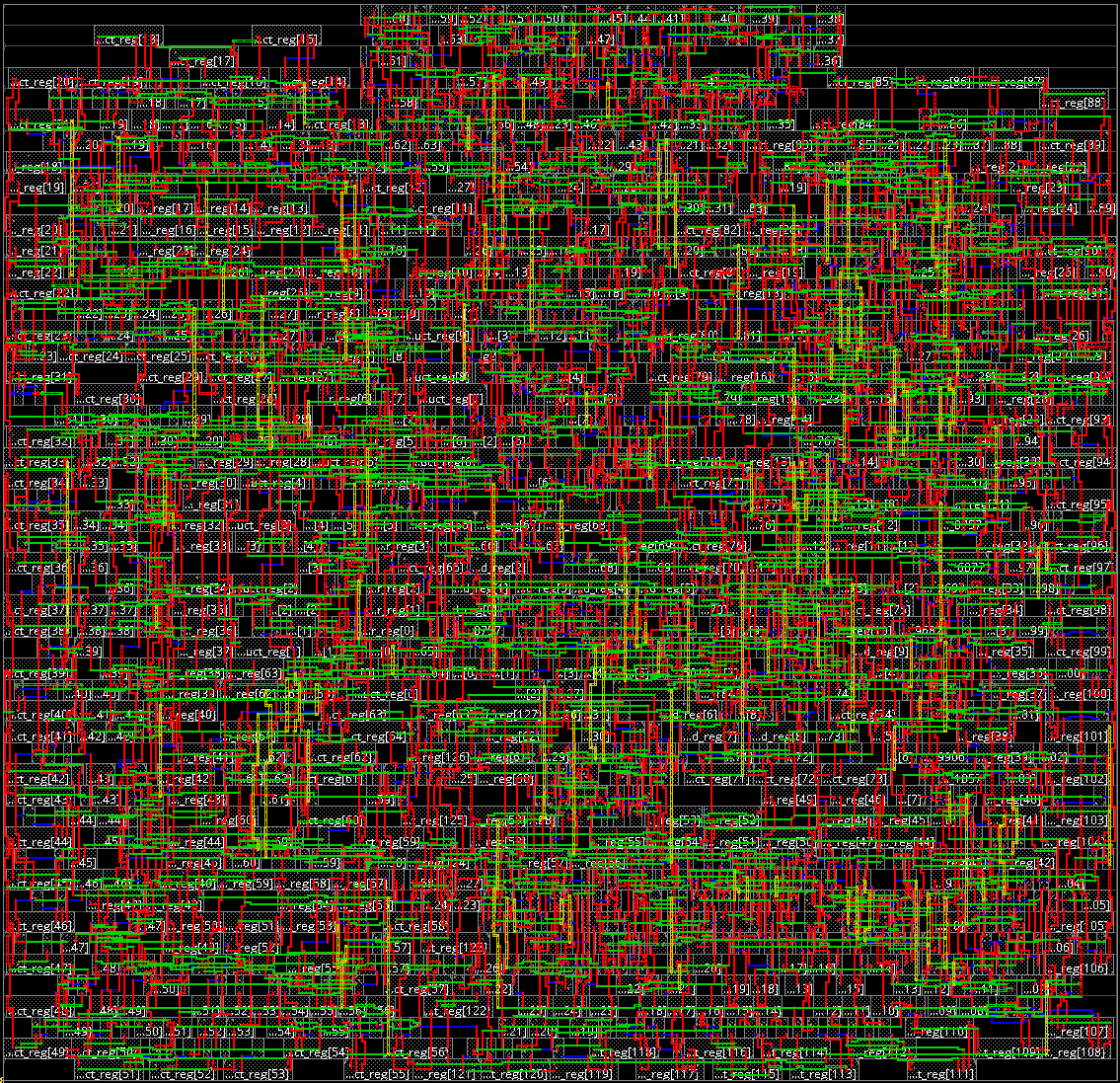}
                \caption{Approximate sequential.}\vspace{3.5mm}
                \label{subfig:Seq_app_64}
            \end{subfigure}
        \end{minipage}
        \caption{The three 64-bit multipliers shown were implemented on a Nangate 45nm Open Cell Library, a generic open-source, standard-cell library \cite{Nangate:11}. Accurate combinatorial and sequential implementations are shown in (\protect\subref{subfig:Com_acc_64}) and (\protect\subref{subfig:Seq_acc_64}), respectively. Our approximate sequential multiplier with halved-carry-chain accumulations is shown in (\protect\subref{subfig:Seq_app_64}).}
        \label{fig:resources_comparison}
    \end{figure}
}

\newcommand{\TABmeasurements}[1]{
\begin{figure}[#1]
	\hspace{-5mm}
	\begin{subfigure}[t]{\columnwidth}
		\hspace{3mm}\trimark{blue_fau} Accurate \quad \capmark{blue_fau} Approximate
		\centering
		\scalebox{0.95}{\begin{tikzpicture}
    \begin{semilogxaxis}[
        view               = {0}{90},
        width              = 0.7\columnwidth,
        height             = 4.5cm,
        colorbar           = true,
        colorbar style     = {
            at             = {(1.09,1)},
            ylabel style   = {rotate = 180},
            ylabel         = {Power $[mW]$},
            ymin           = 3,
            ymax           = 36,
            every y tick scale label/.style = {
                at     = {(yticklabel* cs:1.01,0cm)},
                anchor = near yticklabel
            },
        },
        xmin               = 27,
        xmax               = 1025,
        ytick              = {1, 2, 3, 4, 5},
        ymin               = 0.89,
        ymax               = 5.11,
        ylabel shift       = 11pt,
        xlabel             = {$\text{LUTs}$},
        ylabel             = {$\text{Latency}\ [ns]$},
        grid               = both,
        grid style         = dotted,
        colormap           = {newmap}{
            color(1cm)     = (heatmap_imp_1);
            color(2cm)     = (heatmap_imp_2);
            color(4cm)     = (heatmap_imp_3);
            color(7cm)     = (heatmap_imp_4);
            color(10cm)    = (heatmap_imp_5);
        },
        set layers,
    ]

        
        \addplot3[
            scatter,
            only marks,
            thick,
            mark         = triangle*,
            fill opacity = 0.25,
            mark options = {scale = 1.5},
        ] table [
            x = luts,
            y = ns,
            z = mw,
            col sep = tab,
        ] {results/results_FPGA_acc_seq.dat};
    
        \addplot3[
            scatter,
            only marks,
            thick,
            mark         = *,
            fill opacity = 0.25,
            mark options = {scale = 1.5},
        ] table [
            x = luts,
            y = ns,
            z = mw,
            col sep = tab,
        ] {results/results_FPGA_app_seq.dat};
    
    \end{semilogxaxis}
\end{tikzpicture}}%
		\hspace{-10mm}
		\caption{}
		\label{subfig:fpga_measurements}
	\end{subfigure}\vspace{5mm}
	
	\hspace{-6mm}
	\begin{subfigure}[t]{\columnwidth}
		\hspace{5mm}\trimark{blue_fau} Accurate \quad \capmark{blue_fau} Approximate
		\centering
		\scalebox{0.95}{\begin{tikzpicture}
    \begin{semilogxaxis}[
        view               = {0}{90},
        width              = 0.7\columnwidth,
        height             = 4.5cm,
        colorbar           = true,
        colorbar style     = {
            ylabel style   = {rotate = 180},
            ylabel         = {Power $[mW]$},
            every y tick scale label/.style = {
                at     = {(yticklabel* cs:1.01,0cm)},
                anchor = near yticklabel
            },
        },
        ytick              = {0.25, 0.5, 0.75, 1, 1.25},
        yticklabels        = {0.25, 0.50, 0.75, 1.00, 1.25},
        ymin               = 0.25,
        ymax               = 1.25,
        xlabel             = {$\text{Area}\ [\mu m^2]$},
        ylabel             = {$\text{Latency}\ [ns]$},
        grid               = both,
        grid style         = dotted,
        colormap           = {newmap}{
            color(1cm)     = (heatmap_imp_1);
            color(2cm)     = (heatmap_imp_2);
            color(4cm)     = (heatmap_imp_3);
            color(7cm)     = (heatmap_imp_4);
            color(10cm)    = (heatmap_imp_5);
        },
        set layers,
    ]

        
        \addplot3[
            scatter,
            only marks,
            thick,
            mark         = triangle*,
            fill opacity = 0.25,
            mark options = {scale = 1.5},
        ] table [
            x = um,
            y = ns,
            z = mw,
            col sep = tab,
        ] {results/results_ASIC_acc_seq.dat};
    
        \addplot3[
            scatter,
            only marks,
            thick,
            mark         = *,
            fill opacity = 0.25,
            mark options = {scale = 1.5},
        ] table [
            x = um,
            y = ns,
            z = mw,
            col sep = tab,
        ] {results/results_ASIC_app_seq.dat};
    
    \end{semilogxaxis}
\end{tikzpicture}}%
		\hspace{-10mm}
		\caption{}
		\label{subfig:asic_measurements}
	\end{subfigure}
	\caption{Resources, latency, and power trade-offs for FPGA (\protect\subref{subfig:fpga_measurements}) and ASIC (\protect\subref{subfig:asic_measurements}) implementations. In the sake of fairness the accurate and the approximate implementations were set up to the same clock frequency for each design $n$. The marks represent sequential implementations with different bitwidths $n\op{\in}\{4, 8, 16, 32, 64, 128, 256\}$, each with a splitting point at position $t\op{=}n\op{/}2$. The abscissae are shown in logarithmic scale. The estimations shown in (\protect\subref{subfig:fpga_measurements}) were obtained using Vivado{\tiny\textsuperscript{TM}}. The default balanced synthesis was chosen. The FPGA designs were implemented on a ZYNQ-7 ZC706 Evaluation Board (xc7z045ffg900-2) from Xilinx\textsuperscript{\protect\scalebox{0.5}{\textregistered}} with 218,600 LUTs and 437,200 FFs. The power estimation was performed following a vector-based approach with a set of $2^{16}$ uniform input patterns. The ASIC designs shown in (\protect\subref{subfig:asic_measurements}) were synthesized using Genus Synthesis Solution. The implementation and estimations were obtained using Innovus Implementation System. Both are products of Cadence\textsuperscript{\protect\scalebox{0.5}{\textregistered}}.} 
\end{figure}
}

\newcommand{\FIGpdpmed}[1]{
    \begin{figure}[#1]
        \centering
        {\color{red}\begin{tikzpicture}
    \begin{semilogyaxis}[
        width            = 0.75\columnwidth,
        height           = 4cm,
        xlabel           = {PDP},
        ylabel           = {$\log(\text{MED})$},
        grid             = both,
        grid style       = dotted,
        ytick            = {10e-1,10e+6,10e+13},
        xtick            = {0, 0.05, 0.1, 0.15, 0.2},
        xticklabels      = {0, 0.05, 0.1, 0.15, 0.2},
        enlarge x limits = {abs = 0.2cm},
        enlarge y limits = {abs = 0.2cm},
        xmin             = 0,
        xmax             = 0.2,
        ymin             = 2,
        ymax             = 40000000000000,
        set layers,
    ]
    
        \addplot[
            only marks,
            every mark/.append style = {
                solid,
                draw         = red,
                fill         = red,
                fill opacity = 0.15,
                rotate       = 180,
            },
            mark = triangle*,
        ] table [
            x = PDP,
            y = MED,
            col sep = tab,
        ] {results/MED_PDP.dat};
        
    \end{semilogyaxis}
\end{tikzpicture}}
        \caption{\color{red}Pareto front of the \gls{pdp} and the \gls{med} of multiple approximate sequential multipliers.}\vspace{2mm}
        \label{fig:pdp}
    \end{figure}
}

	\section{Introduction}
    The recent growing interest on \textsc{\Gls{ac}} is partially due to insufficient resources for handling the ever-increasing amount of data created every single day.
    One of the most prominent domains being multimedia applications. 
    In digital image processing, for example, \gls{ac} 
    trades-off imperceptible quality degradation to the human eye with non-functional objectives.
    That is, rather than computing fully accurate results, \gls{ac} allows an algorithm to operate to some degree inaccurately in order to satisfy user-defined goals like reduction of latency, decrease of silicon area usage, or reduction of power dissipation.
    
    From this perspective, and in hand with the fact that arithmetic operations are the backbone of multimedia processing, we propose in this paper a sequential approximate multiplier which operates with reduced carry chains.
    This allows to shorten the critical path within the combinatorial part and increase the operational frequency.
    As discussed in the following, we are aware of only little-to-no work in the literature tackling approximate multiplication following such a sequential approach.

	\section{Related work}\label{sec:related_work}
	Multiple works have proposed the use of approximate building blocks to create larger combinatorial multipliers either by, for example, using approximated compressors or reduced circuitry by manipulating their truth tables \cite{Liu:19.2,Toan:20}.
	Some other works have exploited the capability of Booth encoders for reducing the number of partial products in approximate signed multiplication \cite{Liu:19.1}.
	Such techniques focus on \textit{underdesigning} combinatorial approaches aiming to reduce area, and therefore, power consumption.
	
	%
	
	Little effort has been made in the area of sequential approximate multiplication, overlooking the inherent area savings of such approaches, and possible timing reductions.
	Certainly, the most relevant work experimenting with sequential approaches is the one presented by Chandrasekharan et al. in \cite{Chandrasekharan:19}.
	The authors propose an architecture based on multiple 8-bit sequential multipliers made of adders taken from a collection of approximate arithmetic units found in the literature. 
	These approximate adders were implemented with different configurations, and thus different degrees of inaccuracy.
	%
	%
	The authors of \cite{Chandrasekharan:19} did not present any formal analysis of the errors introduced by such architectures.
	They did, however, obtain and report what they call the \textit{worst-case error} and the \textit{maximum bit-flip} by using a miter.
	Unfortunately, this error estimation approach requires simulation of many input patterns which, in order to obtain reliable results, demands large input sets.
	
	As seen, little consideration has been given to sequential multipliers and to formal techniques for the analysis of errors of such architectures. 
	%
	%

    \begin{table}[t]
        \vspace{-5pt}
		\begin{minipage}[t]{\columnwidth}
			\begin{subtable}[t]{0.5\textwidth}
                \begin{minipage}[t][5.25cm][t]{0.5\columnwidth}
    {\small
    \begin{equation*}
        \begin{aligned}
            &\text{\scriptsize Multiplier}   && \es{4}\:1\:0\:1\:1                 && \es{-18}                                          \text{\color{gray}\tiny 1}  \\[-2pt]
            &\text{\scriptsize Multiplicand} && \es{4}\:1\:1\:0\:1 \eqmark{m1}     && \es{-18}                                          \text{\color{gray}\tiny 2}  \\[-2pt]
            &                                && \es{-5}\left.\begin{aligned}
                                                                                                     & \gc{1\:0\:1\:}1           & \es{-2}\text{\color{gray}\tiny 3}  \\[-2pt]
                                                                                            \gc{0\:} & \gc{0\:0\:0}  \eqmark{m8} & \es{-2}\text{\color{gray}\tiny 4}  \\[-2pt]
                                                                       \tikzmark{P1}\rc{0}\:\gc{0\:} & \gc{1\:0\:1\:}1           & \es{-2}\text{\color{gray}\tiny 5}  \\
                                                                                         \gc{1\:0\:} & \gc{1\:}1                 & \es{-2}\text{\color{gray}\tiny 6}  \\[-2pt]
                                                                                      \gc{1\:0\:1\:} & \gc{1}    \eqmark{m9}     & \es{-2}\text{\color{gray}\tiny 7}  \\[-2pt]
                                                                 \tikzmark{P2}\rc{1}\:\gc{0\:0\:0\:} & \gc{0\:}1                 & \es{-2}\text{\color{gray}\tiny 8}  \\
                                                                            \tikzmark{P3}\gc{0\:0\:} & \gc{1\:0\:}1\:1           & \es{-2}\text{\color{gray}\tiny 9}  \\[-2pt]
                                                                      \tikzmark{P4}\gc{1\:0\:0\:0\:} & \gc{0\:1}   \eqmark{m10}  & \es{-2}\text{\color{gray}\tiny 10} \\[-2pt]
                                                                                   \gc{1\:0\:0\:0\:} & \gc{1\:1\:}1\:1           & \es{-2}\text{\color{gray}\tiny 11}
                                                             \end{aligned}\es{2}\right\rbrace\begin{aligned}
                                                                                                 &\text{\scriptsize Partial} \\[-2pt]
                                                                                                 &\text{\scriptsize products}
                                                                                             \end{aligned}&&\\[-2pt]
            &\text{\scriptsize Product}      && \es{-4}1\:0\:0\:0\:1\:1\:1\:1 \eqmark{m2} && \es{-19}                                        \text{\color{gray}\tiny 12}
        \end{aligned}
        \tikz[overlay,remember picture, dashed]{\draw([xshift=- 8mm, yshift=-0.75mm]m1.south)  -- ([xshift=0mm, yshift=-0.75mm]m1.south);}
        \tikz[overlay,remember picture        ]{\draw([xshift=- 8mm, yshift=-0.75mm]m8.south)  -- ([xshift=2mm, yshift=-0.75mm]m8.south);}
        \tikz[overlay,remember picture        ]{\draw([xshift=- 8mm, yshift=-0.75mm]m9.south)  -- ([xshift=2mm, yshift=-0.75mm]m9.south);}
        \tikz[overlay,remember picture        ]{\draw([xshift=-12mm, yshift=-0.75mm]m10.south) -- ([xshift=4mm, yshift=-0.75mm]m10.south);}
    \end{equation*}}
    \begin{tikzpicture}[overlay,remember picture]
        \scalebox{0.5}{\draw[gray,preaction={-triangle 90,draw,ultra thin,gray}] ([xshift=40pt, yshift=95pt]{pic cs:P1}) to[out=180,in=180] ([xshift=40pt, yshift=47pt]{pic cs:P3});}
        \scalebox{0.5}{\draw[gray,preaction={-triangle 90,draw,ultra thin,gray}] ([xshift=30pt, yshift=60pt]{pic cs:P2}) to[out=180,in=180] ([xshift=30pt, yshift=35pt]{pic cs:P4});}
    \end{tikzpicture}
\end{minipage}

                \parbox{\linewidth}{\vspace*{8mm}\caption{}\label{subtab:ex_acc_com}}
			\end{subtable}\hspace{20pt}
			\begin{subtable}[t]{0.5\textwidth}
                \begin{minipage}[t][5.25cm][t]{0.45\columnwidth}
    {\small
    \begin{equation*}
        \begin{aligned}
            &\text{\scriptsize Multiplier}         & \text{\color{gray}\tiny 1}  \es{-2} &&            &1\:0\:1\:1                  \\[-2pt]
            &\text{\scriptsize Multiplicand}\bs{5} & \text{\color{gray}\tiny 2}  \es{-2} &&            &1\:1\:0\:1\eqmark{m3}       \\[-2pt]
            &                                      & \text{\color{gray}\tiny 3}  \es{-2} &&            &\gc{0\:0\:0\:0\:}           \\[-2pt]
            &                                      & \text{\color{gray}\tiny 4}  \es{-2} &&            &\gc{1\:0\:1\:1\:}\eqmark{m4}\\[-2pt]
            &                                      & \text{\color{gray}\tiny 5}  \es{-2} &&    \rc{0\:}&\gc{1\:0\:1\:1\:}           \\[-2pt]
            &                                      & \text{\color{gray}\tiny 6}  \es{-2} && \eqmark{S1}&\gc{0\:1\:0\:1\:}1          \\[-2pt]
            &                                      & \text{\color{gray}\tiny 7}  \es{-2} &&            &\gc{0\:0\:0\:0\:}\eqmark{m5}\\[-2pt]
            &                                      & \text{\color{gray}\tiny 8}  \es{-2} &&    \rc{0\:}&\gc{0\:1\:0\:1\:}1          \\[-2pt]
            &                                      & \text{\color{gray}\tiny 9}  \es{-2} && \eqmark{S2}&\gc{0\:0\:1\:0\:}1\:1       \\[-2pt]
            &                                      & \text{\color{gray}\tiny 10} \es{-2} &&            &\gc{1\:0\:1\:1\:}\eqmark{m6}\\[-2pt]
            &                                      & \text{\color{gray}\tiny 11} \es{-2} &&    \rc{0\:}&\gc{1\:1\:0\:1\:}1\:1       \\[-2pt]
            &                                      & \text{\color{gray}\tiny 12} \es{-2} && \eqmark{S3}&\gc{0\:1\:1\:0\:}1\:1\:1    \\[-2pt]
            &                                      & \text{\color{gray}\tiny 13} \es{-2} &&            &\gc{1\:0\:1\:1\:}\eqmark{m7}\\[-2pt]
            &                                      & \text{\color{gray}\tiny 14} \es{-2} &&    \rc{1\:}&\gc{0\:0\:0\:1\:}1\:1\:1    \\[-2pt]
            &\text{\scriptsize Product}            & \text{\color{gray}\tiny 15} \es{-2} && \eqmark{S4}&1\:0\:0\:0\:1\:1\:1\:1
        \end{aligned}
        \tikz[overlay,remember picture, dashed]{\draw([xshift=-8 mm, yshift=-0.75mm]m3.south) -- ([xshift=-0.5mm, yshift=-0.75mm]m3.south);}
        \tikz[overlay,remember picture        ]{\draw([xshift=-11mm, yshift=-0.75mm]m4.south) -- ([xshift=-0.5mm, yshift=-0.75mm]m4.south);}
        \tikz[overlay,remember picture        ]{\draw([xshift=-11mm, yshift=-0.75mm]m5.south) -- ([xshift=-0.5mm, yshift=-0.75mm]m5.south);}
        \tikz[overlay,remember picture        ]{\draw([xshift=-11mm, yshift=-0.75mm]m6.south) -- ([xshift=-0.5mm, yshift=-0.75mm]m6.south);}
        \tikz[overlay,remember picture        ]{\draw([xshift=-11mm, yshift=-0.75mm]m7.south) -- ([xshift=-0.5mm, yshift=-0.75mm]m7.south);}
        \tikz[overlay,remember picture]{\scalebox{0.5}{\draw[gray,preaction={-triangle 90,draw,ultra thin,gray}] ([xshift=-57pt, yshift= 41pt]S1.south) |- ([xshift=-48pt, yshift= 26pt]S1.south);}}
        \tikz[overlay,remember picture]{\scalebox{0.5}{\draw[gray,preaction={-triangle 90,draw,ultra thin,gray}] ([xshift=-57pt, yshift=  9pt]S2.south) |- ([xshift=-48pt, yshift=- 6pt]S2.south);}}
        \tikz[overlay,remember picture]{\scalebox{0.5}{\draw[gray,preaction={-triangle 90,draw,ultra thin,gray}] ([xshift=-57pt, yshift=-22pt]S3.south) |- ([xshift=-48pt, yshift=-37pt]S3.south);}}
        \tikz[overlay,remember picture]{\scalebox{0.5}{\draw[gray,preaction={-triangle 90,draw,ultra thin,gray}] ([xshift=-57pt, yshift=-55pt]S4.south) |- ([xshift=-48pt, yshift=-70pt]S4.south);}}
    \end{equation*}}
\end{minipage}

                \hspace*{-5mm}
                \parbox{\linewidth}{\vspace*{8mm}\caption{}\label{subtab:ex_acc_seq}}
			\end{subtable}
        \end{minipage}
		\parbox{\linewidth}{\vspace*{-1mm}\caption{Accurate (\protect\subref{subtab:ex_acc_com}) combinatorial and (\protect\subref{subtab:ex_acc_seq}) sequential multiplication.}}
    \end{table}

\section{Preliminaries}\label{sec:preliminaries}
	
    \begin{figure*}
        \centering
        \begin{subfigure}[t]{0.425\textwidth}
			\centering
			\resizebox{\linewidth}{!}{\begin{tikzpicture}
	[
        >          = stealth,
	    alu/.style = {
            trapezium,
            trapezium angle      = 65,
            shape border rotate  = 180,
            minimum width        = 4cm,
            minimum height       = 1.75cm,
            trapezium stretches  = true,
            append after command = {
                \pgfextra
                    \draw[fill=lightgray!25]    (\tikzlastnode.top    left  corner)
                                             -- (\tikzlastnode.top    right corner)
                                             -- (\tikzlastnode.bottom right corner)
                                             -- ($(\tikzlastnode.bottom right corner)!.800!(\tikzlastnode.bottom side)$)
                                             -- ([yshift=-5.5mm]\tikzlastnode.bottom side)
                                             -- ($(\tikzlastnode.bottom side)!.200!(\tikzlastnode.bottom left corner)$)
                                             -- (\tikzlastnode.bottom left  corner)
                                             -- (\tikzlastnode.top    left  corner);
                \endpgfextra
            },
	    },
	    empty_alu/.style = {
            trapezium,
            trapezium angle      = 65,
            shape border rotate  = 180,
            minimum width        = 4cm,
            minimum height       = 1.75cm,
            trapezium stretches  = true,
	    },
	]
	
	\node[alu] at (0,0) (acc_alu) {};
	\node[align=center] at ([yshift = -0.3cm]acc_alu) (label) {\Large Adder\\[1mm]\large +};
	\node[align=center] at ([xshift = -1.3cm, yshift =  0.3cm]acc_alu) (Cin_MMSP)  {\large $\text{C}_{\text{in}}$};
	\node[align=center] at ([xshift = -0.8cm, yshift = -0.6cm]acc_alu) (Cout_MMSP) {\large $\text{C}_{\text{out}}$};
	
	\node[empty_alu] at (4.75,0) (phantom_alu) {};
	\node at ([xshift = -1.25cm]acc_alu.bottom right corner) (cin_reg_TL) {};
	\node at ([xshift =  0.8cm ]cin_reg_TL) (cin_reg_BR) {};
    
	\node at ([yshift = -1cm   ]acc_alu.top right corner) (reg_MSP_LT) {};
	\node at ([yshift = -2.25cm]acc_alu.top left  corner) (reg_MSP_RB) {};
	\node at ([yshift =  2.5cm ]$(reg_MSP_LT) + (reg_MSP_RB)$) (reg_MSP_C) {};
	\draw[fill=lightgray!25] ([xshift = -0.8cm]reg_MSP_LT) rectangle ([xshift = 0.8cm]reg_MSP_RB);
	\node[align=center] at (reg_MSP_C) (shift_MSP) {\large Shift Reg A};
	
	\node at ([yshift = -1cm   ]phantom_alu.top right corner) (reg_LSP_LT) {};
	\node at ([yshift = -2.25cm]phantom_alu.top left  corner) (reg_LSP_RB) {};
	\node at ([yshift =  2.5cm ]$(reg_LSP_LT) + (reg_LSP_RB) - (5,0)$) (reg_LSP_C) {};
	\draw[fill=lightgray!25] ([xshift = -0.8cm]$(reg_LSP_LT) - (0.25,0)$) rectangle ([xshift = 0.8cm]$(reg_LSP_RB) - (0.25,0)$);
	\node[align=center] at (reg_LSP_C) (shift_LSP) {\large Shift Reg B};
	
	\node at ([xshift = -2.25cm]reg_MSP_LT) (cout_reg_TL) {};
	\node at ([xshift =  0.8 cm]cout_reg_TL) (cout_reg_BR) {};
	\draw[fill=lightgray!25] (cout_reg_TL) rectangle ([yshift = -1.25cm]cout_reg_BR);
	
	
	\draw[-triangle 90] ($(Cout_MMSP) - (0.4,0)$) -- ($(Cout_MMSP) - (3  ,0  )$) |- ($(cout_reg_TL) - (0,0.25)$);
	\draw[-triangle 90] ($(cout_reg_BR) + (0.0,-0.25)$) -- ($(cout_reg_BR) + (0.25,-0.25)$) |- ([xshift = 1.7cm]$(cout_reg_BR) - (1.05,0.6)$);

	\draw[-triangle 90] ([xshift = -6.75cm, yshift = -1.5cm]$(cin_reg_BR)  + (7,0.5)$) |- ([xshift = -6.3cm, yshift = -1.5cm]$(cin_reg_BR) + (7,1)$);
	
	\path[
        solid,
        draw       = black,
        line width = 1mm,
        preaction  = {
            -triangle 90,
            thin,
            draw,
            shorten > = -1mm
        },
    ] (acc_alu.south) -- ([yshift = -0.9cm]acc_alu.south);
    
	\path[
        solid,
        draw       = black,
        line width = 1mm,
        preaction  = {
            -triangle 90,
            thin,
            draw,
            shorten > = -1mm
        },
    ] ([xshift = -0.25cm, yshift = 2.75cm]phantom_alu.south) -- ([xshift = -0.25cm, yshift = -0.9cm]phantom_alu.south);
	
	\path[
        solid,
        draw       = black,
        line width = 1mm,
        preaction  = {
            -triangle 90,
            thin,
            draw,
            shorten > = -1mm
        },
    ] ([yshift = -2.25cm]acc_alu.south) -- ([yshift = -3.25cm]acc_alu.south);
    
	\path[
        solid,
        draw       = black,
        line width = 1mm,
        preaction  = {
            -triangle 90,
            thin,
            draw,
            shorten > = -1mm
        },
    ] ([xshift = -0.25cm, yshift = -2.25cm]phantom_alu.south) -- ([xshift = -0.25cm, yshift = -3.25cm]phantom_alu.south);
    
    \path[
        solid,
        draw       = black,
        line width = 1mm,
        preaction  = {
            -triangle 90,
            thin,
            draw,
            shorten > = -1mm
        },
    ]([yshift = -2.55cm]acc_alu.south) -| ($(cin_reg_TL) + (-1.1,0.5)$  ) -| ($(acc_alu.bottom right corner) + (0.85,0.1)$);
    
	\draw[-triangle 90] ([yshift =-2.55cm]acc_alu.south) -| ($(reg_LSP_LT) - (1.5,0.61)$) -- ($(reg_LSP_LT) - (1.05,0.61)$);
	\draw[-triangle 90] ([xshift = -0.25cm, yshift =-2.55cm]phantom_alu.south) -- ([xshift = 1.5cm, yshift =-2.55cm]phantom_alu.south);

	\path[
        solid,
        draw       = black,
        line width = 1mm,
        preaction  = {
            -triangle 90,
            thin,
            draw,
            shorten > = -1mm
        },
    ] ($(acc_alu.bottom left corner) + (-0.85,1)$) -- ($(acc_alu.bottom left corner) + (-0.85,0.1)$);
    
	
	\draw[black,fill=black] ([yshift = -2.55cm]acc_alu.south) circle (.75ex);
	\draw[black,fill=black] ([xshift = -0.25cm, yshift = -2.55cm]phantom_alu.south) circle (.75ex);
	
	
	\node[align=center] at ([xshift = -6.75cm, yshift = -1.5cm]$(cin_reg_BR) + (7,0.3)$) (foo) {\selectfont $0$};
	
	
	\draw ($(cout_reg_TL) + (0,-0.9)$) -- ($(cout_reg_TL) + (0.15,-1)$) -- ($(cout_reg_TL) + (0,-1.1)$);
	\draw ($(shift_MSP.south west) - (0.85,0.2)$) -- ($(shift_MSP.south west) - (0.7 ,0.1)$) -- ($(shift_MSP.south west) + (-0.85 ,0)$);
    \draw ($(shift_LSP.south west) - (0.86,0.2)$) -- ($(shift_LSP.south west) - (0.71,0.1)$) -- ($(shift_LSP.south west) + (-0.86 ,0)$);
	

    \draw ([xshift =-3.75cm, yshift = 2.2 cm]phantom_alu.south) -- ([xshift =-3.45cm, yshift = 2.5cm]phantom_alu.south);
    \draw ([xshift =-0.4 cm, yshift = 2.2 cm]phantom_alu.south) -- ([xshift =-0.1 cm, yshift = 2.5cm]phantom_alu.south);
    
	\draw ([xshift =-4.5 cm, yshift =-0.6 cm]acc_alu.south)     -- ([xshift =-4.2 cm, yshift =-0.3cm]acc_alu.south);
	\draw ([xshift =-0.15cm, yshift =-0.6 cm]acc_alu.south)     -- ([xshift = 0.15cm, yshift =-0.3cm]acc_alu.south);
	
	\draw ([xshift =-0.15cm, yshift =-3.05cm]acc_alu.south)     -- ([xshift = 0.15cm, yshift =-2.75cm]acc_alu.south);
	\draw ([xshift =-0.4 cm, yshift =-3.05cm]phantom_alu.south) -- ([xshift =-0.1 cm, yshift =-2.75cm]phantom_alu.south);
	
	
	\node[align=center] at ($(phantom_alu.bottom left corner) + (-5.9 , 0.5)$) (foo) {\selectfont $n$};
	\node[align=center] at ($(phantom_alu.bottom left corner) + (-2.55, 0.5)$) (foo) {\selectfont $n$};
	
	\node[align=center] at ($(acc_alu.bottom left corner)     - (6.7 , 2.2)$) (foo) {\selectfont $n$};
	\node[align=center] at ($(acc_alu.bottom left corner)     - (2.3 , 2.2)$) (foo) {\selectfont $n$};
	
	\node[align=center] at ([xshift =-0.325cm, yshift =-2.95cm]acc_alu.south) (foo) {\selectfont $n$};
	\node[align=center] at ([xshift = 4.2cm , yshift =-2.95cm]acc_alu.south) (foo) {\selectfont $n$};
	
	\node[align=center] at ($(acc_alu.bottom left corner) + (-0.95,1.25)$) (MSP_multiplicand) {$B_{lsb} \wedge (a_{n-1} \ldots a_0)$};
	\node[align=center] at ($(phantom_alu.south) + (-0.2,3.02)$) (multiplier) {$b_{n-1} \ldots b_0$};

	\node[align=center] at ($(cout_reg_TL.north west) + (0.3,-0.4)$) (ff_d_out) {$D$};
	\node[align=center] at ($(cout_reg_TL.north east) + (0.5,-0.4)$) (ff_q_out) {$Q$};
	
	\node[align=center] at ([yshift =-3.75cm]acc_alu.south) (MMSP_product) {$p_{2n-1} \ldots p_n$};
	\node[align=center] at ([yshift =-3.75cm]phantom_alu.south) (MLSP_product) {$p_{n-1} \ldots p_0$};
	
	\node[align=center] at ([xshift =-3.45cm, yshift = -2.78cm]phantom_alu.south) (MLSP_product) {$A_{lsb}$};
	\node[align=center] at ([xshift = 1.8 cm, yshift = -2.6 cm]phantom_alu.south) (MLSP_product) {$B_{lsb}$};
	
\end{tikzpicture}}
			\parbox{\linewidth}{\vspace*{-0mm}\caption{}\label{subfig:sch_acc_seq}}
		\end{subfigure}\hfill%
		\begin{subfigure}[t]{0.55\textwidth}
			\centering
			\resizebox{\linewidth}{!}{\input{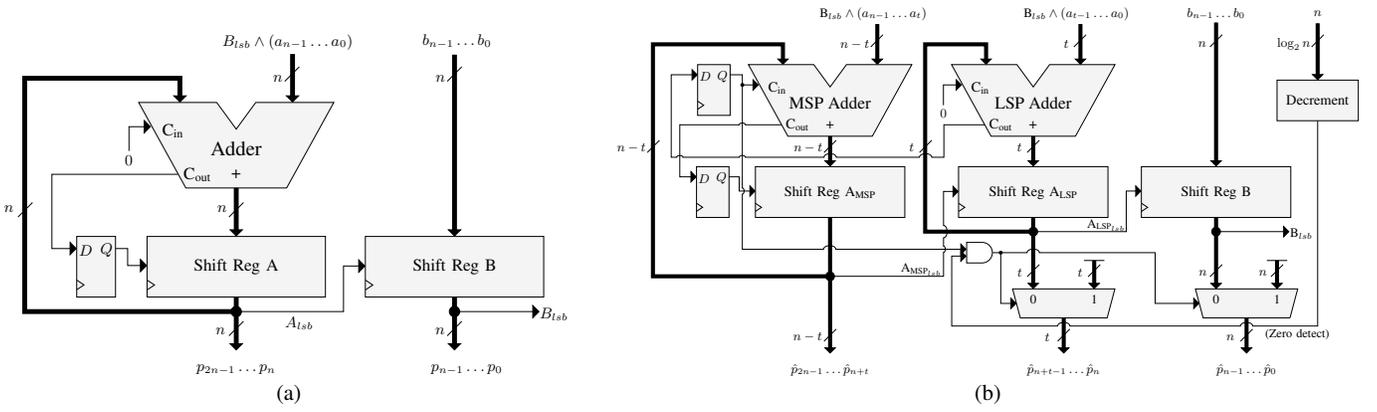}}
			\parbox{\linewidth}{\vspace*{-0mm}\caption{}\label{subfig:sch_app_seq}}
		\end{subfigure}\\[3mm]
		\parbox{\linewidth}{\vspace*{-3mm}\caption{Schematics of sequential circuits implementing the (\protect\subref{subfig:sch_acc_seq}) accurate and the (\protect\subref{subfig:sch_app_seq}) approximate multiplication strategies shown in \Cref{subtab:ex_acc_seq,subfig:ex_app_seq}, respectively. The D flip-flops have asynchronous \textit{clear} inputs. The controllers and clock lines are not shown. The shift registers have synchronous inputs for parallel \textit{load}, \textit{shifting} to the right with left serial input, and \textit{clear} to set to 0---see \mbox{Line 3} of \Cref{subtab:ex_acc_seq,subfig:ex_app_seq}, respectively.  Whilst not shown in (\protect\subref{subfig:sch_acc_seq}), in (\protect\subref{subfig:sch_app_seq}) a decrement unit, which informs the controller about sequence completion, and enables the multiplexing of the least significant $n+t$ bits according to the carry-out of the last accumulation is shown.}
		\label{fig:seq_schematics}}
    \end{figure*}
{}
	
    The so-called grade-school multiplication algorithm multiplies each bit of the multiplicand by the multiplier.
    Representing a multiplication by the radix, each partial product is shifted to the left by the amount equal to the power of the corresponding bit of the multiplicand.
    Consider the example shown in \Cref{subtab:ex_acc_com}.
    \mbox{Line 5} corresponds to the sum of the multiplier multiplied by the 0th and the 1st bits---shifted 0 and 1 positions, respectively---of the multiplicand.
    Due to the addend in the partial addition being shifted to the left, the \gls{lsb} of the augend---shown in black---does not take part in the addition.
    This extra bit is simply concatenated as the \gls{lsb} of the result.
    Thus, only a 4-bit adder is required to perform this accumulation.
    This can be seen by the bits emphasized in blue.
    Finally, the partial accumulated sum shown in \mbox{Line 5} is added to the one from \mbox{Line 8}, as shown in Lines 9 to 11.
    Note once again that the two \glspl{lsb} of the augend---shown in black in Line 9---are not taking part in the addition.
    Therefore, two 4-bit and one 6-bit adders are required to perform the multiplication from \Cref{subtab:ex_acc_com}.
    In a broader sense, an architecture calculating the product of two $n$-bit numbers following this approach has a considerably high complexity and resources-cost:
    There are as many partial products as there are bits in the multiplicand.
    %
    And since two partial products are summed up by a single $(n+\lfloor{2^{k-2}+0.5}\rfloor){\times}n$-bit adder, in total $k\op{=}\log_2n$ additions are required.
    Such a design would need $\sum_{i=1}^{\log_2n}\frac{n}{2^i} = n-1$ adders, scaling linearly with the input bit-width $n$.
    
    \FIGindicesex{t}
    
    On the other hand, a sequential multiplication such as shown in \Cref{subtab:ex_acc_seq}, and a circuit shown in \Cref{subfig:sch_acc_seq}, requires fewer resources.
    Some power savings are expected in comparison with a combinatorial approach because there is less interconnect, which in turn means that the capacitive load on the signals in the design decreases.
    With a lower load, dynamic power consumption can be reduced.
    As seen in the figure, only a single $n{\times}n$-bit adder, one D flip-flop, and two $n$-bit shift registers are needed.
    Consider the example shown in \Cref{subtab:ex_acc_seq}, where \mbox{Line 6} (resp. \mbox{Line 7}) corresponds to the augend (resp. addend) shown in \mbox{Line 3} (resp. \mbox{Line 4}) of \Cref{subtab:ex_acc_com}.
    The result, however, is shifted once to the right---clearly, shifting the augend to the right is equivalent to shifting the addend to the left.
    By doing so, the next partial product may be added directly to the result, as shown in \mbox{Line 10}.
    The bits emphasized in blue represent those bits where the actual addition takes place.
    Observe that, as opposed to the $n-1$ additions performed by a combinatorial approach, a sequential architecture performs only a single addition per clock cycle. 
    
    
    \subsection{Boolean representation}\label{subsec:boolean_representation}
		Formally, let $p\op{:}\Bs^{2n}\op{\rightarrow}\Bs^{2n}$ represent the accurate product of two $n$-bit binary numbers $a$ and $b$ with $a,b\op{\in}\Bs^n$ and $p(a_{n-1},{\ldots},a_0,b_{n-1},{\ldots},b_0)\op{=}(p_{2n{-}1},{\ldots},p_0)$. More specifically, $\mathrm{dec}(p(a,b))\op{=}\mathrm{dec}(a)\op{\cdot}\mathrm{dec}(b)$, where $\mathrm{dec}(x)$ corresponds to the decimal representation of a binary number $x$.
		Let $a_i$ be the $i$-th bit of multiplier $a$, $b_j$ be the $j$-th bit of multiplicand $b$, $S_i^j$ be the $i$-th bit in the $j$-th accumulated sum, and $C_i^j$ the $i$-th carry-bit in the $j$-th carry chain. The values of $S_i^j$ and $C_i^j$ are then determined by:
		
		\begin{align*}
			S_i^j &= \begin{cases}
						a_i \mint b_0,                                                                    & \mathrm{if}\ j = 0\ \mathrm{and}\ i \in [0,n) \\
						0,                                                                             & \mathrm{if}\ j = 0\ \mathrm{and}\ i = n       \\
						S_{1}^{j-1}   \oplus (a_0 \mint b_j),                                             & \mathrm{if}\ j > 0\ \mathrm{and}\ i = 0       \\
						S_{i+1}^{j-1} \oplus C_{i-1}^j \oplus (a_i \mint b_j),                            & \mathrm{if}\ j > 0\ \mathrm{and}\ i \in (0,n) \\
						\fixlacc{C_{n-1}^j,}                                                              & \mathrm{if}\ j > 0\ \mathrm{and}\ i = n,
					\end{cases} 
		\end{align*}
		
		\begin{align*}
			C_i^j &= \begin{cases}
						0,                                                                             & \mathrm{if}\ j = 0                            \\
						S_{1}^{j-1} \mint (a_0 \mint b_j),                                               & \mathrm{if}\ j > 0\ \mathrm{and}\ i = 0       \\
						\left(\left(S_{i+1}^{j-1} \oplus (a_i \mint b_j) \right) \mint C_{i-1}^{j}\right) \\
						\maxt \left(S_{i+1}^{j-1} \mint   a_i \mint b_j  \right),                         & \mathrm{if}\ j > 0\ \mathrm{and}\ i \in (0,n),
					\end{cases}
		\end{align*}
		
		\noindent where $\mint$, $\maxt$, and $\oplus$ represent the logical conjunction, disjunction, and exclusive disjunction, respectively.
		Consequently, the product $p$ is constructed as follows:
		
		\begin{equation}\label{eq:p_acc}
			p_r = \begin{cases}
					\fixmacc{S_0^{r},} & \mathrm{if}\ r \in [0,n{-}1)  \\
					S_{r-n+1}^{n-1},   & \mathrm{if}\ r \in [n{-}1,2n{-}1],
				\end{cases}
		\end{equation}

		
		At clock cycle $j\op{=}0$, the multiplicand $b$ is stored in the shift register B, and all bits in shift register A are set to zero.
		During each shift to the right the \gls{lsb} of register B, i.e., $b_{j-1}$, is disposed to B$_{lsb}$, the \gls{lsb} of register A, i.e., $S_0^j$, is introduced from the left to register B, and the the D flip-flop storing the carry-out C$_{\text{out}}$ from the adder in the previous addition, i.e., $C_{n-1}^{j-1}$, is introduced from the left to register A, see the right angled arrows in \Cref{subtab:ex_acc_seq,subfig:indices}.
		The inputs of the adder are 1) the result of AND\textit{ing} bit B$_{lsb}$ with multiplier $a$, and 2) the previous addition---located in register A, right-shifted once.
		At clock cycle $j\op{=}n$, shift register A contains the $n$ \glspl{msb} and shift register B contains the $n$ \glspl{lsb} of the $2n$-bit product $p$, see \Cref{eq:p_acc}.
		Before introducing a new approximate sequential multiplier, we introduce established error metrics for approximate computations.
		
	
    \subsection{Error metrics}\label{subsec:error_metrics}
%
		
		The following paragraphs summarize error metrics used throughout this paper.
		A formal definition is provided for each metric considering an accurate product $p$ and an approximate product $\hat{p}$.
		These definitions will later be used to derive the formulation of metrics for the error evaluation of the proposed accuracy-configurable sequential multiplier via segmented carry chains.

			%

		When neither a well-accepted error model nor additional knowledge on the design are available, the likelihood of computing an erroneous result is a metric typically used:\\
		
		\begin{itemize}
			\item \textit{\Gls{ber}} is the likelihood of a single bit being erroneous:
		\end{itemize}
		
		\begin{equation}
			\ber(p_i,\hat{p}_i) = \frac{1}{\absv{\Bs^{2n}}}\cdot\sum\limits_{a,b\in \Bs^n}p_i(a,b) \oplus \hat{p}_i(a,b). \label{eqn:def_ber}
		\end{equation}
		
		Arithmetic circuits are generally evaluated based on an \textit{arithmetic error metric}, since the magnitude of its lexicographic deviation can impact the target application greatly:\\
		
		\begin{itemize}
			\item \textit{\Gls{er}} is the likelihood of producing an erroneous result in at least a single bit:
		\end{itemize}
		
		\begin{equation}
			\er(p,\hat{p}) =  \frac{1}{\absv{\Bs^{2n}}}\cdot\sum\limits_{a,b\in \Bs^n} \mathrm{dec}\left(\bigvee\limits_{i=0}^{2n-1}\big(p_i(a,b) \oplus \hat{p}_i(a,b)\big)\right). \label{eqn:def_er}
		\end{equation}
		
%
	
		\begin{itemize}
			\item \textit{\Gls{ed}} is the arithmetic deviation of the approximate output from the accurate result:
		\end{itemize}
		
		\begin{equation}
			\ed\big(p(a,b),\hat{p}(a,b)\big) = \mathrm{dec}\big(p(a,b)\big) - \mathrm{dec}\big(\hat{p}(a,b)\big).\label{eqn:def_ed}
		\end{equation}
		
		\begin{itemize}
			\item \textit{\Gls{mae}} is the largest absolute \gls{ed}:
		\end{itemize}
		
		\begin{equation}
			\mae(p,\hat{p}) = \max\limits_{a,b\in\Bs^n}\left\{\absv{\ed\big(p(a,b),\hat{p}(a,b)\big)}\right\}.\label{eqn:def_mae}
		\end{equation}
		
		\begin{itemize}
			\item \textit{\Gls{med}} is the average \gls{ed} under a given input distribution:
		\end{itemize}
		
		\begin{equation}
			\med(p,\hat{p}) = \frac{1}{\absv{\Bs^{2n}}}\cdot\sum\limits_{a,b\in \Bs^n}\ed\big(p(a,b),\hat{p}(a,b)\big).\label{eqn:def_med}
		\end{equation}
		
		Finally, normalized and relative error metrics allow us to evaluate the arithmetic error across arithmetic units with different bit-widths:\\

		\begin{itemize}
			\item \textit{\Gls{nmed}} is the normalized \gls{med} by the maximum accurate output (cf. \cite{Liu:19.1}):
		\end{itemize}
		
		\begin{equation}
			\nmed(p,\hat{p}) = \frac{\med(p,\hat{p})}{\max\limits_{a,b\in\Bs^n}\{1, \mathrm{dec}\big(p(a,b)\big)\}}.\label{eqn:def_nmed}
		\end{equation}
			
		
		\begin{itemize}
			\item \textit{\Gls{mred}} is the averaged absolute relative \gls{ed} over the accurate result (cf. \cite{Liu:19.1}) under a given input distribution:
		\end{itemize}
		
		\begin{equation}
			\mred(p,\hat{p}) =  \frac{1}{\absv{\Bs^{2n}}}\cdot\sum\limits_{a,b\in \Bs^n}\absv{\frac{\ed(p(a,b),\hat{p}(a,b))}{\max\limits_{a,b\in\Bs^n}\{1, \mathrm{dec}\big(p(a,b)\big)\}}}.\label{eqn:def_mred}
		\end{equation}

\section{Approximate sequential multiplier}\label{sec:proposal}
    In accordance with the advantages of a sequential over a combinatorial design, we propose a circuit applying the concepts of \gls{ac} into the former approach.
    %
    %
    As pointed out in \cite{Chandrasekharan:19}, the authors of \gls{aca} showed that a carry \textit{propagates} only a small fraction of the carry chain in average.
    Based on this observation, approximate addition via segmented carry chains has become a well-investigated methodology \cite{Shafique:15}, resulting in multiple works on its error analysis: most notably works proposing closed form error analysis \cite{Echavarria:16}, and algorithmic approaches \cite{Echavarria:18b}.
    Consequently, the adder within our approximate sequential multiplier is based on carry chain segmentation.
    
    \subsection{Proposed design}\label{subsec:proposed_design}
        \Cref{subfig:sch_app_seq} shows the schematic of the circuit 
        implementing the approximate algorithm proposed in this paper.
        As seen, it is made of two fully accurate adders.
        Segmenting the carry chain of the partial product accumulation with a \gls{lsp} \mbox{$t$-bit} adder and a \gls{msp} \mbox{$(n{-}t)$-bit} adder makes it possible to reduce the latency to \mbox{$\max\{\mathrm{lat}(\text{MSP}),\mathrm{lat}(\text{LSP})\}$}.
        Note that such arithmetic units are not constrained to any approach of addition.
        %
        %
        
        Observe in \Cref{subfig:sch_app_seq} the carry-out of the \gls{lsp} adder connected to the D flip-flop driving the carry-in of the \gls{msp} adder.
        This causes a delay of the carry \textit{propagation} by one clock cycle.
        This situation is described with the example shown in \Cref{subfig:ex_app_seq}.
        The bits marked by the orange rectangle lay on the right side of the splitting point.
        In other words, they represent the \glspl{msb} of the \gls{lsp} adder.
        Examining the resultant accumulation in \mbox{Line 11}, one can observe that a carry \textit{generated} at this location does not immediately \textit{propagate} to the \gls{msp} adder.
        Instead, it becomes the carry-in in the next clock cycle at the same location of the bits marked by the blue rectangle.
        When such a carry is 1 during the last clock cycle, i.e., $C_{t-1}^{n-1}{=}1$, a zero detect signal enables the multiplexers to set all $n{+}t$ \glspl{lsb} to 1, representing the decimal value $2^{n+t}{-}1$, that is, the closest decimal value to the disregarded overflow, thus, considerably reducing the magnitude of the absolute error distance $|\ed|$.
        Such a \textit{fix-to-1} instrumentation reduces the \gls{med} when considering absolute \glspl{ed}.
		However, for error compensation in, for example, cascaded approximate multipliers, it may be disabled to allow for negative \glspl{ed}, and hence, reduce the global \gls{med}.
		
		Our approximate multiplier is formally defined as follows:
        
        {\small
        \begin{align*}
            \hat{S}_i^j &= \begin{cases}
                               a_i \mint b_0,                                                                             & \mathrm{if}\ j = 0\ \mathrm{and}\ i \in [0,n)          \\
                               0,                                                                                         & \mathrm{if}\ j = 0\ \mathrm{and}\ i = n                \\
                               \hat{S}_{i+1}^{j-1} \oplus (a_i \mint b_j),                                                & \mathrm{if}\ j > 0\ \mathrm{and}\ i = 0                \\
                               \hat{S}_{i+1}^{j-1} \oplus (a_i \mint b_j) \oplus \hat{C}_{i-1}^{j-1},                     & \mathrm{if}\ j > 0\ \mathrm{and}\ i = t                \\
                               \hat{S}_{i+1}^{j-1} \oplus \hat{C}_{i-1}^j \oplus (a_i \mint b_j),                         & \mathrm{if}\ j > 0\ \mathrm{and}\ i \in (0,t)\cup(t,n) \\
                               \fixlapp{\hat{C}_{n-1}^j,}                                                                 & \mathrm{if}\ j > 0\ \mathrm{and}\ i = n,
                          \end{cases} \\
            \hat{C}_i^j &= \begin{cases}
                               0,                                                                                         & \mathrm{if}\ j = 0                                     \\
                               \hat{S}_{i+1}^{j-1} \mint (a_i \mint b_j),                                                 & \mathrm{if}\ j > 0\ \mathrm{and}\ i = 0                \\
                               \Big(\left(\hat{S}_{i+1}^{j-1} \oplus (a_i \mint b_j) \right) \mint \hat{C}_{i-1}^{j-1}\Big) \\
                               \maxt \left(\hat{S}_{i+1}^{j-1} \mint a_i \mint b_j  \right),                              & \mathrm{if}\ j > 0\ \mathrm{and}\ i = t                \\
                               \Big(\left(\hat{S}_{i+1}^{j-1} \oplus (a_i \mint b_j) \right) \mint \hat{C}_{i-1}^{j}\Big) \\
                               \maxt \left(\hat{S}_{i+1}^{j-1} \mint a_i \mint b_j  \right),                              & \mathrm{if}\ j > 0\ \mathrm{and}\ i \in (0,t)\cup(t,n).
                           \end{cases}
        \end{align*}}
        
        \noindent Consequently, the approximate product $\hat{p}$ is constructed as follows:
        
        \begin{equation*}\label{eq:p_app}
            \hat{p}_r = \begin{cases}
                            \fixmapp{\hat{S}_0^r,} & \mathrm{if}\ r \in \fixtapp{[0,n{-}1)}\     \mathrm{and}\ \hat{C}_{t-1}^{n-1} = 0 \\
                            1,                                             & \mathrm{if}\ r \in \fixtapp{[0,n{-}1)}\     \mathrm{and}\ \hat{C}_{t-1}^{n-1} = 1 \\
                            \hat{S}_{r-n+1}^{n-1}, & \mathrm{if}\ r \in \fixtapp{[n{-}1,t{+}n)}\ \mathrm{and}\ \hat{C}_{t-1}^{n-1} = 0 \\
                            1,                                             & \mathrm{if}\ r \in \fixtapp{[n{-}1,t{+}n)}\ \mathrm{and}\ \hat{C}_{t-1}^{n-1} = 1 \\
                            \hat{S}_{r-n+1}^{n-1}, & \mathrm{if}\ r \in [t{+}n,2n{-}1].
                        \end{cases}
        \end{equation*}
        
    \subsection{Error analysis}\label{subsec:error_analysis}
		The likelihood of an error being \textit{observable} at the \gls{msp} of the approximate product $\hat{p}$, that is, at any output bit $\hat{p}_{2n-1},\ldots,\hat{p}_{n}\op{\equiv}\hat{S}_n^{n-1},\ldots,\hat{S}_1^{n-1}$, can be computed as follows.
		
		Let $S^j$ be the $j$-th accumulated sum, then:
        
        {\small
        \begin{equation}\label{eq:aer}
            \begin{aligned}
                \er(S^j,\hat{S}^j)&=\rho\bigg(\Big(\hat{S}_t^{j-1} \wedge a_{t-1} \wedge b_{j}\Big) \vee\bigg(\bigvee\limits_{i=0}^{t-2}\Big(\hat{S}_{i+1}^{j-1} \wedge a_i \wedge b_j\Big)\\
                                  &\bigwedge\limits_{l=i+1}^{t-1}\Big(\hat{S}_{l+1}^{j-1} \oplus (a_l \wedge b_j)\Big)\bigg)\bigg),
            \end{aligned}
        \end{equation}}
        
        \noindent when $j\op{>}0$, where $\rho(\chi)$ represents the probability of an event $\chi$ to be true.
        The remaining accumulation $j\op{=}0$ does not introduce errors, $\er(S^0,\hat{S}^0)\op{=}0$.
		The term on the right side of the disjunction in \Cref{eq:aer} represents the event of a carry being \textit{generated} anywhere before the \gls{msb} and being \textit{propagated} until and by the \gls{msb} of the \gls{lsp} of the approximate accumulated sum $\hat{S}^j$, and the term on the left side of the disjunction represents the event of a carry being \textit{generated} directly at the \gls{msb} of the \gls{lsp}.
		
		The remaining product bits $\hat{p}_{n-1},\ldots,\hat{p}_0\op{\equiv}\hat{S}_0^{t+1},\ldots,\hat{S}_0^{n-t-2}$ shall then be computed independently.
		Note that the \gls{er} of the \gls{lsp} of $\hat{p}$ and the \gls{ber} of each remaining output bit consider mutually exclusive events.
		%
		%
		The general disjunction rule shall be applied to avoid computing manifold the occurrence of multiple events.
        According to this rule for $n{-}t$ number of events, it can be shown that:
        
        \begin{align}\label{eq:er}
            \begin{split}
                \er(p,\hat{p}) &= \sum\limits_{k=1}^{n-t}(-1)^{k-1} \cdot \sum\limits_{\mathclap{l \in \mathcal{C}_{k,n-t}}}\rho\Big(\er\left(S^{n-1},\hat{S}^{n-1}\right)_{l_1} \neq 0 \\
                               &\cap \ber\left(S_{0}^{n-t-2},\hat{S}_{0}^{n-t-2}\right)_{l_2} \neq 0 \cap {\ldots} \\
                               &\cap \ber\left(S_{0}^{t+1},\hat{S}_{0}^{t+1}\right)_{l_k} \neq 0\Big),
            \end{split}
        \end{align}
        
        \noindent for $n\op{>}4$ and $t\op{\leq}\frac{n}{2}$, where $\mathcal{C}_{k,n-t}$ represents the set of all ordered $k$-tuples $l_1 {<} {\ldots} {<} l_k$ of $\{1,\ldots,n{-}t\}$.
		When a carry out is \textit{generated} within the \gls{lsp} adder at the last clock cycle $n$, it cannot be \textit{propagated} to the \gls{msp} adder as the carry would only be available in the D flipflop in the next clock cycle. 
		As a remedy, Shift Registers A$_{\text{LSP}}$ and B are fixed to 1 right before the last shift-to-right operation which results in fixing the approximated product bits $\hat{p}_{n+t-1,...,1}$ to 1. 
		The goal is to reduce the overall error. 
		
		%
		In addition to the fact that every error introduced in the accumulations during clock cycles $j\op{<}n-1$ are shifted altogether to the right, the \gls{mae} only occurs when there is a carry propagated at bit position $t{-}1$ in the second last partial accumulation $S^{n-2}$ and no carry at all at the same position in the last partial accumulation $S^{n-1}$. 
		Therefore, the probability of the approximated product $\hat{p}(a,b)$ evaluating to an erroneous result with an \gls{ed} as large as the \gls{mae} is:
		
		\begin{equation*}
			\rho\big(\ed(p(a,b),\hat{p}(a,b)) = \mae(p,\hat{p})\big) = \rho\left(\hat{C}_{t-1}^{n-2} \mint \overline{\hat{C}_{t-1}^{n-1}}\right).
		\end{equation*}
		
		Moreover, as the \gls{mae} occurs when $\hat{C}_{t-1}^{n-2}{=}1$ and $\hat{C}_{t-1}^{n-1}{=}0$, an error is introduced at bit position $t$ in the last partial accumulation, that is, $S^{n-1}_t$, with a magnitude of $2^t$.
		Furthermore, during the calculation of $S^{n-1}$, Shift Register B carries the $n{-}1$ \glspl{lsb} of $\hat{p}$, incrementing the magnitude error introduced in $S^{n-1}$ by $2^{n-1}$.
		Finally, the $t{+}1$ \glspl{lsb} are fully accurate whenever there is not a \textit{fix-to-1} operation, reducing the \gls{mae} by $2^{t+1}$:
		
		\begin{equation}\label{eq:mae}
			\mae(p,\hat{p}) = 2^{n+t-1} - 2^{t+1}.
		\end{equation}
		
		The \gls{med} considers all the erroneous results to calculate the average error:
		
		\begin{equation*}
            \med(p,\hat{p})=\left(\es{7}\sum\limits_{\mathclap{\delta\in[0,\mae(p,\hat{p})]}}\delta\es{5}\right)\cdot\left(\es{10}\sum\limits_{\mathclap{\substack{\forall a,b:\\\ed(p(a,b),\hat{p}(a,b))=\delta}}}\bigl(\mathrm{Pr}(a)\cdot\mathrm{Pr}(b)\bigr)\right),
        \end{equation*}
        
		\noindent where $\mathrm{Pr}$ represents the measured \gls{pdf} of a binary number. Note that $\delta$ iterates over all possible \glspl{ed}, with the \gls{ed} defined as follows:
		
		\begin{align*}
			\begin{split}
				\ed(p(a,b),\hat{p}(a,b)) &= \sum\limits_{i=0}^{2n-1} 2^i \cdot \mathrm{dec}\big(p_i(a,b) \oplus \hat{p}_i(a,b)\big) \\
										 &\cdot \mathrm{sgn}\Big(\mathrm{dec}\big(p_i(a,b)\big)-\mathrm{dec}\big(\hat{p}_i(a,b)\big)\Big),
			\end{split}
		\end{align*}
		
		\noindent where $\mathrm{sgn}(x)\op{=}\frac{x}{|x|}$ when $x\op{\in}\mathbb{Z}^{\neq}$, 0 otherwise.

	\newcommand{\abs}[1]{\lvert #1 \rvert}
\newcommand{\B}{\mathbb{B}}

\section{Results}\label{sec:results}
    \subsection{Error complexity}\label{subsec:error_complexity}

        \begin{thm}\label{thm:ber}
			Computing the \gls{ber} is \#P-complete.
			\begin{proof}
				We will show this by reduction to and from the computation of \gls{er}, which is known to be \#P-complete~\cite{Keszocze:18}.
				%
				
				\begin{itemize}
					\item[$\Rightarrow$] For any given functions $p$ and $\hat{p}$ and an index $i$, computing the \gls{ber} can directly be done by computing $\ber(p_i, \hat{p}_i)=\er(p_i, \hat{p}_i)$. This is very trivial and can easily be seen by realizing that the OR in \Cref{eqn:def_er} fully collapses so that only the single XOR from \Cref{eqn:def_ber} remains. We, therefore, have that $\ber \le \er$.
					
					\item[$\Leftarrow$] The overall \gls{er} can be constructed by computing $2n$ different \gls{ber} values by carefully watching out not to count errors multiple times. We make use of the identity $\ber(p, \hat{p})=\ber(p \oplus \hat{p},0)$ to reformulate the \gls{ber} computation. The \gls{er} can then be determined as:
					\begin{align*}
						\er(p,\hat{p}) &= \sum_{i=1}^m \ber\bigg((p_i\oplus \hat{p}_i) \wedge \bigwedge_{j=1}^{i-1} p_j \Leftrightarrow \hat{p}_j,0\bigg),
					\end{align*}
					showing that $\er \le \ber$.
				\end{itemize}
			
				This concludes the proof that computing \gls{ber} is \#P-complete.
			\end{proof}
		\end{thm}

		\begin{thm}
			Computing the \gls{med} and \gls{mred} is \#P-complete.
			
			\begin{proof}
				The proof for the \emph{average case error} in~\cite[Theorem 6]{Keszocze:18} can be applied with only the minor change of not taking the absolute value of the difference of $p$ and $\hat{p}$ to show that computing the \gls{med} is in \#P.
				
				The authors of~\cite{Keszocze:18} further present a framework for easily determining the complexity of computing error metrics having a certain structure.
				The idea is to employ so-called \emph{rating functions} $\phi$ that describe the error metric to be used in generalized error metrics.
				The \gls{gmed} is defined as:
				\begin{equation*}
					\gmed_{\phi,\zeta} := \zeta \cdot \sum_{x\in \B^{n}} \phi(f(x), \hat{f}(x)),
				\end{equation*}
				\noindent with a scaling factor $\zeta$.
				
				If the function $\phi$ is in the class NC\textsuperscript{1}---roughly speaking: a circuit realizing the function can be created polynomially; see~\cite{Vollmer:07} for details---computing the generalized metric has the complexity of \#P.

				Using the rating function:
				\begin{equation*}
					\phi(a, b) = \frac{\mathrm{dec}(a) - \mathrm{dec}(b)}{\max\{1, \abs{\mathrm{dec}(a)}\}},
				\end{equation*}
				we can write \gls{mred} as:
				\begin{equation*}
					\mred(p,\hat{p}) := \gmed_{\phi,\frac{1}{2^n}}(p,\hat{p}).
				\end{equation*}
				
				As $\phi$ is in NC\textsuperscript{1}, this proves that determining the \gls{mred} has the complexity of \#P.
			\end{proof}
		\end{thm}

    \begin{figure}[]
        \addtocounter{figure}{1}
        \pgfplotsset{yticklabel style = {text width = 3em,align = right}}
        \addtocounter{subfigure}{-2}
        \captionsetup[subfigure]{labelformat=parens}
        \centering
        \begin{tikzpicture}
    \begin{semilogxaxis}[
        width            = 0.75\columnwidth,
        height           = 4cm,
        xlabel           = {},
        ylabel           = {$\er$},
        ylabel style     = {yshift = -2mm},
        grid             = both,
        grid style       = dotted,
        ytick            = {0.0, 0.2, 0.4, 0.6, 0.8, 1},
        xtick            = {4, 8, 16, 32},
        xticklabels      = {4, 8, 16, 32},
        enlarge x limits = {abs = 0.2cm},
        enlarge y limits = {abs = 0.2cm},
        xmin             = 4,
        xmax             = 32,
        ymin             = 0.0,
        ymax             = 1,
        name             = ER_plot,
        set layers,
    ]
    
        \addplot[
            only marks,
            solid,
            mark         = triangle*,
            draw         = my_blue,
            fill         = my_blue,
            fill opacity = 0.15,
        ] table [
            x = n,
            y = ER,
            col sep = tab,
        ] {results/results_metrics_4_16.dat};
        
        \addplot[
            only marks,
            every mark/.append style = {
                solid,
                draw         = my_blue,
                fill         = my_blue,
                fill opacity = 0.15,
                rotate       = 180,
            },
            mark = triangle*,
        ] table [
            x = n,
            y = ER,
            col sep = tab,
        ] {results/results_sim_metrics_18_32.dat};
        
        \foreach \i in {0.05,0.2,0.6,0.65,0.82,0.85,0.88,0.89}
        {
            \edef\temp{\noexpand\filldraw[amethyst,fill opacity = 0.15] (axis cs:8,\i) circle (0.5mm);}\temp
        }
        
        \foreach \i in {0.9309,0.9465,0.9563,0.9602,0.9743,0.9806,0.9901}
        {
            \edef\temp{\noexpand\filldraw[red!75,fill opacity = 0.15] (axis cs:8,\i) circle (0.5mm);}\temp
        }
        
        \foreach \i in {0.9977,0.998,0.9995,0.9997}
        {
            \edef\temp{\noexpand\filldraw[red!75,fill opacity = 0.15] (axis cs:16,\i) circle (0.5mm);}\temp
        }
        
    \end{semilogxaxis}
    
    \node[align=center] at ([xshift=-0.51\columnwidth,yshift=-2.5mm]ER_plot.center) (foo) {\parbox{0\linewidth}{(a)}};
    
\end{tikzpicture}   \\
        \begin{tikzpicture}
    \begin{loglogaxis}[
        width            = 0.75\columnwidth,
        height           = 4cm,
        xlabel           = {},
        ylabel           = {$\mae$},
        ylabel style     = {yshift = -0.2cm},
        grid             = both,
        grid style       = dotted,
        ytick            = {1e+1, 1e+5, 1e+9, 1e+13, 1e+18},
        xtick            = {4, 8, 16, 32},
        xticklabels      = {4, 8, 16, 32},
        enlarge x limits = {abs = 0.2cm},
        enlarge y limits = {abs = 0.2cm},
        xmin             = 4,
        xmax             = 32,
        ymin             = 1e+1,
        ymax             = 1e+18,
        name             = MAE_plot,
        set layers,
    ]
    
        \addplot[
            only marks,
            solid,
            mark         = triangle*,
            draw         = my_blue,
            fill         = my_blue,
            fill opacity = 0.15,
        ] table [
            x = n,
            y = MAE,
            col sep = tab,
        ] {results/results_metrics_4_16.dat};
        
        \addplot[
            only marks,
            solid,
            mark         = triangle*,
            draw         = my_blue,
            fill         = my_blue,
            fill opacity = 0.15,
        ] table [
            x = n,
            y = MAE,
            col sep = tab,
        ] {results/results_sim_metrics_18_32.dat};

        \foreach \i in {2, 10, 18, 50, 82, 114, 242, 370, 498, 626, 1138, 1650, 2162, 4210, 6258, 14450}
        {
            \edef\temp{\noexpand\filldraw[my_green,fill opacity = 0.15] (axis cs:8,\i) circle (0.5mm);}\temp
        }
        
        \foreach \i in {4096,4225,5369,9953,16320,4288,5440,10048,16320,4223,4605,5369,15104,28416}
        {
            \edef\temp{\noexpand\filldraw[red!75,fill opacity = 0.15] (axis cs:8,\i) circle (0.5mm);}\temp
        }
        
        \foreach \i in {2.8e+1,1.02e+3,2.50e+4,1.08e+5,4.56e+5,14.89e+5,53.17e+5,594.85e+5,5949.93e+5}
        {
            \edef\temp{\noexpand\filldraw[my_orange,fill opacity = 0.15] (axis cs:16,\i) circle (0.5mm);}\temp
        }
        
        \foreach \i in {268.44e+6,274.5e+6,280.89e+6,293.47e+6,318.63e+6,469.45e+6,1073.73e+6,520.03e+6,1878.89e+6}
        {
            \edef\temp{\noexpand\filldraw[red!75,fill opacity = 0.15] (axis cs:16,\i) circle (0.5mm);}\temp
        }
        
        \foreach \i in {1.02e+3,4.82e+5,1.43e+8,2.53e+9,3.99e+10,6.88e+11,1.54e+14,2.13e+16,1.62e+18}
        {
            \edef\temp{\noexpand\filldraw[my_orange,fill opacity = 0.15] (axis cs:32,\i) circle (0.5mm);}\temp
        }
        
    \end{loglogaxis}
    
    \node[align=center] at ([xshift=-0.51\columnwidth,yshift=-2.5mm]ER_plot.center) (foo) {\parbox{0\linewidth}{(b)}}; 
    
\end{tikzpicture}  \\
        \begin{tikzpicture}
    \begin{loglogaxis}[
        width            = 0.75\columnwidth,
        height           = 4cm,
        xlabel           = {},
        ylabel           = {$\med$},
        ylabel style     = {yshift = -0.2cm},
        grid             = both,
        grid style       = dotted,
        ytick            = {1e+1,1e+4,1e+7,1e+10,1e+13},
        xtick            = {4, 8, 16, 32},
        xticklabels      = {4, 8, 16, 32},
        enlarge x limits = {abs = 0.2cm},
        enlarge y limits = {abs = 0.2cm},
        xmin             = 4,
        xmax             = 32,
        ymin             = 1e+1,
        ymax             = 1e+13,
        name             = MED_plot,
        set layers,
    ]
    
        \addplot[
            only marks,
            solid,
            mark         = triangle*,
            draw         = my_blue,
            fill         = my_blue,
            fill opacity = 0.15,
        ] table [
            x = n,
            y = MED,
            col sep = tab,
        ] {results/results_metrics_4_16.dat};
        
        \addplot[
           only marks,
            every mark/.append style = {
                solid,
                draw         = my_blue,
                fill         = my_blue,
                fill opacity = 0.15,
                rotate       = 180,
            },
            mark = triangle*,
        ] table [
            x = n,
            y = MED,
            col sep = tab,
        ] {results/results_sim_metrics_18_32.dat};
        
        \foreach \i in {0.3e+2,0.31e+2,6e+2,7.5e+2,12.5e+2,14e+2,37.5e+2,38.5e+2}
        {
            \edef\temp{\noexpand\filldraw[amethyst,fill opacity = 0.15] (axis cs:8,\i) circle (0.5mm);}\temp
        }
        
        \foreach \i in {4.60e+2,2.70e+2,6.00e+1,6.00e+2,7.00e+1}
        {
            \edef\temp{\noexpand\filldraw[pink,fill opacity = 0.15] (axis cs:8,\i) circle (0.5mm);}\temp
        }
        
        \foreach \i in {1.4e+8,4.38e+6,6.68e+6,5.87e+5,1.54e+7,2.44e+7}
        {
            \edef\temp{\noexpand\filldraw[pink,fill opacity = 0.15] (axis cs:16,\i) circle (0.5mm);}\temp
        }
        
        \foreach \i in {12884508.675,639071630.28,8589672.45,165351194.6625,112954192.7175,66999445.11,52826485.5675,42089395.005}
        {
            \edef\temp{\noexpand\filldraw[brown,fill opacity = 0.15] (axis cs:16,\i) circle (0.5mm);}\temp
        }
        
        
    \end{loglogaxis}
    
    \node[align=center] at ([xshift=-0.51\columnwidth,yshift=-2.5mm]ER_plot.center) (foo) {\parbox{0\linewidth}{(c)}}; 
    
\end{tikzpicture}  \\
        \begin{tikzpicture}
    \begin{loglogaxis}[
        width            = 0.75\columnwidth,
        height           = 4cm,
        xlabel           = {},
        ylabel           = {$\nmed$},
        ylabel style     = {yshift = -0.2cm},
        grid             = both,
        grid style       = dotted,
        ytick            = {1e-17,1e-13,1e-9,1e-5,1e-1},
        xtick            = {4, 8, 16, 32},
        xticklabels      = {4, 8, 16, 32},
        enlarge x limits = {abs = 0.2cm},
        enlarge y limits = {abs = 0.2cm},
        xmin             = 4,
        xmax             = 32,
        ymin             = 1e-17,
        ymax             = 1e-1,
        name             = NMED_plot,
        set layers,
    ]
    
        \addplot[
            only marks,
            solid,
            mark         = triangle*,
            draw         = my_blue,
            fill         = my_blue,
            fill opacity = 0.15,
        ] table [
            x = n,
            y = NMED,
            col sep = tab,
        ] {results/results_metrics_4_16.dat};
        
        \addplot[
            only marks,
            every mark/.append style = {
                solid,
                draw         = my_blue,
                fill         = my_blue,
                fill opacity = 0.15,
                rotate       = 180,
            },
            mark = triangle*,
        ] table [
            x = n,
            y = NMED,
            col sep = tab,
        ] {results/results_sim_metrics_18_32.dat};
        
        \foreach \i in {7.69e-6,3.84e-5,1.92e-4,4.38e-4,9.30e-4,2.41e-3,6.34e-3,8.31e-3,5.54e-2}
        {
            \edef\temp{\noexpand\filldraw[my_green,fill opacity = 0.15] (axis cs:8,\i) circle (0.5mm);}\temp
        }
        
        \foreach \i in {0.93e-2,0.9e-2,1.01e-2,2.16e-2,4.11e-2,0.87e-2,1.01e-2,2.22e-2,4.15e-2,0.81e-2,0.78e-2,0.85e-2,2.61e-2,5.46e-2}
        {
            \edef\temp{\noexpand\filldraw[red!75,fill opacity = 0.15] (axis cs:8,\i) circle (0.5mm);}\temp
        }
        
        \foreach \i in {0.007,0.0042,0.0028,0.0016,0.0009,0.0092,0.0053,0.0033,0.0019,0.001}
        {
            \edef\temp{\noexpand\filldraw[pink,fill opacity = 0.15] (axis cs:8,\i) circle (0.5mm);}\temp
        }
        
        \foreach \i in {2.79e-9,8.43e-8,1.68e-6,1.16e-5,3.62e-5,13.3e-5,50.8e-5,703e-5,4420e-5}
        {
            \edef\temp{\noexpand\filldraw[my_orange,fill opacity = 0.15] (axis cs:16,\i) circle (0.5mm);}\temp
        }
        
        \foreach \i in {9.256e-3,13.634e-3,41.636e-3,54.596e-3}
        {
            \edef\temp{\noexpand\filldraw[red!75,fill opacity = 0.15] (axis cs:16,\i) circle (0.5mm);}\temp
        }
        
        \foreach \i in {3.59e-3,1.02e-3,3.43e-4,3.02e-5,5.1e-4,1.24e-3}
        {
            \edef\temp{\noexpand\filldraw[pink,fill opacity = 0.15] (axis cs:16,\i) circle (0.5mm);}\temp
        }
        
        \foreach \i in {1.93e-17,7.92e-15,2.54e-12,4.62e-11,6.78e-10,103e-10,21400e-10,4.16e-4,1.05e-1}
        {
            \edef\temp{\noexpand\filldraw[my_orange,fill opacity = 0.15] (axis cs:32,\i) circle (0.5mm);}\temp
        }
        
        \foreach \i in {3.59e-3,3.42e-4,3.03e-5,5.1e-4,1e-3,4e-4,3.87e-3}
        {
            \edef\temp{\noexpand\filldraw[pink,fill opacity = 0.15] (axis cs:32,\i) circle (0.5mm);}\temp
        }
        
    \end{loglogaxis}
    
    \node[align=center] at ([xshift=-0.51\columnwidth,yshift=-2.5mm]ER_plot.center) (foo) {\parbox{0\linewidth}{(d)}}; 
    
\end{tikzpicture} \\
        \begin{tikzpicture}
    \begin{loglogaxis}[
        width            = 0.75\columnwidth,
        height           = 4cm,
        xlabel           = {$\text{Bitwidth }n$},
        ylabel           = {$\mred$},
        ylabel style     = {yshift = -0.2cm},
        grid             = both,
        grid style       = dotted,
        ytick            = {1e-11, 1e-7, 1e-3, 1e+1, 1e+5},
        xtick            = {4, 8, 16, 32},
        xticklabels      = {4, 8, 16, 32},
        enlarge x limits = {abs = 0.2cm},
        enlarge y limits = {abs = 0.2cm},
        xmin             = 4,
        xmax             = 32,
        ymin             = 1e-11,
        ymax             = 1e+5,
        name             = MRED_plot,
        set layers,
    ]
    
        \addplot[
            only marks,
            solid,
            mark         = triangle*,
            draw         = my_blue,
            fill         = my_blue,
            fill opacity = 0.15,
        ] table [
            x = n,
            y = MRED,
            col sep = tab,
        ] {results/results_metrics_4_16.dat};
        
        \addplot[
            only marks,
            every mark/.append style = {
                solid,
                draw         = my_blue,
                fill         = my_blue,
                fill opacity = 0.15,
                rotate       = 180,
            },
            mark = triangle*,
        ] table [
            x = n,
            y = MRED,
            col sep = tab,
        ] {results/results_sim_metrics_18_32.dat};
        
        \foreach \i in {0.15,0.2,5,6.5,11,12,20,22}
        {
            \edef\temp{\noexpand\filldraw[amethyst,fill opacity = 0.15] (axis cs:8,\i) circle (0.5mm);}\temp
        }
        
        \foreach \i in {3.76e-2,7.88e-2,14.42e-2,16.79e-2,23.28e-2}
        {
            \edef\temp{\noexpand\filldraw[red!75,fill opacity = 0.15] (axis cs:8,\i) circle (0.5mm);}\temp
        }
        
        \foreach \i in {0.0283,0.0218,0.017,0.0117,0.0079,0.0384,0.028,0.0209,0.0141,0.0096}
        {
            \edef\temp{\noexpand\filldraw[pink,fill opacity = 0.15] (axis cs:8,\i) circle (0.5mm);}\temp
        }
        
        \foreach \i in {7.15e-7,2.49e-4,3.37e-3,1.39e-2,9.94e-2,12.4e-2,52e-2,1021e-2,1133e-2}
        {
            \edef\temp{\noexpand\filldraw[my_orange,fill opacity = 0.15] (axis cs:16,\i) circle (0.5mm);}\temp
        }
        
        \foreach \i in {3.85e-2,5.43e-2,15.1e-2,16.98e-2,22.43e-2}
        {
            \edef\temp{\noexpand\filldraw[red!75,fill opacity = 0.15] (axis cs:16,\i) circle (0.5mm);}\temp
        }
        
        \foreach \i in {1.80e-2,7.55e-3,5.06e-4,3.45e-3}
        {
            \edef\temp{\noexpand\filldraw[pink,fill opacity = 0.15] (axis cs:16,\i) circle (0.5mm);}\temp
        }
        
        \foreach \i in {7.01e-12,2.52e-9,1.16e-6,1.42e-5,1.83e-4,23.2e-4,5350e-4,5.79e+1,6.92e+4}
        {
            \edef\temp{\noexpand\filldraw[my_orange,fill opacity = 0.15] (axis cs:32,\i) circle (0.5mm);}\temp
        }
        
    \end{loglogaxis}
    
    \node[align=center] at ([xshift=-0.51\columnwidth,yshift=-2.5mm]ER_plot.center) (foo) {\parbox{0\linewidth}{(e)}}; 
    
\end{tikzpicture} \\
        \addtocounter{figure}{-1}
        \caption{Relations: \capmark{my_green} \cite{Liu:19.2}; \capmark{pink} \cite{Toan:20}; \capmark{my_orange} \cite{Liu:19.1}; \capmark{red!75} \cite{Liu:18}; \capmark{amethyst} \cite{Guo:20}; \capmark{brown} \cite{Ebrahimi:20}. The symbol \trimark{my_blue} marks our findings determined using exhaustive experimentation. The symbol \trimarkinv{my_blue} marks our findings evaluated using \gls{mc} simulations. Note that multiple points taken from the same source are shown as the referenced works report multiple variations of their architectures for the same bitwidths. Similarly, the multiple markings showcasing our design correspond to designs with different chosen splitting points $t$, $t\op{\in} \{2,\ldots,n/2\}$ in the carry chain.}
        \label{fig:plots_ems}
    \end{figure}

	\subsection{Error estimation}\label{subsec:error_estimation}
        We have empirically 
        found that the general cases of $\hat{S}_i^j$ and $\hat{C}_i^j$, that is:
        
        \begin{align}
            \hat{S}_i^j &= \hat{S}_{i+1}^{j-1} \oplus \hat{C}_{i-1}^j \oplus (a_i \mint b_j),\label{eq:hatSij} \\
            \hat{C}_i^j &= \Big(\left(\hat{S}_{i+1}^{j-1} \oplus (a_i \mint b_j) \right) \mint \hat{C}_{i-1}^{j}\Big) \maxt \left(\hat{S}_{i+1}^{j-1} \mint a_i \mint b_j  \right),\label{eq:hatCij}
        \end{align}
        
        \noindent have well-conditioned controlabilities regardless of their high fanout-reconvergence.
        Simply put, small changes in input variables $\hat{S}_{i+1}^{j-1}$, $\hat{C}_{i-1}^j$, $a_i$, and $b_j$ only slightly affect the probabilities $\rho\big(\hat{S}_i^j\big)$ and $\rho\big(\hat{C}_i^j\big)$, which are required for the calculation of the error metrics previously introduced.
        In an attempt to tackle the \#P-completeness of the \gls{er}, \gls{med}, \gls{nmed}, and \gls{mred} metrics, we propose an approximation of the probabilities $\rho\big(\hat{S}_i^j\big)$ and $\rho\big(\hat{C}_i^j\big)$.
        Consider \Cref{eq:hatSij,eq:hatCij} in their disjunctive normal forms:
        
        \begin{align*}
            \begin{split}
                \hat{S}_i^j =     &\Big(\overline{\hat{S}_{i+1}^{j-1}} \mint           \hat{C} _{i-1}^j \mint \overline{a_i} \Big) \maxt \Big(\overline{\hat{S}_{i+1}^{j-1}} \mint           \hat{C} _{i-1}^j \mint \overline{b_j} \Big) \maxt \Big(\overline{\hat{S}_{i+1}^{j-1}} \mint \overline{\hat{C}_{i-1}^j} \mint a_i \mint b_j \Big) \\
                            \maxt &\Big(          \hat{S} _{i+1}^{j-1} \mint \overline{\hat{C}_{i-1}^j} \mint \overline{a_i} \Big) \maxt \Big(          \hat{S} _{i+1}^{j-1} \mint \overline{\hat{C}_{i-1}^j} \mint \overline{b_j} \Big) \maxt \Big(          \hat{S} _{i+1}^{j-1} \mint           \hat{C} _{i-1}^j \mint a_i \mint b_j \Big),\\[2.5mm]
                \hat{C}_i^j =     &\Big(\hat{C}_{i-1}^j \mint a_i \mint b_j\Big) \maxt \Big(\hat{S}_{i+1}^{j-1} \mint a_i \mint b_j\Big) \maxt \Big(\hat{S}_{i+1}^{j-1} \mint \hat{C}_{i-1}^j\Big).
            \end{split}
        \end{align*}
        
        Observe that we may define:
        
        \begin{align*}
            \hat{\rho}\left(\hat{S}_{i}^{j}\big\rvert_{\mathcal{V}}\right) &= \hat{\rho}\left(\overline{\hat{S}_{i+1}^{j-1}}\big\rvert_{\{\overline{a}_{i}\} \cup \mathcal{V}}\right){\cdot}\hat{\rho}\left(\hat{C}_{i-1}^{j}\big\rvert_{\{\overline{a}_{i}\} \cup \mathcal{V}}\right){\cdot}\hat{\rho}\left(\overline{a}_{i}\big\rvert_{\mathcal{V}}\right)\\
                &+\hat{\rho}\left(\hat{S}_{i+1}^{j-1}\big\rvert_{\{\overline{a}_{i}\} \cup \mathcal{V}}\right){\cdot}\hat{\rho}\left(\overline{\hat{C}_{i-1}^{j}}\big\rvert_{\{\overline{a}_{i}\} \cup \mathcal{V}}\right){\cdot}\hat{\rho}\left(\overline{a}_{i}\big\rvert_{\mathcal{V}}\right)\\
                &+\hat{\rho}\left(\overline{\hat{S}_{i+1}^{j-1}}\big\rvert_{\{a_i\} \cup \mathcal{V}}\right){\cdot}\hat{\rho}\left(\overline{\hat{C}_{i-1}^{j}}\big\rvert_{\{a_i\} \cup \mathcal{V}}\right){\cdot}\hat{\rho}\left(a_i\big\rvert_{\mathcal{V}}\right){\cdot}\hat{\rho}\left(b_{j}\right)\\
                &+\hat{\rho}\left(\hat{S}_{i+1}^{j-1}\big\rvert_{\{a_i\} \cup \mathcal{V}}\right){\cdot}\hat{\rho}\left(\hat{C}_{i-1}^{j}\big\rvert_{\{a_i\} \cup \mathcal{V}}\right){\cdot}\hat{\rho}\left(a_i\big\rvert_{\mathcal{V}}\right){\cdot}\hat{\rho}\left(b_{j}\right)\\
                &+\hat{\rho}\left(\hat{S}_{i+1}^{j-1}\right),\\[2.5mm]
            \hat{\rho}\left(\hat{C}_i^j\big\rvert_{\mathcal{V}}\right) &= \hat{\rho}\left(\hat{C}_{i-1}^{j}\big\rvert_{\{a_i\} \cup \mathcal{V}}\right){\cdot}\hat{\rho}\left(a_i\big\rvert_{\mathcal{V}}\right){\cdot}\hat{\rho}\left(b_j\right)\\
                &+\hat{\rho}\left(\hat{S}_{i+1}^{j-1}\big\rvert_{\{a_i\} \cup \mathcal{V}}\right){\cdot}\hat{\rho}\left(a_i\big\rvert_{\mathcal{V}}\right){\cdot}\hat{\rho}\left(b_j\right)\\
                &+\hat{\rho}\left(\hat{S}_{i+1}^{j-1}\big\rvert_{\mathcal{V}}\right){\cdot}\hat{\rho}\left(\hat{C}_{i-1}^{j}\big\rvert_{\mathcal{V}}\right),
        \end{align*}
        
        \noindent where $F\big\rvert_\mathcal{V}$ represents the cofactor\footnote{Recall that the cofactor $f\rvert_{x_i}$ of a function $f$ with respect to $x_i$ is the function obtained by assuming $x_i\op{=}1$. Similarly, $f\rvert_{\overline{x}_i}$, is the function obtained by setting $x_i\op{=}0$.} of $F$ w.r.t. every $a_i\op{\in}\mathcal{V}$.
        Note that cofactor $\hat{C}_{i-1}^{j}\big\rvert_{\overline{b_j}}\op{=}1$ may never occur and $\overline{\hat{C}_{i-1}^{j}}\big\rvert_{b_j}\op{=}1$ always holds true, thus making unnecessary considering correlations with $b_j$.
        As the reader may observe, correlations between $\hat{S}_i^j$ and $\hat{C}_i^j$ are disregarded as we only consider cofactors w.r.t. $a_i$, and not among themselves.

\begin{figure}[t!]
	\hspace{-5mm}
	\begin{subfigure}[t]{\columnwidth}
		\hspace{3mm}\trimark{blue_fau} Accurate \quad \capmark{blue_fau} Approximate
		\centering
		\scalebox{0.95}{\begin{tikzpicture}
    \begin{semilogxaxis}[
        view               = {0}{90},
        width              = 0.7\columnwidth,
        height             = 4.5cm,
        colorbar           = true,
        colorbar style     = {
            at             = {(1.09,1)},
            ylabel style   = {rotate = 180},
            ylabel         = {Power $[mW]$},
            ymin           = 3,
            ymax           = 36,
            every y tick scale label/.style = {
                at     = {(yticklabel* cs:1.01,0cm)},
                anchor = near yticklabel
            },
        },
        xmin               = 27,
        xmax               = 1025,
        ytick              = {1, 2, 3, 4, 5},
        ymin               = 0.89,
        ymax               = 5.11,
        ylabel shift       = 11pt,
        xlabel             = {$\text{LUTs}$},
        ylabel             = {$\text{Latency}\ [ns]$},
        grid               = both,
        grid style         = dotted,
        colormap           = {newmap}{
            color(1cm)     = (heatmap_imp_1);
            color(2cm)     = (heatmap_imp_2);
            color(4cm)     = (heatmap_imp_3);
            color(7cm)     = (heatmap_imp_4);
            color(10cm)    = (heatmap_imp_5);
        },
        set layers,
    ]

        
        \addplot3[
            scatter,
            only marks,
            thick,
            mark         = triangle*,
            fill opacity = 0.25,
            mark options = {scale = 1.5},
        ] table [
            x = luts,
            y = ns,
            z = mw,
            col sep = tab,
        ] {results/results_FPGA_acc_seq.dat};
    
        \addplot3[
            scatter,
            only marks,
            thick,
            mark         = *,
            fill opacity = 0.25,
            mark options = {scale = 1.5},
        ] table [
            x = luts,
            y = ns,
            z = mw,
            col sep = tab,
        ] {results/results_FPGA_app_seq.dat};
    
    \end{semilogxaxis}
\end{tikzpicture}}%
		\hspace{-10mm}
		\caption{}
		\label{subfig:fpga_measurements}
	\end{subfigure}\vspace{5mm}
	
	\hspace{-6mm}
	\begin{subfigure}[t]{\columnwidth}
		\hspace{5mm}\trimark{blue_fau} Accurate \quad \capmark{blue_fau} Approximate
		\centering
		\scalebox{0.95}{\begin{tikzpicture}
    \begin{semilogxaxis}[
        view               = {0}{90},
        width              = 0.7\columnwidth,
        height             = 4.5cm,
        colorbar           = true,
        colorbar style     = {
            ylabel style   = {rotate = 180},
            ylabel         = {Power $[mW]$},
            every y tick scale label/.style = {
                at     = {(yticklabel* cs:1.01,0cm)},
                anchor = near yticklabel
            },
        },
        ytick              = {0.25, 0.5, 0.75, 1, 1.25},
        yticklabels        = {0.25, 0.50, 0.75, 1.00, 1.25},
        ymin               = 0.25,
        ymax               = 1.25,
        xlabel             = {$\text{Area}\ [\mu m^2]$},
        ylabel             = {$\text{Latency}\ [ns]$},
        grid               = both,
        grid style         = dotted,
        colormap           = {newmap}{
            color(1cm)     = (heatmap_imp_1);
            color(2cm)     = (heatmap_imp_2);
            color(4cm)     = (heatmap_imp_3);
            color(7cm)     = (heatmap_imp_4);
            color(10cm)    = (heatmap_imp_5);
        },
        set layers,
    ]

        
        \addplot3[
            scatter,
            only marks,
            thick,
            mark         = triangle*,
            fill opacity = 0.25,
            mark options = {scale = 1.5},
        ] table [
            x = um,
            y = ns,
            z = mw,
            col sep = tab,
        ] {results/results_ASIC_acc_seq.dat};
    
        \addplot3[
            scatter,
            only marks,
            thick,
            mark         = *,
            fill opacity = 0.25,
            mark options = {scale = 1.5},
        ] table [
            x = um,
            y = ns,
            z = mw,
            col sep = tab,
        ] {results/results_ASIC_app_seq.dat};
    
    \end{semilogxaxis}
\end{tikzpicture}}%
		\hspace{-10mm}
		\caption{}
		\label{subfig:asic_measurements}
	\end{subfigure}
	\caption{Resources, latency, and power trade-offs for FPGA (\protect\subref{subfig:fpga_measurements}) and ASIC (\protect\subref{subfig:asic_measurements}) implementations. In the sake of fairness the accurate and the approximate implementations were set up to the same clock frequency for each design $n$. The marks represent sequential implementations with different bitwidths $n\op{\in}\{4, 8, 16, 32, 64, 128, 256\}$, each with a splitting point at position $t\op{=}n\op{/}2$. The abscissae are shown in logarithmic scale. The estimations shown in (\protect\subref{subfig:fpga_measurements}) were obtained using Vivado{\tiny\textsuperscript{TM}}. The default balanced synthesis was chosen. The FPGA designs were implemented on a ZYNQ-7 ZC706 Evaluation Board (xc7z045ffg900-2) from Xilinx\textsuperscript{\protect\scalebox{0.5}{\textregistered}} with 218,600 LUTs and 437,200 FFs. The power estimation was performed following a vector-based approach with a set of $2^{16}$ uniform input patterns. The ASIC designs shown in (\protect\subref{subfig:asic_measurements}) were synthesized using Genus Synthesis Solution. The implementation and estimations were obtained using Innovus Implementation System. Both are products of Cadence\textsuperscript{\protect\scalebox{0.5}{\textregistered}}.} 
\end{figure}
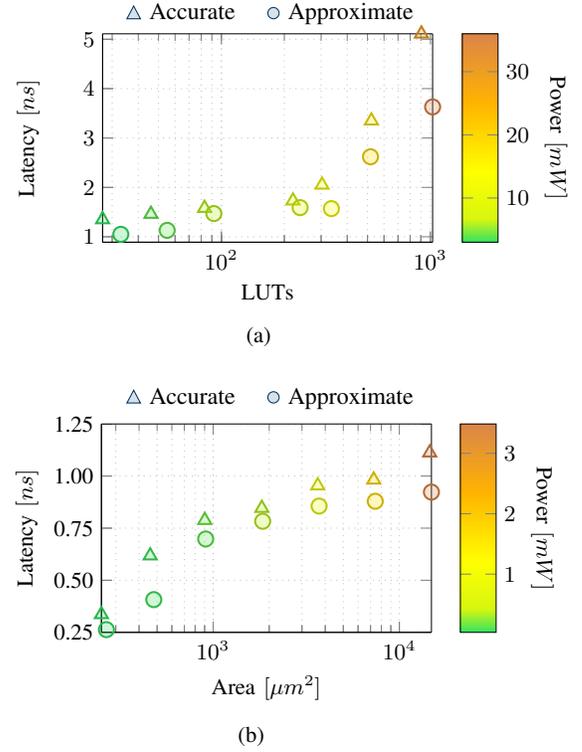

    \subsection{Error evaluation}\label{subsec:error_evaluation}
        Unfortunately, we may not evaluate the \gls{er}, \gls{med}, \gls{nmed}, and \gls{mred} metrics using the formulae presented in \Cref{subsec:error_metrics,subsec:error_analysis} due to their \#P-completeness.
        Therefore, in the sake of fairness when comparing our empirical findings with those from the related work, we performed exhaustive simulations for approximate multipliers with bitwidths $n\op{\leq}16$.
        We opted for \gls{mc} simulations for larger designs using input sets with $2^{32}$ uniformly distributed input patterns.
        \Cref{fig:plots_ems} shows the findings of such evaluations.
        Note that the results shown in \Cref{subfig:plot_MAE} were obtained using the closed-form formula from \Cref{eq:mae}.
        The abscissae of \Cref{fig:plots_ems} were scaled logarithmically, similarly for the ordinates of \Cref{subfig:plot_MAE,subfig:plot_MED,subfig:plot_NMED,subfig:plot_MRED}.
        We compare our findings with those found in the literature \cite{Liu:18,Liu:19.1,Liu:19.2,Toan:20,Guo:20,Ebrahimi:20}.
        As can be seen, not many attempts have been made to evaluate the \gls{er} and \gls{med} of multipliers with bitwidths larger than 16.
        However, Liu et al. do report their findings after simulating their 32-bit approximate combinatorial multiplier for \gls{mae}, \gls{nmed}, and \gls{mred} in \cite{Liu:19.1}.
        On the one hand, one may notice that our designs do not dominate every design found in the literature.
        However, we may also observe that our approach is within the same range of accuracy as the state-of-the-art methodologies.
        On the other hand, our design takes advantage of the inherent resource savings of sequential over combinatorial multipliers, as we will discuss in the following.

	\subsection{Evaluation of Latency and Power Tradeoffs}\label{subsec:hdw_measurements}
        Our design is platform-agnostic, however, due to the massive parallel processing and reconfiguration capabilities of FPGA technology delivering a great price-performance ratio for the algorithms commonly used in multimedia processing, we report results for ASIC as well as for FPGA implementations.
        In the latter, apart from the available dedicated \gls{dsp} slices, one can place arithmetic processing elements based on \glspl{lut} wherever they are most effective without structural restrictions.
		
        %
        When there are more multipliers in the design than available \gls{dsp} slices, the critical path on the FPGA may traverse one of the multipliers implemented on the \glspl{lut}, thus making the regular FPGA's fabric responsible of the overall latency.
        Therefore, part of our experimentation was performed in \glspl{lut}.
        
        We show in \Cref{subfig:fpga_measurements} the number of LUTs, latency, and power consumption estimated after implementing an Accurate Sequential Multiplier and our Approximate Sequential Multiplier on an FPGA.
        In average, we observed an area overhead of our proposed approximate sequential multiplier w.r.t. a fully accurate sequential multiplier, and consequently, a slight power dissipation overhead of only 3.6\%.
        Yet, when comparing the accurate and the approximate sequential multiplier designs, one may also observe in \Cref{subfig:fpga_measurements} that even though our approximate design has a slight area overhead w.r.t. the accurate sequential multiplier for each value of $n$, it always has a shorter latency by 19.15\% in average and up to 29\%---for $n\op{=}256$.
        Interestingly, we observed---not shown in the figure---that fully combinatorial multipliers with bitwidths smaller than 8 consume fewer resources than sequential approaches.
        This is due to the area overhead of the sequential architectures for such small arithmetic units.
		This, however, amortizes for higher bitwidths with up to 99\%---i.e., the 256-bit architecture---of area savings and reduced dynamic power consumption of the sequential w.r.t. a combinatorial design.
        
        Finally, we show in \Cref{subfig:asic_measurements} the latency, area, and power consumption after implementing a Sequential Accurate Multiplier and our Sequential Approximate Multiplier on a Nangate 45nm Open Cell Library from Silicon Integration Initiative (Si2).
        %
        %
        %
        %
        As can be seen, for the ASIC target, the slightly noticeable resource overhead for the 4- and 8-bit designs vanishes for architectures with greater bitwidths.
        Similarly to our findings after implementing our design on an FPGA, the latency is always reduced due to the shortened carry chains by 16.1\% in average and up to 34.14\%---for $n\op{=}8$.
        Coincidentally, the power dissipation overhead for ASICs is fairly similar to the one observed in \Cref{subfig:fpga_measurements}, only 3.6\%.
        The area overhead, on the other hand, is under 3\%.

	\section{Conclusion}\label{sec:conclusions}
	In this paper, we introduced an approximate sequential multiplier with segmented carry chain and variable accuracy.
	In order to better model and understand the accuracy degradation of our design we formally defined closed-form formulae for the most common error metrics.
	We further discussed the complexity of these error metrics presenting proofs for the \#P-completeness of the \gls{ber}, \gls{med}, and \gls{mred}.
	Finally, in order to ease further comparisons, we implemented multiple ASIC as well as FPGA designs with different bitwidths and splitting points using Verilog and VHDL.
	By means of exhaustive simulations we found that the approach herein presented has an accuracy degradation well within the same ranges as those proposed in the literature.
	At the same time, estimations using well known synthesizers showed that our approximate-multiplication technique has significant latency improvements---of up to 29 and 34.14\% when targeting FPGAs and ASICs, respectively---due to the shortened critical paths achieved by the segmentation of the carry chains with negligible power and area overheads.
	

	\bibliographystyle{IEEEtran}
	\bibliography{Approximate_multiplier}

\begin{thebibliography}{10}
\providecommand{\url}[1]{#1}
\csname url@samestyle\endcsname
\providecommand{\newblock}{\relax}
\providecommand{\bibinfo}[2]{#2}
\providecommand{\BIBentrySTDinterwordspacing}{\spaceskip=0pt\relax}
\providecommand{\BIBentryALTinterwordstretchfactor}{4}
\providecommand{\BIBentryALTinterwordspacing}{\spaceskip=\fontdimen2\font plus
\BIBentryALTinterwordstretchfactor\fontdimen3\font minus
  \fontdimen4\font\relax}
\providecommand{\BIBforeignlanguage}[2]{{%
\expandafter\ifx\csname l@#1\endcsname\relax
\typeout{** WARNING: IEEEtran.bst: No hyphenation pattern has been}%
\typeout{** loaded for the language `#1'. Using the pattern for}%
\typeout{** the default language instead.}%
\else
\language=\csname l@#1\endcsname
\fi
#2}}
\providecommand{\BIBdecl}{\relax}
\BIBdecl

\bibitem{Liu:19.2}
W.~{Liu}, T.~{Zhang}, E.~{McLarnon}, M.~{O'Neill}, P.~{Montuschi}, and
  F.~{Lombardi}, ``Design and analysis of majority logic based approximate
  adders and multipliers,'' \emph{IEEE Transactions on Emerging Topics in
  Computing}, 2019.

\bibitem{Toan:20}
N.~{Van Toan} and J.~{Lee}, ``{FPGA-based Multi-Level Approximate Multipliers
  for High-Performance Error-Resilient Applications},'' \emph{IEEE Access},
  2020.

\bibitem{Liu:19.1}
\BIBentryALTinterwordspacing
W.~Liu, T.~Cao, P.~Yin, Y.~Zhu, C.~Wang, E.~E.~S. Jr., and F.~Lombardi,
  ``Design and analysis of approximate redundant binary multipliers,''
  \emph{{IEEE} Trans. Computers}, vol.~68, no.~6, pp. 804--819, 2019. [Online].
  Available: \url{https://doi.org/10.1109/TC.2018.2890222}
\BIBentrySTDinterwordspacing

\bibitem{Chandrasekharan:19}
A.~Chandrasekharan, D.~Große, and R.~Drechsler, \emph{Design Automation
  Techniques for Approximation Circuits: Verification, Synthesis and
  Test}.\hskip 1em plus 0.5em minus 0.4em\relax {Springer International
  Publishing}, 01 2019.

\bibitem{Shafique:15}
\BIBentryALTinterwordspacing
M.~Shafique, W.~Ahmad, R.~Hafiz, and J.~Henkel, ``A low latency generic
  accuracy configurable adder,'' in \emph{Proceedings of the 52nd Annual Design
  Automation Conference, San Francisco, CA, USA, June 7-11, 2015}, 2015, pp.
  86:1--86:6. [Online]. Available:
  \url{https://doi.org/10.1145/2744769.2744778}
\BIBentrySTDinterwordspacing

\bibitem{Echavarria:16}
\BIBentryALTinterwordspacing
J.~Echavarria, S.~Wildermann, A.~Becher, J.~Teich, and D.~Ziener, ``{FAU: Fast
  and error-optimized approximate adder units on LUT-Based FPGAs},'' in
  \emph{2016 International Conference on Field-Programmable Technology, {FPT}
  2016, Xi'an, China, December 7-9, 2016}, 2016, pp. 213--216. [Online].
  Available: \url{https://doi.org/10.1109/FPT.2016.7929536}
\BIBentrySTDinterwordspacing

\bibitem{Echavarria:18b}
\BIBentryALTinterwordspacing
J.~Echavarria, S.~Wildermann, E.~Potwigin, and J.~Teich, ``Efficient arithmetic
  error rate calculus for visibility reduced approximate adders,''
  \emph{Embedded Systems Letters}, vol.~10, no.~2, pp. 37--40, 2018. [Online].
  Available: \url{https://doi.org/10.1109/LES.2017.2760922}
\BIBentrySTDinterwordspacing

\bibitem{Keszocze:18}
\BIBentryALTinterwordspacing
O.~Keszocze, M.~Soeken, and R.~Drechsler, ``The complexity of error metrics,''
  \emph{Information Processing Letters}, vol. 139, pp. 1--7, 2018. [Online].
  Available:
  \url{http://www.sciencedirect.com/science/article/pii/S0020019018301352}
\BIBentrySTDinterwordspacing

\bibitem{Vollmer:07}
H.~Vollmer, \emph{Introduction to Circuit Complexity: A Uniform Approach},
  2007th~ed., ser. Texts in Theoretical Computer Science. An {EATCS} Series,
  P.~Viaroli, P.~Lasserre, and P.~Campostrini, Eds.\hskip 1em plus 0.5em minus
  0.4em\relax Springer, 1999.

\bibitem{Liu:18}
\BIBentryALTinterwordspacing
W.~Liu, J.~Xu, D.~Wang, C.~Wang, P.~Montuschi, and F.~Lombardi, ``Design and
  evaluation of approximate logarithmic multipliers for low power
  error-tolerant applications,'' \emph{{IEEE} Trans. on Circuits and Systems},
  vol. 65-I, no.~9, pp. 2856--2868, 2018. [Online]. Available:
  \url{https://doi.org/10.1109/TCSI.2018.2792902}
\BIBentrySTDinterwordspacing

\bibitem{Guo:20}
Y.~{Guo}, H.~{Sun}, and S.~{Kimura}, ``{Small-Area and Low-Power FPGA-Based
  Multipliers using Approximate Elementary Modules},'' in \emph{25th Asia and
  South Pacific Design Automation Conference, {ASP-DAC} 2020, Beijing, China,
  January 13-16, 2020}, 2020.

\bibitem{Ebrahimi:20}
Z.~{Ebrahimi}, S.~{Ullah}, and A.~{Kumar}, ``{LeAp: Leading-one Detection-based
  Softcore Approximate Multipliers with Tunable Accuracy},'' in \emph{25th Asia
  and South Pacific Design Automation Conference, {ASP-DAC} 2020, Beijing,
  China, January 13-16, 2020}, 2020.

\end{thebibliography}
	
\end{document}